\documentclass[10pt]{article}
\usepackage{epsfig,amsmath,graphicx,amssymb,overpic,cite,amsthm}

\def\be{\begin{equation}}
\def\ee{\end{equation}}
\def\bee{\begin{eqnarray}}
\def\ene{\end{eqnarray}}
\def\bes{\begin{subequations}}
\def\ees{\end{subequations}}
\def\no{\nonumber}
\def\Re{{\rm Re}\,}

\def\d{\displaystyle}

\begin{document}

\baselineskip=12pt
\renewcommand {\thefootnote}{\dag}
\renewcommand {\thefootnote}{\ddag}
\renewcommand {\thefootnote}{ }

\pagestyle{plain}

\begin{center}
\baselineskip=16pt \leftline{} \vspace{-.3in} {\Large \bf An initial-boundary value problem of
the general three-component  nonlinear Schr\"odinger equation with a $4\times 4$ Lax pair on a finite interval } \\[0.1in]
\end{center}

\vspace{0.01in}
\begin{center}
 Zhenya Yan\footnote{{\it Email address}: zyyan@mmrc.iss.ac.cn}
 \\[0.08in]
{\it \small Key Laboratory of Mathematics Mechanization, Institute
of Systems Science, AMSS, \\ Chinese Academy of Sciences, Beijing 100190, China\\
School of Mathematical Sciences, University of Chinese Academy of Sciences, Beijing 100049, China}
\end{center}

{\baselineskip=15pt


\begin{tabular}{p{16cm}}
 \hline \\
\end{tabular}

\vspace{-0.18in}

\begin{abstract} \small \baselineskip=15pt

     We investigate the initial-boundary value problem for the general three-component nonlinear Schr\"odinger (gtc-NLS) equation with a $4\times 4$ Lax pair on a finite interval by extending the Fokas unified approach. The solutions of the gtc-NLS equation can be expressed in terms of the solutions of a $4\times 4$ matrix Riemann-Hilbert (RH) problem formulated in the complex $k$-plane. Moreover, the relevant jump matrices of the RH problem can be explicitly found via the three spectral functions arising from the initial data, the Dirichlet-Neumann boundary data. The global relation is also established to deduce two distinct but equivalent types of representations (i.e., one by using the large $k$ of asymptotics of the eigenfunctions and another one in terms of the Gelfand-Levitan-Marchenko (GLM) method) for the Dirichlet and Neumann boundary value problems. Moreover,
     the relevant formulae for boundary value problems on the finite interval can reduce to ones on the half-line
 as the length of the interval approaches to infinity. Finally, we also give the linearizable boundary conditions for the GLM representation.


\vspace{0.1in} \noindent {\it Keywords:} Riemann-Hilbert problem; General three-component nonlinear Schr\"odinger equation; Initial-boundary value problem;  Global relation;  Maps between Dirichlet and Neumann problems; Gelfand-Levitan-Marchenko representation \vspace{0.2in}

\end{abstract}

\vspace{-0.05in}
\begin{tabular}{p{16cm}}
  \hline \\
\end{tabular}

\vspace{-0.15in}

\section{Introduction}

\quad In the theory of integrable systems, the powerful inverse scattering transform (IST)~\cite{ist,ist2,soliton} (also called nonlinear Fourier transform) was presented to analytically study the initial value problems of the integrable nonlinear wave  equations starting from the spectral analysis of their associated systems of linear eigenvalue equations (also known as the Lax pair~\cite{lax}). After that, some significant extensions of the IST were gradually developed. For instance, Deift and Zhou~\cite{rh} developed the IST to present the nonlinear steepest descent method to explicitly explore the long-time asymptotics of the Cauchy problems of (1+1)-dimensional integrable nonlinear evolution equations in terms of RH problems. Fokas~\cite{f1} extended the idea of the IST to put forward a unified method studying {\it boundary} value problems of both linear and integrable nonlinear PDEs with Lax pairs~\cite{f2,f3,f4,f5}.
Especially, the Fokas' method can be used to study integrable nonlinear PDEs in terms of  the simultaneous spectral analysis of both parts of the Lax pairs and the global relations among spectral functions. This approach obviously differs from the standard IST in which the spectral analysis of only one part of the Lax pairs was considered~\cite{f4}.

The Fokas' unified method has been used to explore boundary value problems of some physically significant integrable nonlinear evolution equations (NLEEs) with $2\times 2$ Lax pairs on the half-line and the finite interval (e.g., the nonlinear Schr\"odinger equation~\cite{f1, nls1,nls2,nls3,nls4}, the sine-Gordon equation~\cite{sg,sg2}, the KdV equation~\cite{kdv}, the mKdV equation~\cite{mkdv0, mkdv1,mkdv2}, the derivative nonlinear Schr\"odinger equation~\cite{dnls1,dnls2}, Ernst equations~\cite{ernst,ernst2}, and etc.~\cite{glm1,glm2,glm3,glm4,m1,m2,m3})
and ones with $3\times 3$ Lax pairs on the half-line and the finite interval (e.g., ~\cite{le12}, the Degasperis-Procesi equation~\cite{ds}, the Sasa-Satsuma equation~\cite{ss}, the coupled nonlinear Schr\"odinger equations~\cite{cnls1,cnls2,cnls3,cnls4}, and the Ostrovsky-Vakhnenko equation~\cite{os}).

 To the best of our knowledge, there was no report on the initial-boundary value (IBV) problems of integrable NLEEs with $4\times 4$ Lax pairs on the half-line or the finite interval before. The aim of this paper is to develop a methodology for analyzing the IBV problems for integrable NLEEs with $4\times 4$ Lax pairs on a finite interval. The extension will contain some novelties from $2\times 2$ and $3\times 3$ to $4\times 4$ matrix Lax pairs, but
the two key steps of this method~\cite{f1,f2,f3,f4} keep invariant: (i) Finding an integral representation of the solution in terms of a matrix RH problem formulated in the complex $k$-plane ($k$ is a spectral parameter of the associated Lax pair). The integral representation in general
contains the unknown boundary data such that this expression of the solution is not effective yet; (ii) Applying a global relation to consider the unknown boundary values. The representation of the unknown boundary values in general involves the solution of a nonlinear problem. But,
this problem for the linearizable boundary conditions can be ignored since the unknown boundary values can be avoided in terms of only algebraic operations.

In this paper, we will exhibit how steps (i) and (ii) can be actualized for the integrable general three-component nonlinear Schr\"odinger (gtc-NLS) equation with a $4\times 4$ Lax pair~\cite{gnls}
\bee
\label{pnls}
\left\{\begin{array}{l}
\d i q_{1t}+ q_{1xx}-2\left[\alpha_{11}|q_1|^{2}+\alpha_{22}|q_2|^2+\alpha_{33}|q_3|^2
 +2\,\Re(\alpha_{12}\bar{q}_1q_2+\alpha_{13}\bar{q}_1q_3+\alpha_{23}\bar{q}_2q_3)\right]q_1=0, \vspace{0.1in}\\
\d i q_{2t}+ q_{2xx}-2\left[\alpha_{11}|q_1|^{2}+\alpha_{22}|q_2|^2+\alpha_{33}|q_3|^2
 +2\,\Re(\alpha_{12}\bar{q}_1q_2+\alpha_{13}\bar{q}_1q_3+\alpha_{23}\bar{q}_2q_3)\right]q_2=0, \vspace{0.1in}\\
\d i q_{3t}+ q_{3xx}-2\left[\alpha_{11}|q_1|^{2}+\alpha_{22}|q_2|^2+\alpha_{33}|q_3|^2
 +2\,\Re(\alpha_{12}\bar{q}_1q_2+\alpha_{13}\bar{q}_1q_3+\alpha_{23}\bar{q}_2q_3)\right]q_3=0,
 \end{array}\right.
 \ene
 where the complex-valued vector fields $q_j=q_j(x,t),\, j=1,2,3$ are the sufficiently smooth functions defined in the finite region
 $\Omega=\{(x,t)| x\in [0, L],\, t\in [0, T]\}$, with $L>0$ being the length of the interval and $T>0$ being the fixed finite time, the overbar denotes the complex conjugate, $\Re(\cdot)$ denotes the real part, and the six coefficients $\alpha_{ij}$'s $(1\leq i\leq j\leq 3)$ combine a $3\times 3$ Hermitian-unitary matrix
  \bee
\mathcal{M}=\left(\begin{array}{ccc} \alpha_{11} & \alpha_{12} & \alpha_{13}  \\
                            \bar{\alpha}_{12} & \alpha_{22} & \alpha_{23}   \\
                            \bar{\alpha}_{13} & \bar{\alpha}_{23} &   \alpha_{33}  \end{array}\right),\quad
                            \mathcal{M}=\mathcal{M}^{\dag}, \quad \mathcal{M}^2=\mathbb{I}.
  \ene
The gtc-NLS equations contain the group velocity dispersion (GVD, i.e., $q_{jxx}$), self-phase modulation (SPM, e.g., $|q_j|^2q_j$), cross-phase modulation (XPM, e.g., $|q_j|^2q_{s},\, j\not=s$), pair-tunneling  modulation (PTM, e.g., $q_j^2\bar{q}_s,\, j\not=s$),
and three-tunneling  modulation (TTM, e.g., $q_1\bar{q}_2q_3$). System (\ref{pnls}) admits the distinct cases for the six parameters $\alpha_{ij},\, (1\leq i\leq j\leq 3)$ such as
the three-component focusing NLS equation for $\alpha_{jj}=-1$ and $\alpha_{ij}=0$ with $i<j$, the three-component defocusing NLS equation for $\alpha_{jj}=1$ and $\alpha_{ij}=0$ with $i<j$, the three-component mixed NLS equation for $(\alpha_{11}=-1,\, \alpha_{22}=\alpha_{33}=1$) or $(\alpha_{11}=1,\, \alpha_{22}=\alpha_{33}=-1$) and $\alpha_{ij}=0$ with $i<j$, and other general three-component NLS equation. Recently, the three-component defocusing NLS equation with nonzero boundary conditions was studied via the IST~\cite{3nls}.

We would like to investigate the gtc-NLS equation (\ref{pnls}) with the initial-boundary value problems
 \bee\label{ibv}
 \begin{array}{lll}
 {\rm Initial\,\, conditions:} & q_j(x, t=0)=q_{0j}(x),&  j=1,2,3, \vspace{0.1in} \\
 {\rm Dirichlet \,\, boundary \,\, conditions:} & q_j(x=0, t)=u_{0j}(t), & q_j(x=L, t)=v_{0j}(t), \,\,\, j=1,2,3, \vspace{0.1in} \\
 {\rm Neumann \,\, boundary \,\, conditions:} &  q_{jx}(x=0, t)=u_{1j}(t),& q_{jx}(x=L, t)=v_{1j}(t), \,\,\, j=1,2,3,
 \end{array} \ene
where the initial data $q_{0j}(x),\,(j=1,2,3)$, and Dirichlet and Neumann boundary data $u_{0j}(t), \, v_{0j}(t)$ and $u_{1j}(t),\, v_{1j}(t),\, j=1,2,3$ are sufficiently smooth and compatible at points $(x,t)=(0, 0),\, (L, 0)$, respectively.

The rest of this paper is organized as follows. In Sec. 2, we investigate the  spectral analysis of the associated $4\times 4$ Lax pair of Eq.~(\ref{pnls}), such as the eigenfunctions, the jump matrices, and the global relation. Sec. 3 gives the corresponding $4\times 4$ matrix RH problem  by means of the jump matrices obtained in Sec. 2. The global relation is used to establish the map between the Dirichlet and Neumann boundary values in Sec. 4. Particularly, the relevant formulae for boundary value problems on the finite interval can reduce to ones on the half-line
 as the length of the interval approaches to infinity. In Sec. 5, we present the Gelfand-Levitan-Marchenko (GLM) representation of the eigenfunctions in terms of the global relation. Moreover, we also show that the GLM representation is equivalent to one in Sec. 4. Finally, we also give the linearizable boundary conditions for the GLM
representation.

\section{The spectral analysis of a $4\times 4$ Lax pair}

\subsection*{\it 2.1.\,  The exact one-form}

\quad The gtc-NLS system (\ref{pnls}) can be regarded as the compatible condition of a $4 \times 4$ Lax pair~\cite{gnls}
\bee \label{lax}
\left\{\begin{array}{l}
                \psi_x+ik\sigma_4\psi=U(x,t)\psi,    \vspace{0.1in}  \\
                \psi_t+2ik^2\sigma_4\psi=V(x,t,k)\psi,
                 \end{array}\right.
    \ene
where $\psi=\psi(x,t,k)$ is a complex 4$\times$4 matrix-valued or $4\times 1$ column vector-valued  eigenfunction, $k\in \mathbb{C}$ is an iso-spectral parameter, $\sigma_4={\rm diag}(1,1,1,-1)$, and the $4 \times 4$ matrices $U$ and $V$ are defined by
\bee
U(x,t)=\left(\begin{array}{cccc}
            0 & 0 & 0 & q_1(x,t) \vspace{0.05in}\\
            0&  0 & 0 & q_2(x,t) \vspace{0.05in}\\
             0&  0 & 0 & q_3(x,t) \vspace{0.05in}\\
           p_1(x,t) & p_2(x,t) & p_3(x,t) & 0
            \end{array}\right),\quad  V(x,t,k)=2kU(x,t)+V_0(x,t),
            \ene
with $p_1(x,t)=\alpha_{11}\bar{q}_1+\bar{\alpha}_{12}\bar{q}_2+\bar{\alpha}_{13}\bar{q}_3,\,
p_2(x,t)=\alpha_{12}\bar{q}_1+\alpha_{22}\bar{q}_2+\bar{\alpha}_{23}\bar{q}_3,\,
 p_3(x,t)=\alpha_{13}\bar{q}_1+\alpha_{23}\bar{q}_2+\alpha_{33}\bar{q}_3,$ and
\bee V_0(x,t)=-i(U_{x}+U^2)\sigma_4=-i\left(\begin{array}{cccc}
           q_1p_1 & q_1p_2 & q_1p_3  & -q_{1x} \vspace{0.1in}\\
           q_2p_1 & q_2p_2 & q_2p_3 & -q_{2x} \vspace{0.1in}\\
           q_3p_1 & q_3p_2 & q_3p_3 & -q_{3x} \vspace{0.1in}\\
         p_{1x} & p_{2x} & p_{3x} & -(q_1p_1+q_2p_2+q_3p_3)
            \end{array}\right),
                     \ene

Define a new eigenfunction $\mu(x,t,k)$ by
 \bee\label{mud}
 \mu(x,t,k)=\psi(x,t,k)e^{i(kx+2k^2t)\sigma_4},
 \ene
such that the Lax pair (\ref{lax}) becomes the equivalent form for $\mu(x,t,k)$
\bee\label{mulax}
    \left\{     \begin{array}{l}
                 \mu_x+ik[\sigma_4,\mu]= U(x,t)\mu,    \vspace{0.1in}            \\
                 \mu_t+2ik^2[\sigma_4,\mu]=V(x,t,k)\mu,
                 \end{array}\right.
    \ene
where $[\sigma_4, \mu]\equiv\sigma_4\mu-\mu\sigma_4$. Let $\hat{\sigma}_4$ denote the commutator with respect to $\sigma_4$ and the operator acting on a $4\times 4$ matrix $X$ by
$\hat{\sigma}_4X=[\sigma_4, X]$ such that $e^{\hat{\sigma}_4}X=e^{\sigma_4}Xe^{-\sigma_4}$, then the Lax pair (\ref{mulax}) can be written as a full derivative form
\bee
d\left[e^{i(kx+2k^2t)\hat{\sigma}_4}\mu(x,t,k)\right]=W(x,t,k),
\label{dform}
\ene
where the exact one-form $W(x,t,k)$ is of the form
\bee \label{w}
 W(x,t,k)=e^{i(kx+2k^2t)\hat{\sigma}_4}[U(x,t)dx+V(x,t,k)dt]\mu(x,t,k).
 \ene

\subsection*{\it 2.2. \, The definition and boundedness of eigenfunctions $\mu_j's$ }

\quad For any point $(x,t)$ in the region $\Omega=\{(x,t)| x\in [0, L],\, t\in [0, T]\}$,  let $\{\gamma_j\}_1^4$ be four contours connecting
fours vertexes $(x_1, t_1)=(0, T),\, (x_2, t_2)=(0, 0),\, (x_3, t_3)=(L, 0),\, (x_4, t_4)=(L, T)$ to $(x,t)$, respectively (see Fig.~\ref{ga}).
Therefore we get the following inequalities on these contours:
\bee\label{gammad}
 \begin{array}{rll}
 \gamma_1: (0, T)\to (x,t), & x-x' \geq 0, & t-\tau \leq 0, \vspace{0.1in}\\
 \gamma_2: (0, 0)\to (x,t), & x-x' \geq 0, & t-\tau \geq 0, \vspace{0.1in}\\
 \gamma_3: (L, 0)\to (x,t), & x-x' \leq 0, & t-\tau \geq 0, \vspace{0.1in}\\
 \gamma_4: (L, T)\to (x,t), & x-x' \leq 0, & t-\tau \leq 0,
 \end{array}
\ene

\begin{figure}[!t]
\begin{center}
{\scalebox{0.7}[0.7]{\includegraphics{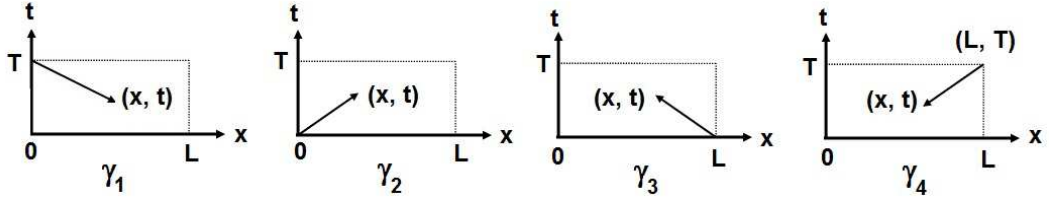}}}
\end{center}
\vspace{-0.25in}\caption{Contours $\gamma_j \, (j=1,2,3,4)$ from points $(x_j, t_j)$ to $(x,t)$ in the region $\Omega=\{(x,t)| x\in [0, L],\, t\in [0, T]\}$. }
\label{ga}
\end{figure}

 By means of the Volterra integral equations, it follows from Eqs.~(\ref{dform}) and (\ref{w}) that we introduce the four eigenfunctions $\{\mu_j\}_1^4$ on the four contours $\{\gamma_j\}_1^4$
\bee \label{mus}
\begin{array}{l}
\mu_j(x,t,k)=\d\mathbb{I}+\int_{(x_j, t_j)}^{(x,t)}e^{-i(kx+2k^2t)\hat{\sigma}_4}W_j(x',\tau,k),
\end{array}
\ene
 where $\mathbb{I}={\rm diag}(1,1,1,1)$, the integral is over a piecewise smooth curve from $(x_j, t_j)$ to $(x,t)$, and $W_j(x,t,k)$ is given by Eq.~(\ref{w}) with $\mu(x,t,k)$ replaced by $\mu_j(x,t,k)$. Since the one-form $W_j$ are closed, thus $\mu_j$
 are independent of the path of integration. If we take the paths of integration to be parallel to the $x$ and $t$ axes, then the integral Eq.~(\ref{mus}) reduces to
\bee\label{musg}
\begin{array}{l}
\mu_j(x,t,k)=\d \mathbb{I}+\int_{x_j}^x e^{-ik(x-x')\hat{\sigma}_4}(U\mu_j)(x',t,k)dx'+e^{-ik(x-x_j)\hat{\sigma}_4}\int_{t_j}^te^{-2ik^2(t-\tau)\hat{\sigma}_4} (V\mu_j)(x_j,\tau,k)d\tau,
   \end{array}
\ene

It follows from Eq.~(\ref{musg}) that the four columns of the matrix $\mu_j(x,t,k)$ contain the following exponentials
\bes
\label{muc}
\bee
& [\mu_j]_s: \,\, e^{2ik(x-x')+4ik^2(t-\tau)},\,\, j=1,2,3,4; s=1,2,3, \hspace{2.0in} \vspace{0.1in}\\
&[\mu_j]_4: \,\, e^{-2ik(x-x')-4ik^2(t-\tau)},\, e^{-2ik(x-x')-4ik^2(t-\tau)}, \, e^{-2ik(x-x')-4ik^2(t-\tau)},\, j=1,2,3,4
 \ene
 \ees
To analyze the bounded regions of the eigenfunctions $\{\mu_j\}_1^4$, we need to use the curve $\{k\in \mathbb{C} |  (\Re f(k))(\Re g(k))=0,\, f(k)=ik,\, g(k)=ik^2\}$ to separate the complex $k$-plane into four regions (see Fig.~\ref{kplane}):
\bee
\begin{array}{l}
D_1=\{k\in\mathbb{C} \,|\, \Re f(k)<0 \,\, {\rm and} \,\,  \Re g(k)<0\}, \vspace{0.1in} \\
D_2=\{k\in\mathbb{C} \,|\, \Re f(k)<0 \,\, {\rm and} \,\,  \Re g(k)>0\}, \vspace{0.1in}\\
D_3=\{k\in\mathbb{C} \,|\,\Re f(k)>0 \,\, {\rm and} \,\, \Re g(k)<0\}, \vspace{0.1in}\\
D_4=\{k\in\mathbb{C} \,|\, \Re f(k)>0\,\, {\rm and} \,\,  \Re g(k)>0\},
\end{array}
\label{d}
\ene
which implies that $D_1$ and $D_3$ ($D_2$ and $D_4$) are symmetric about the origin.

Thus it follows from Eqs.~(\ref{gammad}), (\ref{muc}) and (\ref{d}) that the regions, where the different columns of eigenfunctions $\{\mu_j\}_1^4$ are bounded and analytic in the complex $k$-plane, are presented below:
\bee \label{muregion}
\left\{\begin{array}{l}
 \mu_1: (f_- \cap g_+,\, f_- \cap g_+,\, f_- \cap g_+,\, f_+ \cap g_-)=: (D_2, D_2, D_2, D_3), \vspace{0.1in} \\
 \mu_2: (f_- \cap g_-,\, f_- \cap g_-,\, f_- \cap g_-,\, f_+ \cap g_+)=: (D_1, D_1, D_1, D_4), \vspace{0.1in}\\
 \mu_3: (f_+ \cap g_-,\, f_+ \cap g_-,\, f_+ \cap g_-,\, f_- \cap g_+)=: (D_3, D_3, D_3, D_2), \vspace{0.1in}\\
 \mu_4: (f_+ \cap g_+,\, f_+ \cap g_+,\, f_+ \cap g_+,\, f_- \cap g_-)=: (D_4, D_4, D_4, D_1),
\end{array}\right.
\ene
where $f_+=: \Re f(k)>0,\, f_-=:\Re f(k)<0,\, g_+=: \Re g(k)>0$, and $ g_-=:\Re g(k)<0$.

\begin{figure}[!t]
\begin{center}
{\scalebox{0.2}[0.2]{\includegraphics{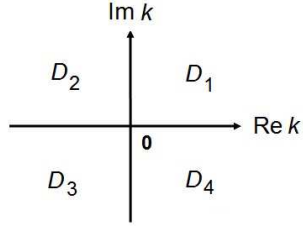}}}
\end{center}
\vspace{-0.25in}\caption{The regions $D_n\, (n=1,2,3,4)$ separating the complex $k$-plane. }
\label{kplane}
\end{figure}

\subsection*{\it 2.3. \, The definition of the new matrix-valued functions $M_n$'s}

\quad To construct the jump matrix in a RH problem, we introduce the solutions $M_n(x,t,k)$ of Eq.~(\ref{mulax})
\bee \label{mn}
(M_n)_{sj}(x,t,k)=\delta_{sj}+\int_{(\gamma^n)_{sj}}\left(e^{-i(kx+2k^2t)\hat{\sigma}_4}W_n(x',\tau,k)\right)_{sj}, \quad k\in D_n,\quad  s,j=1,2,3,4,
\ene
where $W_n(x,t,k)$ isdefined by Eq.~(\ref{w}) with $\mu(x,t,k)$ replaced with $M_n(x,t,k)$, and the contours $(\gamma^n)_{sj}$'s are given by
\bee\label{gamma}
(\gamma^n)_{sj}=\left\{ \begin{array}{l}
\gamma_1,\,\,\, {\rm if} \,\,  \Re f_s(k)> \Re f_j(k) \,\,  {\rm and} \,\, \Re g_s(k) \leq \Re g_j(k), \vspace{0.1in} \\
\gamma_2,\,\,\, {\rm if}  \,\, \Re f_s(k)> \Re f_j(k) \,\, {\rm and} \,\, \Re g_s(k) > \Re g_j(k), \vspace{0.1in}\\
\gamma_ 3,\,\,\, {\rm if} \,\,  \Re f_s(k) \leq \Re f_j(k) \,\, {\rm and}\,\,  \Re g_s(k) \geq \Re g_j(k), \vspace{0.1in}\\
\gamma_4, \,\,\, {\rm if} \,\,  \Re f_s(k)\leq \Re f_j(k)  \,\, {\rm and} \,\, \Re g_s(k)\leq \Re g_j(k),
\end{array}\right.
\ene
for $k\in D_n$, where $f_{1,2,3}(k)=-f_{4}(k)=-ik,\, g_{1,2,3}(k)=-g_{4}(k)=-2ik^2$.

Notice that to distinguish $(\gamma^n)_{sj}$'s to be the contour $\gamma_3$ or $\gamma_4$ for the special cases, $\Re f_s(k)=\Re f_j(k)$ and $\Re g_s(k)=\Re g_j(k)$, we choose them in these cases as $\gamma_3$ (or $\gamma_4)$ which must appear in the matrix $\gamma^{n}$, otherwise, we choose them in all these cases as the same $\gamma_3$ (or $\gamma_4)$.

The definition (\ref{gamma}) of $(\gamma^n)_{sj}$  implies that $\gamma^n\, (n=1,2,3,4)$ are explicitly given by
\bee\begin{array}{rr}
\gamma^1=\left(
\begin{array}{cccc}
 \gamma_4 & \gamma_4 & \gamma_4 & \gamma_2 \\
 \gamma_4 & \gamma_4 & \gamma_4 & \gamma_2 \\
 \gamma_4 & \gamma_4 & \gamma_4 & \gamma_2 \\
 \gamma_4 & \gamma_4 & \gamma_4 & \gamma_4
 \end{array}
\right), &
\gamma^2=\left(
\begin{array}{cccc}
 \gamma_3 & \gamma_3 & \gamma_3 & \gamma_1 \\
 \gamma_3 & \gamma_3 & \gamma_3 & \gamma_1 \\
 \gamma_3 & \gamma_3 & \gamma_3 & \gamma_1 \\
 \gamma_3 & \gamma_3 & \gamma_3 & \gamma_3
 \end{array}
\right), \vspace{0.1in} \\
\gamma^3=\left(
\begin{array}{cccc}
 \gamma_3 & \gamma_3 & \gamma_3 & \gamma_3 \\
 \gamma_3 & \gamma_3 & \gamma_3 & \gamma_3 \\
 \gamma_3 & \gamma_3 & \gamma_3 & \gamma_3 \\
 \gamma_1 & \gamma_1 & \gamma_1 & \gamma_3
 \end{array}
\right), &
\gamma^4=\left(
\begin{array}{cccc}
 \gamma_4 & \gamma_4 & \gamma_4 & \gamma_4 \\
 \gamma_4 & \gamma_4 & \gamma_4 & \gamma_4 \\
 \gamma_4 & \gamma_4 & \gamma_4 & \gamma_4 \\
 \gamma_2 & \gamma_2 & \gamma_2 & \gamma_4
 \end{array}
\right),
\end{array}\ene

\vspace{0.1in}
\noindent {\bf Proposition 2.1.} {\it For the matrix-valued functions $M_n(x,t,k)\, (n=1,2,3,4)$ defined by Eq.~(\ref{mn}) for $k\in \bar{D}_n$ and $(x,t)\in \Omega$, and any fixed point $(x,t)$, $M_n(x,t,k)$'s are the bounded and analytic functions of $k\in D_n$ away from a possible discrete set of singularity $\{k_j\}$ at which the Fredholm determinants vanish. Moreover, $M_n(x,t,k)$'s admit the bounded and continuous extensions to $\bar{D}_n$ and
\bee\label{ml}
 M_n(x,t,k)=\mathbb{I}+O\left(\frac{1}{k}\right),\,\,k\in D_n,\,\ k\to\infty,\,\,  n=1,2,3,4.
\ene}

\noindent {\bf Proof.} Similar to the proof for the $3\times 3$ Lax pair in ~\cite{le12}, we can also proof
the bounedness and analyticity of $M_n$. The substitution of
\bee\no
\mu(x,t,k)=M_n(x,t,k)=M_n^{(0)}(x,t,k)+\sum_{j=1}^{\infty}\frac{M_n^{(j)}(x,t,k)}{k^j},\quad k \to \infty,
\ene
into the $x$-part of the Lax pair (\ref{mulax}) yields Eq.~(\ref{ml}). $\square$ \\

The above-defined matrix-valued functions $M_n$'s can be used to formulate a $4\times 4$ matrix Riemann-Hilbert problem.

\subsection*{\it 2.4.\, The  spectral functions and jump matrices}

\quad We introduce the spectral functions $S_n(k)\, (n=1,2,3, 4)$ by
\bee
 S_n(k)=M_n(x=0,t=0,k), \quad k\in D_n,\quad n=1,2,3,4.
\ene
Let $M(x,t,k)$ denote the sectionally analytic function on the Riemann $k$-spere which is equivalent to $M_n(x,t,k)$ for $k\in D_n$. Then $M(x,t,k)$ solves the jump equations
\bee\label{jumpc}
 M_n(x,t,k)=M_m(x,t,k)J_{mn}(x,t,k),\quad k\in \bar{D}_n\cap \bar{D}_m, \quad n,m=1,2,3,4,\quad n\not= m,
\ene
with the jump matrices $J_{mn}(x,t,k)$ defined by
\bee\label{jump}
J_{mn}(x,t,k)=e^{-i(kx+2k^2t)\hat{\sigma}_4}(S_m^{-1}(k)S_n(k)).
\ene

\subsection*{\it 2.5. \,  The minors or the transpose of the adjugates of eigenfunctions}

\quad To conveniently calculate the spectral functions $S_n(k)$ in the following sections, we need to use the cofactor matrix $X^A$ (or the transpose of the adjugate) of a $4\times 4$ matrix $X$ defined as
\bee\no
{\rm adj}(X)^T=X^A=\left(\begin{array}{rrrr}
 m_{11}(X) & -m_{12}(X) &  m_{13}(X) &  -m_{14}(X) \vspace{0.05in}\\
 -m_{21}(X) & m_{22}(X) &  -m_{23}(X) &  m_{24}(X) \vspace{0.05in}\\
 m_{31}(X) & -m_{32}(X) &  m_{33}(X) &  -m_{34}(X) \vspace{0.05in}\\
 -m_{41}(X) & m_{42}(X) &  -m_{43}(X) &  m_{44}(X)
\end{array}\right),
\ene
where $m_{ij}(X)$ denotes the $(ij)$th minor of $X$ and $(X^A)^TX ={\rm adj}(X) X=\det X$.

It follows from the Lax pair (\ref{lax}) that the eigenfunction $\{\mu_j^A\}_1^4$ of the matrices $\{\mu_j(x,t,k)\}_1^4$ satisfy the Lax equation
\bee\label{mualax}
    \left\{   \begin{array}{l}
                 \mu_{x}^A-ik[\sigma_4,\mu^A]= -U^T(x,t)\mu^A,    \vspace{0.1in}            \\
                 \mu_{t}^A-2ik^2[\sigma_4,\mu^A]=-V^T(x,t,k)\mu^A,
                 \end{array} \right.
                 \ene
whose solutions can be written as the form
\bee
\mu_j^A(x,t,k)=\mathbb{I}-\int_{\gamma_j}e^{i[k(x-x')+2k^2(t-\tau)]\hat{\sigma}_4}\left[U^T(x', \tau)dx'+V^T(x', \tau, k)d\tau\right]
\mu_j^A(x', \tau, k),\quad j=1,2,3,4,
\ene
in terms of the Volterra integral equations.

It is easy to check that the regions of boundedness of $\mu_j^A$:
\bee \no \left\{\begin{array}{l}
 \mu_1^A(x,t,k) {\rm \,\, is \,\, bounded\,\, for\,\,} k\in (D_3, D_3, D_3, D_2), \vspace{0.1in}\\
 \mu_2^A(x,t,k) {\rm \,\, is \,\, bounded\,\, for\,\,} k\in (D_4, D_4, D_4, D_1), \vspace{0.1in}\\
 \mu_3^A(x,t,k) {\rm \,\, is \,\, bounded\,\, for\,\,} k\in (D_2, D_2, D_2, D_3), \vspace{0.1in}\\
 \mu_4^A(x,t,k) {\rm \,\, is \,\, bounded\,\, for\,\,} k\in (D_1, D_1, D_1, D_4).
\end{array}\right.
\ene
which are symmetric ones of $\mu_j$ about the $\Re k$-axis (cf. Eq.~(\ref{muregion})).

\subsection*{\it 2.6.\, Symmetries of eigenfunctions}

\quad Let
\bee
\check{U}(x,t, k)=-ik\sigma_4+U(x,t),\quad \check{V}(x,t, k)=-2ik^2\sigma_4+V(x,t,k).
\ene
in the Lax pair (\ref{lax}). Then we have
\bee\label{mpr}
P\overline{\check{U}(x,t, \bar{k})}P=-\check{U}(x,t,k)^T, \quad P\overline{\check{V}(x,t, \bar{k})}P=-\check{V}(x,t,k)^T,
\ene
where the symmetric matrix $P$ is taken as
\bee\label{mp}
P=\left(\begin{array}{cccc} \alpha_{11} & \bar{\alpha}_{12} & \bar{\alpha}_{13} & 0 \\
                            \alpha_{12} & \alpha_{22} & \bar{\alpha}_{23} & 0 \\
                            \alpha_{13} & \alpha_{23} &   \alpha_{33} & 0  \\
                            0 & 0 & 0 & -1
                           \end{array}\right),
 \quad P^2=\mathbb{I},\quad P=P^{\dag},
\ene

Notice that the symmetric matrix $P$ used here  differs from the diag ones used in $3\times 3$ Lax pairs~\cite{cnls1,cnls2,cnls3,cnls4}.

Similar to the proof in Ref.~\cite{nls3}, based on Eq.~(\ref{mualax}) and (\ref{mpr}) we have the following proposition:

\vspace{0.1in}
\noindent {\bf Proposition 2.2.} {\it The matrix-valued eigenfunctions $\psi(x,t,k)$ of the Lax pair (\ref{lax}) and $\mu_j(x,t,k)$ of the Lax pair (\ref{mulax}) both possess the same symmetric relations
\bee\label{symmetry}
\begin{array}{l}
\psi^{-1}(x,t,k)=P\overline{\psi(x,t,\bar{k})}^TP,\quad
 \mu_j^{-1}(x,t,k)=P\overline{\mu_j(x,t,\bar{k})}^TP,\quad j=1,2,3,4,
\end{array}
\ene }
\begin{figure}[!t]
\begin{center}
{\scalebox{0.3}[0.3]{\includegraphics{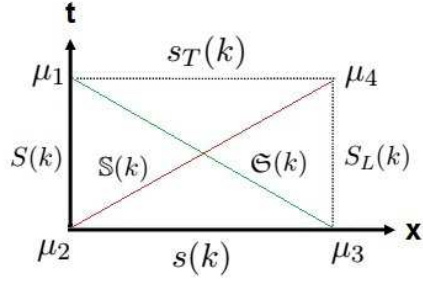}}}
\end{center}
\vspace{-0.3in}\caption{The relations among $\mu_j(x,t,k),\, j=1,2,3, 4$. }
\label{mu}
\end{figure}

Moreover, In the domains where $\mu_j$ is bounded, we have
\bee
 \mu_j(x,t,k)=\mathbb{I}+O\left(\frac{1}{k}\right),\quad k\to \infty, \quad j=1,2,3, 4
\ene
and
\bee
 {\rm det} [\mu_j(x,t,k)]=1, \quad j=1,2,3, 4
\ene
since the traces of the matrices  $U(x,t, k)$ and $V(x,t,k)$ are zero.

\subsection*{\it 2.7. \, The relations between spectral functions and jump matrices $J_{mn}$}

\quad Since these functions $\mu_j$ are dependent, thus we can define three $4\times 4$ matrix-valued functions $S(k),\, s(k)$ and $\mathbb{S}(k)$
between $\mu_2$ and $\mu_j,\, j=1,3,4$ in the form (cf. Fig.~\ref{mu})
\bee\label{mu1234}\left\{\begin{array}{l}
\mu_1(x,t,k)=\mu_2(x,t,k)e^{-i(kx+2k^2t)\hat{\sigma}_4}S(k), \vspace{0.1in}\\
\mu_3(x,t,k)=\mu_2(x,t,k)e^{-i(kx+2k^2t)\hat{\sigma}_4}s(k), \vspace{0.1in}\\
\mu_4(x,t,k)=\mu_2(x,t,k)e^{-i(kx+2k^2t)\hat{\sigma}_4}\mathbb{S}(k),
\end{array} \right.
\ene

Evaluating  system (\ref{mu1234}) at $(x,t)=(0,0)$ and the three equations in system (\ref{mu1234}) at $(x, t)=(0, T), (L, 0), (L, T)$, respectively, we have
\bee\label{sss}
\left\{\begin{array}{l}
 S(k)=\mu_1(0,0,k)=e^{2ik^2T\hat{\sigma}_4}\mu_2^{-1}(0,T,k),\vspace{0.1in}\\
 s(k)=\mu_3(0,0,k)=e^{ikL\hat{\sigma}_4}\mu_2^{-1}(L,0,k),\vspace{0.1in}\\
\mathbb{S}(k)=\mu_4(0,0,k)=e^{i(kL+2k^2T)\hat{\sigma}_4}\mu_2^{-1}(L,T,k),
\end{array} \right.
\ene

Except for the defined three relations, it follows from Eqs.~(\ref{mu1234}) and (\ref{sss}) that we can find other three  relations:

(i) the relation between $\mu_3(x,t,k)$ and $\mu_4(x,t,k)$
\bee \no
\begin{array}{rl}
\mu_4(x,t,k)=&\!\!\! \mu_3(x,t,k)e^{-i[k(x-L)+2k^2(t-T)]\hat{\sigma}_4}\mu_3^{-1}(L, T, k)=\mu_3(x,t,k)e^{-i[k(x-L)+2k^2t]\hat{\sigma}_4}S_L(k),
\end{array}
\ene
with
\bee \label{sl}
\label{sr} S_L(k)=\mu_4(L, 0,k)=e^{2ik^2T\hat{\sigma}_4}\mu_3^{-1}(L, T, k),
\ene
(ii) the relation between $\mu_1(x,t,k)$ and $\mu_4(x,t,k)$
\bee\no \begin{array}{rl}
\mu_3(x,t,k)=&\!\!\! \mu_1(x,t,k)e^{-i(kx+2k^2t)\hat{\sigma}_4}\mathfrak{S}(k),
\end{array}
\ene
with
\bee \label{gss}
\mathfrak{S}(k)=S^{-1}(k)s(k),
\ene
and (iii) the relation between $\mu_1(x,t,k)$ and $\mu_4(x,t,k)$
\bee\no \begin{array}{rl}
\mu_4(x,t,k)=&\!\!\! \mu_1(x,t,k)e^{-i(kx+2k^2t)\hat{\sigma}_4}s_T(k),
\end{array}
\ene
with
\bee\label{st}
s_T(k)=S^{-1}(k)\mathbb{S}(k),
\ene

It follows from Eqs.~(\ref{sss}) and (\ref{sl}) that we have the relation
\bee
\mathbb{S}(k)=s(k)e^{ikL\hat{\sigma}_4}S_L(k),
\ene
The map of these relations among $\mu_j$ is exhibited in Fig.~\ref{mu}.

According to the definition (\ref{musg}) of $\mu_j$, Eq.~(\ref{sss}) and (\ref{sl}) imply that
\bee\label{musg1}
\begin{array}{rl}
s(k)=&\!\!\! \d \mathbb{I}-\int_0^L e^{ikx'\hat{\sigma}_4}(U\mu_3)(x',0,k)dx' =\left[\mathbb{I}+\int_0^L e^{ikx'\hat{\sigma}_4}(U\mu_2)(x', 0,k)dx'\right]^{-1} ,  \vspace{0.1in}\\
S(k)=&\!\!\!\d \mathbb{I}-\int_0^T e^{2ik^2\tau\hat{\sigma}_4}(V\mu_1)(0, \tau,k)dx' =\left[\mathbb{I}+\int_0^T e^{2ik^2\tau\hat{\sigma}_4}(V\mu_2)(0,\tau,k)d\tau\right]^{-1},  \vspace{0.1in}\\
S_L(k)=&\!\!\! \d \mathbb{I}-\int_0^T e^{2ik^2\tau\hat{\sigma}_4}(V\mu_4)(L,\tau,k)d\tau=\left[\mathbb{I}+\int_0^T e^{2ik^2\tau\hat{\sigma}_4}(V\mu_3)(L,\tau,k)d\tau\right]^{-1}, \vspace{0.1in}\\
\mathbb{S}(k)=&\!\!\!\d \mathbb{I}-\int_0^L e^{ikx'\hat{\sigma}_4}(U\mu_4)(x',0,k)dx'
      -e^{ikL\hat{\sigma}_4}\int_0^T e^{2ik^2\tau\hat{\sigma}_4}(V\mu_4)(L,\tau,k)d\tau\vspace{0.1in}\\
=&\!\!\!\d \left[\mathbb{I}+e^{2ik^2T\hat{\sigma}_4}\int_0^L e^{ikx'\hat{\sigma}_4}(U\mu_2)(x',T,k)dx'+\int_0^T e^{2ik^2\tau\hat{\sigma}_4}(V\mu_2)(0,\tau,k)d\tau\right]^{-1},
  \end{array}
\ene
which leads to $\mathfrak{S}(k)$ and $s_T(k)$ in terms of Eqs.~(\ref{gss}) and (\ref{st}), where $\mu_{j_2}(0,t,k),\ j_2=1,2,\, \mu_{j_3}(L, t, k),\, j_3=3,4,\, \mu_{j_1}(x,0,k),\, j_1=2,3,4,\,
\mu_2(x,T,k),\, 0<x<L,\, 0<t<T$ are defined by the integral equations
\bee \no\begin{array}{l}
\mu_1(0,t,k)=\d \mathbb{I}+\int_T^te^{-2ik^2(t-\tau)\hat{\sigma}_4} (V\mu_1)(0,\tau,k)d\tau,\vspace{0.1in}\\
\mu_2(0,t,k)=\d \mathbb{I}+\int_0^te^{-2ik^2(t-\tau)\hat{\sigma}_4} (V\mu_2)(0,\tau,k)d\tau,\vspace{0.1in}\\
\mu_3(L,t,k)=\d \mathbb{I}+\int_0^te^{-2ik^2(t-\tau)\hat{\sigma}_4} (V\mu_3)(L,\tau,k)d\tau, \vspace{0.1in}\\
\mu_4(L,t,k)=\d \mathbb{I}+\int_T^te^{-2ik^2(t-\tau)\hat{\sigma}_4} (V\mu_4)(L,\tau,k)d\tau,  \vspace{0.1in}\\
\mu_2(x,0,k)=\d \mathbb{I}+\int_0^x e^{ikx'\hat{\sigma}_4}(U\mu_2)(x',0,k)dx',\vspace{0.1in}\\
\mu_3(x,0,k)=\d \mathbb{I}+\int_L^x e^{ikx'\hat{\sigma}_4}(U\mu_3)(x',0,k)dx',\vspace{0.1in}\\
\mu_4(x,0,k)=\d \mathbb{I}+\int_L^x e^{ikx'\hat{\sigma}_4}(U\mu_4)(x',0,k)dx'
 -e^{-ik(x-L)\hat{\sigma}_4}\int_0^Te^{2ik^2\tau\hat{\sigma}_4} (V\mu_4)(L,\tau,k)d\tau, \vspace{0.1in}\\
\mu_2(x,T,k)=\d \mathbb{I}+\int_{0}^x e^{-ik(x-x')\hat{\sigma}_4}(U\mu_2)(x',T,k)dx'+e^{-ikx\hat{\sigma}_4}\int_{0}^Te^{-2ik^2(T-\tau)\hat{\sigma}_4} (V\mu_2)(0,\tau,k)d\tau,
  \end{array}
\ene

It follows from the properties of $\mu_j$ and $\mu_j^A$ that the functions  $\{S(k),\, s(k),\, \mathbb{S}(k),\, S_L(k)\}$ and \\ $\{S^A(k),\, s^A(k),\, \mathbb{S}^A(k),\, S_L^A(k)\}$ have the following boundedness:
 \bee\no
\left\{\begin{array}{l}
 S(k){\rm \,\, is \,\, bounded\,\, for\,\,} k\in  (D_2\cup D_4, D_2\cup D_4, D_2\cup D_4, D_1\cup D_3), \vspace{0.1in} \\
 s(k){\rm \,\, is \,\, bounded\,\, for\,\,} k\in  (D_3\cup D_4, D_3\cup D_4, D_3\cup D_4, D_1\cup D_2), \vspace{0.1in}\\
 \mathbb{S}(k){\rm \,\, is \,\, bounded\,\, for\,\,} k\in  (D_4, D_4, D_4, D_1), \vspace{0.1in}\\
 S_L(k){\rm \,\, is \,\, bounded\,\, for\,\,} k\in  (D_2\cup D_4, D_2\cup D_4, D_2\cup D_4, D_1\cup D_3), \vspace{0.1in}\\
 S^A(k){\rm \,\, is \,\, bounded\,\, for\,\,} k\in  (D_1\cup D_3, D_1\cup D_3, D_1\cup D_3, D_21\cup D_4), \vspace{0.1in} \\
 s^A(k){\rm \,\, is \,\, bounded\,\, for\,\,} k\in  (D_1\cup D_2, D_1\cup D_2, D_1\cup D_2, D_3\cup D_4), \vspace{0.1in}\\
 \mathbb{S}^A(k){\rm \,\, is \,\, bounded\,\, for\,\,} k\in  (D_2, D_2, D_2, D_3), \vspace{0.1in}\\
 S_L^A(k){\rm \,\, is \,\, bounded\,\, for\,\,} k\in  (D_2\cup D_4, D_2\cup D_4, D_2\cup D_4, D_1\cup D_3),
\end{array} \right.
\ene

\vspace{0.1in}
\noindent {\bf Proposition 2.3.} {\it The matrix-valued functions $S_n(x,t,k)\, (n=1,2,3,4)$ defined by
\bee\label{mns}
 M_n(x,t,k)=\mu_2(x,t,k)e^{-i(kx+2k^2t)\hat{\sigma}_4}S_n(k), \quad k\in D_n,
 \ene
with $M_n$ given by Eq.~(\ref{mn}) can be determined by the entries of the data $S(k)=(S_{ij})_{4\times 4},\, s(k)=(s_{ij})_{4\times 4}$, and $\mathbb{S}(k)=(\mathbb{S}_{ij})_{4\times 4}$  given by Eq.~(\ref{sss}) as follows:
\bee\label{sn}
\begin{array}{ll}
S_1(k)=\left(\begin{array}{cccc}
  \mathbb{S}_{11} &  \mathbb{S}_{12} & \mathbb{S}_{13} & 0 \vspace{0.1in}\\
 \mathbb{S}_{21} &  \mathbb{S}_{22} & \mathbb{S}_{23}  & 0 \vspace{0.1in}\\
\mathbb{S}_{31} &  \mathbb{S}_{32} &  \mathbb{S}_{33} & 0 \vspace{0.1in}\\
\mathbb{S}_{41} &  \mathbb{S}_{42}  &  \mathbb{S}_{43} & \dfrac{1}{m_{44}(\mathbb{S})}
 \end{array}\right), \qquad
S_2(k)=\left(\begin{array}{cccc}
 s_{11} &  s_{12} & s_{13} & \dfrac{S_{14}}{(S^Ts^A)_{44}} \vspace{0.1in}\\
 s_{21} &  s_{22} &  s_{23} &\dfrac{S_{24}}{(S^Ts^A)_{44}} \vspace{0.1in}\\
  s_{31} &  s_{32}  &  s_{33} & \dfrac{S_{34}}{(S^Ts^A)_{44}} \vspace{0.1in}\\
  s_{41} &  s_{42} &  s_{43} & \dfrac{S_{44}}{(S^Ts^A)_{44}}
 \end{array}
\right), \vspace{0.15in}\\
S_3(k)=\left(\begin{array}{cccc}
 S_3^{(11)} &  S_3^{(12)} & S_3^{(13)} & s_{14} \vspace{0.1in}\\
S_3^{(21)} &  S_3^{(22)} & S_3^{(23)}  & s_{24} \vspace{0.1in}\\
S_3^{(31)} &  S_3^{(32)} & S_3^{(33)} & s_{34} \vspace{0.1in}\\
S_3^{(41)} &  S_3^{(42)} & S_3^{(43)} & s_{44}
 \end{array}
\right),\quad
S_4(k)=\left(\begin{array}{cccc}
\dfrac{n_{11,44}(\mathbb{S})}{\mathbb{S}_{44}} & \dfrac{n_{12,44}(\mathbb{S})}{\mathbb{S}_{44}} & \dfrac{n_{13,44}(\mathbb{S})}{\mathbb{S}_{44}}
  & \mathbb{S}_{14} \vspace{0.1in}\\
\dfrac{n_{21,44}(\mathbb{S})}{\mathbb{S}_{44}} & \dfrac{n_{22,44}(\mathbb{S})}{\mathbb{S}_{44}}& \dfrac{n_{23,44}(\mathbb{S})}{\mathbb{S}_{44}}
  &  \mathbb{S}_{24} \vspace{0.1in}\\
\dfrac{n_{31,44}(\mathbb{S})}{\mathbb{S}_{44}} & \dfrac{n_{32,44}(\mathbb{S})}{\mathbb{S}_{44}} & \dfrac{n_{33,44}(\mathbb{S})}{\mathbb{S}_{44}} &
\mathbb{S}_{34} \vspace{0.1in}\\
0 & 0 &  0 & \mathbb{S}_{44}\end{array}
\right),
\end{array}
\ene
where $n_{i_1j_1,i_2j_2}(X)$ denotes the determinant of the sub-matrix generated by choosing the cross elements of $i_{1,2}$th rows and $j_{1,2}$th columns of $X$, and
\bee\no
\left\{\begin{array}{l}
S_3^{(1l)}=\d\frac{m_{24}(S)n_{1l,24}(s)-m_{34}(S)n_{1l,34}(s)+m_{44}(S)n_{1l,44}(s)}{(s^TS^A)_{44}}, \vspace{0.1in}\\
S_3^{(2l)}=\d\frac{m_{14}(S)n_{2l,14}(s)-m_{34}(S)n_{2l,34}(s)+m_{44}(S)n_{2l,44}(s)}{(s^TS^A)_{44}}, \vspace{0.1in}\\
S_3^{(3l)}=\d\frac{m_{14}(S)n_{3l,14}(s)-m_{24}(S)n_{3l,24}(s)+m_{44}(S)n_{3l,44}(s)}{(s^TS^A)_{44}}, \vspace{0.1in}\\
S_3^{(4l)}=\d\frac{m_{14}(S)n_{4l,14}(s)-m_{24}(S)n_{4l,24}(s)+m_{34}(S)n_{4l,34}(s)}{(s^TS^A)_{44}},
\end{array}\right. \quad l=1,2,3,
\ene
}

\noindent {\bf Proof.}\, We introduce the matrix-valued functions $R_n(k), S_n(k), T_n(k)$, and $P_n(k),\, n=1,2,3,4)$ by $M_n(x,t,k)$ and $\mu_j(x,t,k)$
\bee\label{rstp}
\left\{\begin{array}{l}
M_n(x,t,k)=\mu_1(x,t,k)e^{-i(kx+2k^2t)\hat{\sigma}_4}R_n(k), \vspace{0.1in}\\
M_n(x,t,k)=\mu_2(x,t,k)e^{-i(kx+2k^2t)\hat{\sigma}_4}S_n(k), \vspace{0.1in}\\
M_n(x,t,k)=\mu_3(x,t,k)e^{-i(kx+2k^2t)\hat{\sigma}_4}T_n(k), \vspace{0.1in}\\
M_n(x,t,k)=\mu_4(x,t,k)e^{-i(kx+2k^2t)\hat{\sigma}_4}P_n(k),
\end{array}\right.
\ene

It follows from Eq.~(\ref{rstp}) that we have the relations
\bee
\left\{\begin{array}{l}
R_n(k)=e^{2ik^2T\hat{\sigma}_4}M_n(0,T,k),\vspace{0.1in}\\
S_n(k)=M_n(0,0,k),\vspace{0.1in}\\
T_n(k)=e^{ikL\hat{\sigma}_4}M_n(L,0,k),\vspace{0.1in} \\
P_n(k)=e^{i(kL+2k^2T)\hat{\sigma}_4}M_n(L,T,k),
\end{array}\right.
\ene
and
\bee \label{sneq}
\left\{\begin{array}{l}
S(k)=\mu_1(0,0,k)=S_n(k)R_n^{-1}(k), \vspace{0.1in}\\
s(k)=\mu_3(0,0,k)=S_n(k)T_n^{-1}(k), \vspace{0.1in}\\
\mathbb{S}(k)=\mu_4(0,0,k)=S_n(k)P_n^{-1}(k),
\end{array}\right.
\ene
which can in general obtain the functions $\{R_n, S_n, T_n, P_n\}$ for the given functions $\{s(k), S(k), \mathbb{S}(k)\}$.

Moreover, we can also determine some entries of $\{R_n, S_n, T_n, P_n\}$ in terms of Eqs.~(\ref{mn}) and (\ref{rstp})
\bee\label{rp}
\left\{
\begin{array}{l}
(R_n(k))_{ij}=0, \,\,\,\,\, {\rm if} \,\,\, (\gamma^n)_{ij}=\gamma_1, \vspace{0.1in} \\
(S_n(k))_{ij}=0,  \,\,\,\,\, {\rm if} \,\,\,  (\gamma^n)_{ij}=\gamma_2, \vspace{0.1in}\\
(T_n(k))_{ij}=\delta_{ij}, \,\,\, {\rm if} \,\,\,  (\gamma^n)_{ij}=\gamma_3, \vspace{0.1in}\\
(P_n(k))_{ij}=\delta_{ij},  \,\,\, {\rm if} \,\,\,  (\gamma^n)_{ij}=\gamma_4,
\end{array}
\right.
 \ene
Thus it follows from systems (\ref{sneq}) and  (\ref{rp}) that we can find Eq.~(\ref{sn}). $\square$

\subsection*{\it 2.8.\, The residue conditions for $M_n$}

\quad Since $\mu_2(x,t,k)$ is an entire function, it follows from Eq.~(\ref{mns}) that $M_n(x,t,k)$ only have singularities at the points where the $S_n(k)$'s have singularities. We find from the expressions of $S_n(k)$ given by  Eq.~(\ref{sn}) that the possible singularities of $M_n$ are as follows:

\begin{itemize}

\item {} $[M]_4$ could admit poles in $D_1$ at the zeros of $m_{44}(\mathbb{S})(k)$;
\item {} $[M]_4$ could have poles in $D_2$ at the zeros of $(S^Ts^A)_{44}(k)$;
\item {} $[M]_l,\, l=1,2,3$ could be of poles in $D_3$ at the zeros of  $(s^TS^A)_{44}(k)$;
\item {} $[M]_l,\, l=1,2,3$ could have poles in $D_4$ at the zeros of $\mathbb{S}_{44}(k)$.
\end{itemize}

We introduce the above possible zeros by $\{k_j\}_1^N$ and suppose that they satisfy the following assumption.

\vspace{0.1in}
\noindent {\bf Assumption 2.4.} {\it We assume that
\begin{itemize}

\item {} $m_{44}(\mathbb{S})(k)$ has $n_1$ possible simple zeros in $D_1$ denoted by $\{k_j\}_1^{n_1}$;

\item {} $(S^Ts^A)_{44}(k)$ has $n_2-n_1$ possible simple zeros in $D_2$ denoted by $\{k_j\}_{n_1+1}^{n_2}$;

\item {} $(s^TS^A)_{44}(k)$ has $n_3-n_2$ possible simple zeros in $D_3$ denoted by $\{k_j\}_{n_2+1}^{n_3}$;

\item {} $\mathbb{S}_{44}(k)$ has $N-n_3$ possible simple zeros in $D_4$ denoted by $\{k_j\}_{n_3+1}^N$;
\end{itemize}
and that none of these zeros coincide. Moreover, none of these functions are assumed to have zeros on the boundaries od the $D_n$'s ($n=1,2,3,4)$.}

We can deduce the residue conditions at these zeros in the following expressions:

\vspace{0.1in}
\noindent {\bf Proposition 2.5. } {\it Let $\{M_n\}_1^4$ be the eigenfunctions given by Eq.~(\ref{mn}) and suppose that the set $\{k_j\}_1^N$ of singularities is as the above-mentioned Assumption 2.4. Then we have the following residue conditions for $M_n$:
\bee
 \label{rm1a} \begin{array}{rl}
\d{\rm Res}_{k=k_j}
[M_1]_4\!\!=&\!\!\!\!\d \frac{n_{12,23}(\mathbb{S})(k_j)[M_1(k_j)]_1\! -\! n_{11,23}(\mathbb{S})(k_j)[M_1(k_j)]_2
\!+\!n_{11,22}(\mathbb{S})(k_j)[M_1(k_j)]_3}{\dot{m}_{44}(\mathbb{S})(k_j)m_{34}(\mathbb{S})(k_j)}e^{2\theta(k_j)}, \vspace{0.1in}\\
        & \quad {\rm for}\quad 1\leq j\leq n_1,\quad k\in D_1,
        \end{array}
\ene

\bee \label{rm2a}
\begin{array}{rl}
\d{\rm Res}_{k=k_j}[M_2]_4\!\!=&\!\!\!\!\d\dfrac{[M_2(k_j)]_1[S_{14}(k_j)n_{22,43}(s)(k_j)\!-\!S_{24}(k_j)n_{12,43}(s)(k_j)\!+\!S_{44}(k_j)n_{12,23}(s)(k_j)]}
          {\dot{(S^Ts^A)}_{44}(k_j)m_{34}(s)(k_j)e^{-2\theta(k_j)}}    \vspace{0.1in}\\
  &\!\!\!\!\d -\dfrac{[M_2(k_j)]_2[S_{14}(k_j)n_{21,43}(s)(k_j)\!-\!S_{24}(k_j)n_{11,43}(s)(k_j)\!+\!S_{44}(k_j)n_{11,23}(s)(k_j)]}
          {\dot{(S^Ts^A)}_{44}(k_j)m_{34}(s)(k_j)e^{-2\theta(k_j)}}    \vspace{0.1in}\\
  &\!\!\!\!+\dfrac{[M_2(k_j)]_3[S_{14}(k_j)n_{21,42}(s)(k_j)\!-\!S_{24}(k_j)n_{11,42}(s)(k_j)\!+\!S_{44}(k_j)n_{11,22}(s)(k_j)]}
          {\dot{(S^Ts^A)}_{44}(k_j)m_{34}(s)(k_j)e^{-2\theta(k_j)}},\vspace{0.1in}\\
        & \quad {\rm for}\quad n_1+1\leq j\leq n_2,\quad k\in D_2,
\end{array}
\ene

\bee
 \label{rm3a} \begin{array}{rl}
 {\rm Res}_{k=k_j}[M_3]_l\!\!=&\!\!\!\! \dfrac{m_{14}(S)(k_j)n_{4l,14}(s)(k_j)\!-\!m_{24}(S)(k_j)n_{4l,24}(s)(k_j)\!+\!m_{34}(S)(k_j)n_{4l,34}(s)(k_j)}
          {\dot{(s^TS^A)}_{44}(k_j)s_{44}(k_j)e^{2\theta(k_j)}} \vspace{0.1in}\\
          &\times[M_3k_j)]_4,  \quad {\rm for}\quad n_2+1\leq j\leq n_3,\quad k\in D_3,\quad l=1,2,3,
  \end{array}
\ene
\bee
 \label{rm4a}
 {\rm Res}_{k=k_j}[M_4]_l= -\dfrac{\mathbb{S}_{4l}(k_j)}
   {\dot{\mathbb{S}}_{44}(k_j)}[M_4(k_j)]_4e^{-2\theta(k_j)},\quad {\rm for}\quad n_3+1\leq j\leq N,\quad k\in D_4, \quad l=1,2,3,
   \ene
where the overdot stands for the derivative with resect to the parameter $k$ and $\theta=\theta(k)=-i(kx+2k^2t)$.}

\vspace{0.1in}
\noindent {\bf Proof.}\, It follows from Eqs.~(\ref{mns}) and (\ref{sn}) that the four columns of $M_1$ are given by
 the matrices $\mu_2$ and $S_1(k)$
\bes \label{m1}\bee
\label{m1a}  &
 [M_1]_1=[\mu_2]_1\mathbb{S}_{11}+[\mu_2]_2\mathbb{S}_{21}+[\mu_2]_3\mathbb{S}_{31}+[\mu_2]_4\mathbb{S}_{41}e^{-2\theta}, \vspace{0.1in}\\
\label{m1b} &[M_1]_2=[\mu_2]_1\mathbb{S}_{12}+[\mu_2]_2\mathbb{S}_{22}+[\mu_2]_3\mathbb{S}_{32}+[\mu_2]_4\mathbb{S}_{42}e^{-2\theta}, \vspace{0.1in}\\
\label{m1c} &[M_1]_3=[\mu_2]_1\mathbb{S}_{13}+[\mu_2]_2\mathbb{S}_{23}+[\mu_2]_3\mathbb{S}_{33}+[\mu_2]_4\mathbb{S}_{43}e^{-2\theta}, \vspace{0.1in}\\
\label{m1d} &[M_1]_4=\d\frac{[\mu_2]_4}{m_{44}(\mathbb{S})},
  \ene\ees
the four columns of $M_2$ are given by  the matrices $\mu_2$ and $S_2(k)$
\bes \label{m2}\bee
\label{m2a} &[M_2]_1=[\mu_2]_1 s_{11} +[\mu_2]_2s_{21}+[\mu_2]_3s_{31}+[\mu_2]_4s_{41}e^{-2\theta}, \vspace{0.1in}\\
\label{m2b} &[M_2]_2=[\mu_2]_1 s_{12} +[\mu_2]_2s_{22}+[\mu_2]_3s_{32}+[\mu_2]_4s_{42}e^{-2\theta}, \vspace{0.1in}\\
\label{m2c} &[M_2]_3=[\mu_2]_1 s_{13} +[\mu_2]_2s_{23}+[\mu_2]_3s_{33}+[\mu_2]_4s_{43}e^{-2\theta}, \vspace{0.1in}\\
\label{m2d} &[M_2]_4=\d\frac{[\mu_2]_1S_{14}}{(S^Ts^A)_{44}}e^{2\theta}+\frac{[\mu_2]_2S_{24}}{(S^Ts^A)_{44}}e^{2\theta}
                +\frac{[\mu_2]_3S_{34}}{(S^Ts^A)_{44}}e^{2\theta}+\frac{[\mu_2]_4S_{44}}{(S^Ts^A)_{44}},
  \ene\ees
the four columns of $M_3$ are given by  the matrices $\mu_2$ and $S_3(k)$
\bes \label{m3}\bee
\label{m3a} &[M_3]_1=[\mu_2]_1 S_3^{(11)} +[\mu_2]_2S_3^{(21)}+[\mu_2]_3S_3^{(31)}+[\mu_2]_4S_3^{(41)}e^{-2\theta}, \vspace{0.1in}\\
\label{m3b} &[M_3]_2=[\mu_2]_1 S_3^{(12)} +[\mu_2]_2S_3^{(22)}+[\mu_2]_3S_3^{(32)}+[\mu_2]_4S_3^{(42)}e^{-2\theta}, \vspace{0.1in} \\
\label{m3c} &[M_3]_3=[\mu_2]_1 S_3^{(13)} +[\mu_2]_2S_3^{(23)}+[\mu_2]_3S_3^{(33)}+[\mu_2]_4S_3^{(43)}e^{-2\theta},  \vspace{0.1in}\\
\label{m3d} &[M_3]_4=[\mu_2]_1s_{14}e^{2\theta}+[\mu_2]_2s_{24}e^{2\theta}+[\mu_2]_3s_{34}e^{2\theta}+[\mu_2]_4s_{44},\qquad
  \ene\ees
and the four columns of $M_4$ are given by  the matrices $\mu_2$ and $S_4(k)$
\bes \label{m4}\bee
\label{m4a} &[M_4]_1=\d
[\mu_2]_1 \frac{n_{11,44}(\mathbb{S})}{\mathbb{S}_{44}}+[\mu_2]_2 \frac{n_{21,44}(\mathbb{S})}{\mathbb{S}_{44}} +[\mu_2]_3 \frac{n_{31,44}(\mathbb{S})}{\mathbb{S}_{44}}, \vspace{0.1in}\\
\label{m4b} &[M_4]_2=\d
 [\mu_2]_1 \frac{n_{12,44}(\mathbb{S})}{\mathbb{S}_{44}}+[\mu_2]_2 \frac{n_{22,44}(\mathbb{S})}{\mathbb{S}_{44}} +[\mu_2]_3 \frac{n_{32,44}(\mathbb{S})}{\mathbb{S}_{44}}, \vspace{0.1in}\\
\label{m4c} &[M_4]_3=\d
 [\mu_2]_1 \frac{n_{13,44}(\mathbb{S})}{\mathbb{S}_{44}}+[\mu_2]_2 \frac{n_{23,44}(\mathbb{S})}{\mathbb{S}_{44}} +[\mu_2]_3 \frac{n_{33,44}(\mathbb{S})}{\mathbb{S}_{44}}, \vspace{0.1in}\\
\label{m4d} &[M_4]_4=[\mu_2]_1\mathbb{S}_{14}e^{2\theta}+[\mu_2]_2\mathbb{S}_{24}e^{2\theta}+[\mu_2]_3\mathbb{S}_{34}e^{2\theta}+[\mu_2]_4\mathbb{S}_{44},
  \ene\ees

For the case that $k_j\in D_1$ is a simple zero of $m_{44}(\mathbb{S})(k)$, it follows from Eqs.~(\ref{m1a})-(\ref{m1c}) that we have $[\mu_2]_j,\, j=1,2,4$ and then substitute them into Eq.~(\ref{m1d}) to yield
\bee \nonumber
  [M_1]_4=\d \frac{n_{12,23}(\mathbb{S})[M_1]_1-n_{11,23}(\mathbb{S})[M_1]_2+n_{11,22}(\mathbb{S})[M_1]_3}
           {m_{34}(\mathbb{S})m_{44}(\mathbb{S})}e^{2\theta}-
           \frac{[\mu_2]_3}{m_{34}(\mathbb{S})}e^{2\theta},
\ene
whose residue at $k_j$ yields Eq.~(\ref{rm1a})  for $k_j\in D_1$, respectively.

Similarly, we solve Eqs~(\ref{m2a})-(\ref{m2c}) for $[\mu_2]_j,\, j=1,2,4$ and then substitute them into Eq~(\ref{m2d}) to yield
\bee \nonumber
\begin{array}{rl}
[M_2]_4=&\!\!\!\dfrac{[M_2]_1[S_{14}n_{22,43}(s)-S_{24}n_{12,43}(s)+S_{44}n_{12,23}(s)]}
          {(S^Ts^A)_{44}m_{34}(s)}e^{2\theta}    \vspace{0.1in}\\
  &\!\!\!-\dfrac{[M_2]_2[S_{14}n_{21,43}(s)-S_{24}n_{11,43}(s)+S_{44}n_{11,23}(s)]}
          {(S^Ts^A)_{44}m_{34}(s)}e^{2\theta}    \vspace{0.1in}\\
  &\!\!\!\d +\dfrac{[M_2]_3[S_{14}n_{21,42}(s)-S_{24}n_{11,42}(s)+S_{44}n_{11,22}(s)]}
          {(S^Ts^A)_{44}m_{34}(s)}e^{2\theta}   -\frac{[\mu_2]_3}{m_{34}(s)}e^{2\theta},
\end{array}
\ene
whose residues at $k_j$ yields Eq.~(\ref{rm2a}) for $k_j\in D_2$, respectively. Similarly, we can show Eq.~(\ref{rm3a}) for $k_j\in D_3$ and Eq.~(\ref{rm4a}) for $k_j\in D_4$ by analyzing Eqs.~(\ref{m3a})-(\ref{m4d}).  $\square$

\subsection*{\it 2.9.\, The global relation}

\quad The definitions of the above-mentioned spectral functions $S(k), s(k), S_L(k)$, and $\mathbb{S}(k)$ imply that they are dependent. It follows from Eqs.~(\ref{mu1234}) and (\ref{sr}) that
\bee \label{srr} \begin{array}{rl}
\mu_4(x,t,k)=&\mu_2(x,t,k)e^{-i(kx+2k^2t)\hat{\sigma}_4}\mathbb{S}(k) \vspace{0.1in} \\
         =&\mu_2(x,t,k)e^{-i(kx+2k^2t)\hat{\sigma}_4}[s(k)e^{ikL\hat{\sigma}_4}S_L(k)] \vspace{0.1in}\\
         =&\mu_1(x,t,k)e^{-i(kx+2k^2t)\hat{\sigma}_4}[S^{-1}(k)s(k)e^{ikL\hat{\sigma}_4}S_L(k)],
        \end{array}
\ene
which leads to the  global relation
\bee\label{srr2}
c(T,k)=\mu_4(0, T, k)=e^{-2ik^2T\hat{\sigma}_4}[S^{-1}(k)s(k)e^{ikL\hat{\sigma}_4}S_L(k)],
\ene
by evaluating Eq.~(\ref{srr}) at the point $(x,t)=(0, T)$ and using $\mu_1(0, T, k)=\mathbb{I}$.

\section{The $4\times 4$ matrix Riemann-Hilbert problem}

\quad By using the district contours $\gamma_j\, (j=1,2,3,4)$, the integral solutions of the revised Lax pair (\ref{mulax}), and $S_n$ due to $\{S(k), s(k), \mathbb{S}(k), S_L(k)\}$, we have defined the sectionally analytic function $M_n(x,t,k)\, (n=1,2,3,4)$, which solves a
$4\times 4$ matrix Riemann-Hilbert (RH) problem. This RH problem can be formulated on basis of the initial and boundary data of the functions $q_1(x,t)$, $q_2(x,t)$ and $q_3(x,t)$. Thus the solution of Eq.~(\ref{pnls}) for all values of $x,t$ can be refound by solving the RH problem.

\vspace{0.1in}
\noindent {\bf Theorem 3.1.}\, {\it Let $(q_1(x,t), q_2(x,t), q_3(x,t))$ be a solution of Eq.~(\ref{pnls}) in the interval domain  $\Omega=\{(x,t)| x\in [0, L],\, t\in [0, T]\}$. Then it can be reconstructed from the initial data defined by
\bee\no
 q_j(x, t=0)=q_{0j}(x),&  j=1,2,3,
\ene
and Dirichlet and Neumann boundary values defined by
 \bee\no
 \begin{array}{lll}
  {\it Dirichlet \,\, boundary \,\, data:} & q_j(x=0, t)=u_{0j}(t), & q_j(x=L, t)=v_{0j}(t), \,\,\, j=1,2,3, \vspace{0.1in} \\
 {\it Neumann \,\, boundary \,\, data:} &  q_{jx}(x=0, t)=u_{1j}(t),& q_{jx}(x=L, t)=v_{1j}(t), \,\,\, j=1,2,3,
 \end{array} \ene

We can use the initial and boundary data to define the jump matrices $J_{mn}(x, t, k),\, (n, m = 1,..., 4)$ given by Eq.~(\ref{jump}) as well as the spectral functions $S(k), \, s(k)$ and  $\mathbb{S}(k)$ defined by Eq. (\ref{sss}). Assume that the possible zeros $\{k_j\}^N_1$ of the functions $m_{44}(\mathbb{S})(k)$, $(S^Ts^A)_{44}(k)$, $(s^TS^A)_{44}(k)$, and $\mathbb{S}_{44}(k)$ are as in Assumption 2.4. Then the solution $(q_1(x,t),\, q_2(x,t), q_3(x,t))$ of Eq.~(\ref{pnls}) is given by $M(x,t,k)$ in the form
\bee\label{solu}
q_j(x,t)=\d 2i\lim_{k\to \infty}(kM(x,t,k))_{j4},\quad j=1,2,3,
\ene
where $M(x,t,k)$ satisfies the following $4\times 4$ matrix Riemann-Hilbert problem:

\begin{itemize}
\item {} $M(x,t,k)$ is sectionally meromorphic on the Riemann $k$-sphere with jumps
 across the contours $\bar{D}_n\cup \bar{D}_m$, $(n, m = 1,..., 4)$ (see Fig.~\ref{kplane}).

\item {} Across the contours $\bar{D}_n\cup \bar{D}_m\, (n, m = 1,..., 4)$, $M(x, t, k)$ satisfies the
 jump condition (\ref{jumpc}).

\item {} The residue conditions of $M(x,t,k)$ are satisfied in Proposition 2.5.

\item {} $M(x, t, k) = \d\mathbb{I}+O(1/k)$ as $k\to\infty$.

\end{itemize}
}

\noindent {\bf Proof.}\, System~(\ref{solu}) can be deduced from the large $k$ asymptotics of the eigenfunctions.
We can follow the similar one in Refs.~\cite{f3, nls3}  to show the rest proof of the Theorem. $\square$

\section{The nonlinearizable boundary conditions}

\quad The key difficulty of initial-boundary value problems is to find the boundary values for a well-posed problem.
 All boundary value conditions are required for the definition of $S(k)$ and $S_L(k)$, and hence for the formulate the
 RH problem. Our main conclusion exhibits the unknown boundary condition on basis of the prescribed boundary condition and the initial condition
 in terms of the solution of a system of nonlinear integral equations.

\subsection*{\it 4.1.\, The generalized global relation}

\quad By evaluating Eqs.~(\ref{srr}) and (\ref{srr2}) at the point $(x,t)=(0, t)$, we have
\bee\no
 c(t,k)=\mu_2(0,t,k)e^{-2ik^2t\hat{\sigma}_4}[s(k)e^{ikL\hat{\sigma}_4}S_L(k)],
 \ene
which and Eq.~(\ref{sl}) lead to
\bee \label{gr}
\begin{array}{rl}
  c(t,k)=&\!\!\mu_2(0,t,k)e^{-2ik^2t\hat{\sigma}_4}[s(k)e^{ikL\hat{\sigma}_4}e^{2ik^2t\hat{\sigma}_4}\mu_3^{-1}(L,t,k)], \vspace{0.1in}\\
=&\!\!\mu_2(0,t,k)[e^{-2ik^2t\hat{\sigma}_4}s(k)][e^{ikL\hat{\sigma}_4}\mu_3^{-1}(L,t,k)],
\end{array}
\ene
Thus, the column vectors $[c(t,k)]_j,\, j=1,2,3$ are analytic and bounded in $D_4$ away from the possible zeros of $\mathbb{S}_{44}(k)$ and of order $O(\frac{1+e^{-2ikL}}{k})$ as $k\to \infty$, and the column vector $[c(t,k)]_4$ is analytic and bounded in $D_1$ away from the possible zeros of $m_{44}(\mathbb{S})(k)$ and of order $O(\frac{1+e^{2ikL}}{k})$ as $k\to \infty$,

\subsection*{\it 4.2.\,  Asymptotic behaviors of eigenfunctions}

\quad It follows from the Lax pair (\ref{mulax}) that the eigenfunctions $\{\mu_j\}_1^4$ possess the following asymptotics as $k\to\infty$
\bee\label{mua}
\begin{array}{rl}
\mu_j(x,t,k)=&\!\!\!\!\!\d\mathbb{I}+\sum_{i=1}^2
\frac{1}{k^i}\left(\begin{array}{cccc} \mu_{j,11}^{(i)} & \mu_{j,12}^{(i)} & \mu_{j,13}^{(i)} & \mu_{j,14}^{(i)} \vspace{0.1in} \\
                                      \mu_{j,21}^{(i)} & \mu_{j,22}^{(i)} & \mu_{j,23}^{(i)} & \mu_{j,24}^{(i)} \vspace{0.1in}\\
                                      \mu_{j,31}^{(i)} & \mu_{j,32}^{(i)} & \mu_{j,33}^{(i)} & \mu_{j,34}^{(i)} \vspace{0.1in}\\
                                      \mu_{j,41}^{(i)} & \mu_{j,42}^{(i)} & \mu_{j,43}^{(i)} & \mu_{j,44}^{(i)}
  \end{array}\right) +O(\frac{1}{k^3}) \vspace{0.2in}\\
 =&\!\!\! \d\mathbb{I}+\frac{1}{k}\left(\begin{array}{cccc}
 \int_{(x_j, t_j)}^{(x,t)}\Delta_{11}^{(1)} &\int_{(x_j, t_j)}^{(x,t)}\Delta_{12}^{(1)} &\int_{(x_j, t_j)}^{(x,t)}\Delta_{13}^{(1)} & \d-\frac{i}{2}q_1 \vspace{0.1in}\\
  \int_{(x_j, t_j)}^{(x,t)}\Delta_{21}^{(1)} & \int_{(x_j, t_j)}^{(x,t)}\Delta_{22}^{(1)} &\int_{(x_j, t_j)}^{(x,t)}\Delta_{23}^{(1)} & \d-\frac{i}{2}q_2 \vspace{0.1in}\\
  \int_{(x_j, t_j)}^{(x,t)}\Delta_{31}^{(1)} & \int_{(x_j, t_j)}^{(x,t)}\Delta_{32}^{(1)} & \int_{(x_j, t_j)}^{(x,t)}\Delta_{33}^{(1)} & \d-\frac{i}{2}q_3 \vspace{0.1in}\\
    \d \frac{i}{2}p_1 & \d\frac{i}{2}p_2 & \d\frac{i}{2}p_3 & \int_{(x_j, t_j)}^{(x,t)}\Delta_{44}^{(1)}
  \end{array}\right) \vspace{0.2in}\\
      &+\d\frac{1}{k^2}\left(\begin{array}{cccc}
     \int_{(x_j, t_j)}^{(x,t)}\Delta_{11}^{(2)} &\int_{(x_j, t_j)}^{(x,t)}\Delta_{12}^{(2)} &\int_{(x_j, t_j)}^{(x,t)}\Delta_{13}^{(2)} & \mu_{j,14}^{(2)} \vspace{0.1in}\\
      \int_{(x_j, t_j)}^{(x,t)}\Delta_{21}^{(2)} & \int_{(x_j, t_j)}^{(x,t)}\Delta_{22}^{(2)} &\int_{(x_j, t_j)}^{(x,t)}\Delta_{23}^{(2)} & \mu_{j,24}^{(2)} \vspace{0.1in}\\
     \int_{(x_j, t_j)}^{(x,t)}\Delta_{31}^{(2)} & \int_{(x_j, t_j)}^{(x,t)}\Delta_{32}^{(2)} &\int_{(x_j, t_j)}^{(x,t)}\Delta_{33}^{(2)} & \mu_{j,34}^{(2)} \vspace{0.1in}\\
                                      \mu_{j,41}^{(2)} & \mu_{j,42}^{(2)} & \mu_{j,43}^{(2)} & \int_{(x_j, t_j)}^{(x,t)}\Delta_{44}^{(2)}
  \end{array}\right)+O(\frac{1}{k^3}),
  \end{array}
\ene
where we have introduced the following functions
\bee\no
\left\{\begin{array}{rl}
\Delta_{jl}^{(1)}=&\!\!\d\frac{i}{2}q_jp_ldx+\frac{1}{2}(q_jp_{lx}-q_{jx}p_l)dt, \quad j,l=1,2,3,\vspace{0.1in} \\
\Delta_{44}^{(1)}=&\!\! \d-\frac{i}{2}\sum_{j=1}^3q_jp_jdx+\frac{1}{2}\sum_{j=1}^3(p_jq_{jx}-p_{jx}q_j)dt,
\end{array}\right.
\ene
and
\bee\no
\left\{\begin{array}{rl}
\mu_{j,l4}^{(2)}=&\!\!\! \d\frac{1}{4}q_{lx}+\frac{1}{2i}q_l\int_{(x_j,t_j)}^{(x,t)}\Delta_{44}^{(1)}, \quad l=1,2,3,\vspace{0.1in} \\
\mu_{j,4l}^{(2)}=&\!\!\!\d\frac{1}{4}p_{lx}+\frac{i}{2}\sum_{s=1}^3p_s\int_{(x_j,t_j)}^{(x,t)}\Delta_{sl}^{(1)}, \quad l=1,2,3, \vspace{0.1in} \\
\Delta_{sl}^{(2)}=&\!\!\!\d\left[\frac{1}{4}q_sp_{lx}+\frac{i}{2}q_s\sum_{n=1}^3p_n\int_{(x_j,t_j)}^{(x,t)}\Delta_{nl}^{(1)}\right]dx \vspace{0.1in} \\
   &\!\!\! \d +\left\{\frac{1}{4}\left[q_sp_{lx}+iq_{sx}p_{lx}-iq_sp_l\sum_{j=1}^3q_jp_j\right]
   +\frac{1}{2}\sum_{n=1}^3(q_sp_{nx}-q_{sx}p_n)\int_{(x_j,t_j)}^{(x,t)}\Delta_{nl}^{(1)}\right\}dt,\, s,l=1,2,3,  \vspace{0.1in} \\

\Delta_{44}^{(2)}=&\!\!\!\!\d \left[\frac{1}{4}\sum_{l=1}^3p_lq_{lx}-\frac{i}{2}\sum_{l=1}^3p_lq_{l}\int_{(x_j,t_j)}^{(x,t)}\Delta_{44}^{(1)}\right]dx \vspace{0.1in} \\
 &+\d \left\{\frac{1}{4}\left[\sum_{l=1}^3(p_lq_{lx}-ip_{lx}q_{lx})+i\left(\sum_{l=1}^3p_lq_{l}\right)^2\right]+ \frac{1}{2}\sum_{l=1}^3(p_lq_{lx}-p_{lx}q_{l})\int_{(x_j,t_j)}^{(x,t)}\Delta_{44}^{(1)}\right\}dt,
 \end{array}\right.
\ene
The functions $\{\mu^{(i)}_{jl}=\mu^{(i)}_{jl}(x,t)\}_1^4,\, i=1, 2$ are independent of $k$.

We define the function $\{\Psi_{ij}(t,k)\}_{i,j=1}^4$  as
\bee\label{mu2asy}
\begin{array}{l}
\mu_2(0, t,k)=(\Psi_{sj}(t, k))_{4\times 4}=\d \mathbb{I}+\sum_{l=1}^2\frac{1}{k^l}\left(\begin{array}{cccc}
 \Psi_{11}^{(l)}(t) & \Psi_{12}^{(l)}(t) & \Psi_{13}^{(l)}(t) & \Psi_{14}^{(l)}(t) \vspace{0.1in} \\
 \Psi_{21}^{(l)}(t) & \Psi_{22}^{(l)}(t) & \Psi_{23}^{(l)}(t) & \Psi_{24}^{(l)}(t) \vspace{0.1in} \\
 \Psi_{31}^{(l)}(t) & \Psi_{32}^{(l)}(t) & \Psi_{33}^{(l)}(t) & \Psi_{34}^{(l)}(t) \vspace{0.1in} \\
 \Psi_{41}^{(l)}(t) & \Psi_{42}^{(l)}(t) & \Psi_{43}^{(l)}(t) & \Psi_{44}^{(l)}(t)
 \end{array}\right) +O(\frac{1}{k^3}),
\end{array}
\ene

Based on the asymptotic of Eq.~(\ref{mua}) and the boundary data at $x=0$, we find
\bee\left\{\!\!\begin{array}{l}
\Psi_{14}^{(1)}(t)=-\d\frac{i}{2}u_{01}(t), \quad \Psi_{24}^{(1)}(t)=-\frac{i}{2}u_{02}(t),
\quad \Psi_{34}^{(1)}(t)=-\frac{i}{2}u_{03}(t), \vspace{0.1in} \\
\Psi_{41}^{(1)}(t)=\d\frac{i}{2}\left[\alpha_{11}\bar{u}_{01}(t)+\bar{\alpha}_{12}\bar{u}_{02}(t)+\bar{\alpha}_{13}\bar{u}_{03}(t)\right], \vspace{0.1in} \\
\Psi_{42}^{(1)}(t)=\d\frac{i}{2}\left[\alpha_{12}\bar{u}_{01}(t)+\alpha_{22}\bar{u}_{02}(t)+\bar{\alpha}_{23}\bar{u}_{03}(t)\right], \vspace{0.1in} \\
\Psi_{43}^{(1)}(t)=\d\frac{i}{2}\left[\alpha_{13}\bar{u}_{01}(t)\!+\!\alpha_{23}\bar{u}_{02}(t)\!+\!\alpha_{33}\bar{u}_{03}(t)\right],
\vspace{0.1in} \\
\Psi_{14}^{(2)}=\d\frac{1}{4}u_{11}+\frac{1}{2i}u_{01}\Psi_{44}^{(1)},\quad
\Psi_{24}^{(2)}=\d\frac{1}{4}u_{12}+\frac{1}{2i}u_{02}\Psi_{44}^{(1)},\quad
\Psi_{34}^{(2)}=\d\frac{1}{4}u_{13}+\frac{1}{2i}u_{03}\Psi_{44}^{(1)},  \vspace{0.1in} \\
\Psi_{44}^{(1)}=\d\frac{1}{2}\int^t_0\Big\{u_{11}\left[\alpha_{11}\bar{u}_{01}(t)\!+\!\bar{\alpha}_{12}\bar{u}_{02}(t)\!+\!
  \bar{\alpha}_{13}\bar{u}_{03}(t)\right]\!+\!u_{12}\left[\alpha_{12}\bar{u}_{01}(t)\!+\!\alpha_{22}\bar{u}_{02}(t)\!+\!
  \bar{\alpha}_{23}\bar{u}_{03}(t)\right] \vspace{0.1in}  \\
 \quad\quad +u_{13}\left[\alpha_{13}\bar{u}_{01}(t)+\alpha_{23}\bar{u}_{02}(t)+\alpha_{33}\bar{u}_{03}(t)\right]
 -u_{01}\left[\alpha_{11}\bar{u}_{11}(t)+\bar{\alpha}_{12}\bar{u}_{12}(t)+\bar{\alpha}_{13}\bar{u}_{13}(t)\right]  \vspace{0.1in} \\
\quad\quad  -u_{02}\left[\alpha_{12}\bar{u}_{11}(t)+\alpha_{22}\bar{u}_{12}(t)+\bar{\alpha}_{23}\bar{u}_{13}(t)\right]
 -u_{03}\left[\alpha_{13}\bar{u}_{11}(t)+\alpha_{23}\bar{u}_{12}(t)+\alpha_{33}\bar{u}_{13}(t)\right]\Big\}dt,
\end{array}\right.
\ene

Thus we have the the boundary data at $x=0$:
\bee\label{ud}
\left\{\begin{array}{l}
u_{01}(t)=2i\Psi_{14}^{(1)}(t),\quad  u_{02}(t)=2i\Psi_{24}^{(1)}(t), \quad  u_{03}(t)=2i\Psi_{34}^{(1)}(t),  \vspace{0.1in} \\
u_{11}(t)= 4\Psi_{14}^{(2)}(t)+2iu_{01}(t)\Psi_{44}^{(1)}(t),  \vspace{0.1in} \\
u_{12}(t)= 4\Psi_{24}^{(2)}(t)+2iu_{02}(t)\Psi_{44}^{(1)}(t),  \vspace{0.1in} \\
u_{13}(t)= 4\Psi_{34}^{(2)}(t)+2iu_{03}(t)\Psi_{44}^{(1)}(t),
\end{array}\right.
\ene

Similarly, we assume that the asymptotic formula of $\mu_3(L, t, k)=\{\phi_{ij}(t,k)\}_{i,j=1}^4$ is of the from
\bee\label{mu3asy}
\begin{array}{l}
\d \mu_3(L, t,k)=(\phi_{sj}(t, k))_{4\times 4}= \mathbb{I}+\sum_{l=1}^2\frac{1}{k^l}\left(\begin{array}{cccc}
 \phi_{11}^{(l)}(t) & \phi_{12}^{(l)}(t) & \phi_{13}^{(l)}(t) & \phi_{14}^{(l)}(t) \vspace{0.05in}\\
 \phi_{21}^{(l)}(t) & \phi_{22}^{(l)}(t) & \phi_{23}^{(l)}(t) & \phi_{24}^{(l)}(t) \vspace{0.05in}\\
  \phi_{31}^{(l)}(t) & \phi_{32}^{(l)}(t) & \phi_{33}^{(l)}(t) & \phi_{34}^{(l)}(t) \vspace{0.05in}\\
   \phi_{41}^{(l)}(t) & \phi_{42}^{(l)}(t) & \phi_{43}^{(l)}(t) & \phi_{44}^{(l)}(t)
 \end{array}
\right) +O(\frac{1}{k^3}),
\end{array}
\ene

By using the asymptotic of Eq.~(\ref{mua}) and the boundary data at $x=L$, we find
\bee\left\{\begin{array}{l}
\phi_{14}^{(1)}(t)=-\d\frac{i}{2}v_{01}(t), \quad \phi_{24}^{(1)}(t)=-\frac{i}{2}v_{02}(t), \quad \phi_{34}^{(1)}(t)=-\frac{i}{2}v_{03}(t), \vspace{0.1in} \\
\phi_{41}^{(1)}(t)=\d\frac{i}{2}\left[\alpha_{11}\bar{v}_{01}(t)+\bar{\alpha}_{12}\bar{v}_{02}(t)+\bar{\alpha}_{13}\bar{v}_{03}(t)\right], \vspace{0.1in} \\
\phi_{42}^{(1)}(t)=\d\frac{i}{2}\left[\alpha_{12}\bar{v}_{01}(t)+\alpha_{22}\bar{v}_{02}(t)+\bar{\alpha}_{23}\bar{v}_{03}(t)\right], \vspace{0.1in} \\
\phi_{43}^{(1)}(t)=\d\frac{i}{2}\left[\alpha_{13}\bar{v}_{01}(t)+\alpha_{23}\bar{v}_{02}(t)+\alpha_{33}\bar{v}_{03}(t)\right], \vspace{0.1in} \\
\phi_{14}^{(2)}=\d\frac{1}{4}v_{11}+\frac{1}{2i}v_{01}\phi_{44}^{(1)},\quad
\phi_{24}^{(2)}=\d\frac{1}{4}v_{12}+\frac{1}{2i}v_{02}\phi_{44}^{(1)},\quad
\phi_{34}^{(2)}=\d\frac{1}{4}v_{12}+\frac{1}{2i}v_{03}\phi_{44}^{(1)}, \vspace{0.1in} \\
\phi_{44}^{(1)}=\d\frac{1}{2}\int^t_0\Big\{v_{11}\left[\alpha_{11}\bar{v}_{01}(t)\!+\!\bar{\alpha}_{12}\bar{v}_{02}(t)\!+\!
  \bar{\alpha}_{13}\bar{v}_{03}(t)\right]\!+\!v_{12}\left[\alpha_{12}\bar{v}_{01}(t)\!+\!\alpha_{22}\bar{v}_{02}(t)\!+\!
  \bar{\alpha}_{23}\bar{v}_{03}(t)\right] \vspace{0.1in}  \\
 \quad\quad +v_{13}\left[\alpha_{13}\bar{v}_{01}(t)+\alpha_{23}\bar{v}_{02}(t)+\alpha_{33}\bar{v}_{03}(t)\right]
 -v_{01}\left[\alpha_{11}\bar{v}_{11}(t)+\bar{\alpha}_{12}\bar{v}_{12}(t)+\bar{\alpha}_{13}\bar{v}_{13}(t)\right]  \vspace{0.1in} \\
\quad\quad  -v_{02}\left[\alpha_{12}\bar{u}_{11}(t)+\alpha_{22}\bar{v}_{12}(t)+\bar{\alpha}_{23}\bar{v}_{13}(t)\right]
 -v_{03}\left[\alpha_{13}\bar{v}_{11}(t)+\alpha_{23}\bar{v}_{12}(t)+\alpha_{33}\bar{v}_{13}(t)\right]\Big\}dt,
\end{array}\right.
\ene
which generates the following expressions for the boundary values at $x=L$
\bee\label{vd}
\left\{\begin{array}{l}
v_{01}(t)=2i\phi_{14}^{(1)}(t),\quad  v_{02}(t)=2i\phi_{24}^{(1)}(t), \quad  v_{03}(t)=2i\phi_{34}^{(1)}(t),  \vspace{0.1in} \\
v_{11}(t)= 4\phi_{14}^{(2)}(t)+2iv_{01}(t)\phi_{44}^{(1)}(t),  \vspace{0.1in} \\
v_{12}(t)= 4\phi_{24}^{(2)}(t)+2iv_{02}(t)\phi_{44}^{(1)}(t),  \vspace{0.1in} \\
v_{13}(t)= 4\phi_{34}^{(2)}(t)+2iv_{03}(t)\phi_{44}^{(1)}(t),
\end{array}\right.
\ene

For the vanishing initial values, it follows from Eq.~(\ref{gr}) that we have the following asymptotic of the global relation $c_{j4}(t,k)$ and $c_{4j}(t,k), j=1,2,3$.

\vspace{0.1in}
\noindent {\bf Proposition 4.1.} {\it Let the initial and Dirichlet boundary conditions be compatible at points $x=0, L$ (i.e.,
$q_{0j}(0)=u_{0j}(0)$ at $x=0$ and  $q_{0j}(L)=v_{0j}(0)$ at $x=L$,\, $j=1,2,3$). Then,
the global relation (\ref{gr}) with the vanishing initial data implies that the large $k$ behaviors of
 $c_{j4}(t,k)$ and $c_{4j}(t,k), j=1,2,3$ are of the form }
\bes\bee
&\label{c14} 
\begin{array}{rl}
c_{14}(t,k)=&\d \frac{\Psi_{14}^{(1)}}{k}+\frac{\Psi_{14}^{(2)}+\Psi_{14}^{(1)}\bar{\phi}_{44}^{(1)}}{k^2}
+O\left(\frac{1}{k^3}\right) \vspace{0.08in}\\ &\d-\left\{\frac{\alpha_{11}\bar{\phi}_{41}^{(1)}+\bar{\alpha}_{12}\bar{\phi}_{42}^{(1)}+\bar{\alpha}_{13}\bar{\phi}_{43}^{(1)}}{k} +\frac{1}{k^2}\left[\alpha_{11}\bar{\phi}_{41}^{(2)}+\bar{\alpha}_{12}\bar{\phi}_{42}^{(2)}+\bar{\alpha}_{13}\bar{\phi}_{43}^{(2)} \right.\right.\vspace{0.08in}\\
&\d+\Psi_{11}^{(1)}\left(\alpha_{11}\bar{\phi}_{41}^{(1)}+\bar{\alpha}_{12}\bar{\phi}_{42}^{(1)}
 +\bar{\alpha}_{13}\bar{\phi}_{43}^{(1)}\right)
+\Psi_{12}^{(1)}\left(\alpha_{12}\bar{\phi}_{41}^{(1)}+\alpha_{22}\bar{\phi}_{42}^{(1)}
+\bar{\alpha}_{23}\bar{\phi}_{43}^{(1)}\right) \qquad\quad\vspace{0.08in}\\
& \d +\left.\left. \Psi_{13}^{(1)}\left(\alpha_{13}\bar{\phi}_{41}^{(1)}+\alpha_{23}\bar{\phi}_{42}^{(1)}+\alpha_{33}\bar{\phi}_{43}^{(1)}\right)\right]
  +O\left(\frac{1}{k^3}\right)\right\}e^{2ikL},\quad k\to\infty,
\end{array} \\
&\label{c24} \begin{array}{rl}
c_{24}(t,k)=&\d \frac{\Psi_{24}^{(1)}}{k}+\frac{\Psi_{24}^{(2)}+\Psi_{24}^{(1)}\bar{\phi}_{44}^{(1)}}{k^2}
+O\left(\frac{1}{k^3}\right) \vspace{0.08in}\\
&\d -\left\{\frac{\alpha_{12}\bar{\phi}_{41}^{(1)}+\alpha_{22}\bar{\phi}_{42}^{(1)}+\bar{\alpha}_{23}\bar{\phi}_{43}^{(1)}}{k} +\frac{1}{k^2}\left[\alpha_{12}\bar{\phi}_{41}^{(2)}+\alpha_{22}\bar{\phi}_{42}^{(2)}+\bar{\alpha}_{23}\bar{\phi}_{43}^{(2)} \right.\right.\vspace{0.08in}\\
&\d+\Psi_{21}^{(1)}\left(\alpha_{11}\bar{\phi}_{41}^{(1)}+\bar{\alpha}_{12}\bar{\phi}_{42}^{(1)}
 +\bar{\alpha}_{13}\bar{\phi}_{43}^{(1)}\right)
+\Psi_{22}^{(1)}\left(\alpha_{12}\bar{\phi}_{41}^{(1)}+\alpha_{22}\bar{\phi}_{42}^{(1)}
+\bar{\alpha}_{23}\bar{\phi}_{43}^{(1)}\right) \qquad\quad \vspace{0.08in}\\
& \d +\left.\left. \Psi_{23}^{(1)}\left(\alpha_{13}\bar{\phi}_{41}^{(1)}+\alpha_{23}\bar{\phi}_{42}^{(1)}+\alpha_{33}\bar{\phi}_{43}^{(1)}\right)\right]
  +O\left(\frac{1}{k^3}\right)\right\}e^{2ikL},\quad k\to\infty,
\end{array}\\
&\label{c34} \begin{array}{rl}
c_{34}(t,k)=&\d \frac{\Psi_{34}^{(1)}}{k}+\frac{\Psi_{34}^{(2)}+\Psi_{34}^{(1)}\bar{\phi}_{44}^{(1)}}{k^2}
+O\left(\frac{1}{k^3}\right) \vspace{0.08in}\\
&\d-\left\{\frac{\alpha_{13}\bar{\phi}_{41}^{(1)}+\alpha_{23}\bar{\phi}_{42}^{(1)}+\alpha_{33}\bar{\phi}_{43}^{(1)}}{k} +\frac{1}{k^2}\left[\alpha_{13}\bar{\phi}_{41}^{(2)}+\alpha_{23}\bar{\phi}_{42}^{(2)}+\alpha_{33}\bar{\phi}_{43}^{(2)} \right.\right.\vspace{0.08in}\\
&\d+\Psi_{31}^{(1)}\left(\alpha_{11}\bar{\phi}_{41}^{(1)}+\bar{\alpha}_{12}\bar{\phi}_{42}^{(1)}
 +\bar{\alpha}_{13}\bar{\phi}_{43}^{(1)}\right)
+\Psi_{32}^{(1)}\left(\alpha_{12}\bar{\phi}_{41}^{(1)}+\alpha_{22}\bar{\phi}_{42}^{(1)}
+\bar{\alpha}_{23}\bar{\phi}_{43}^{(1)}\right) \qquad\quad \vspace{0.08in}\\
& \d +\left.\left. \Psi_{33}^{(1)}\left(\alpha_{13}\bar{\phi}_{41}^{(1)}+\alpha_{23}\bar{\phi}_{42}^{(1)}+\alpha_{33}\bar{\phi}_{43}^{(1)}\right)\right]
  +O\left(\frac{1}{k^3}\right)\right\}e^{2ikL},\quad k\to\infty,
\end{array}
\ene
\ees

\bee
\label{c41} \begin{array}{rl}
c_{41}(t,k)=&\d -\left\{\frac{\alpha_{11}\bar{\phi}_{14}^{(1)}+\alpha_{12}\bar{\phi}_{24}^{(1)}+\alpha_{13}\bar{\phi}_{34}^{(1)}}{k}
+\frac{1}{k^2}\left[\alpha_{11}\bar{\phi}_{14}^{(2)}+\alpha_{12}\bar{\phi}_{24}^{(2)}+\alpha_{13}\bar{\phi}_{34}^{(2)} \right.\right. \vspace{0.08in}\\
&\d \left.\left.+\Psi_{44}^{(1)}\left(\alpha_{11}\bar{\phi}_{14}^{(1)}+\alpha_{12}\bar{\phi}_{24}^{(1)}+\alpha_{13}\bar{\phi}_{34}^{(1)}\right)\right]
+O\left(\frac{1}{k^3}\right)\right\}e^{-2ikL} \vspace{0.08in}\\
&\d+\frac{1}{k}\left[(\alpha_{11}^2+|\alpha_{12}|^2+|\alpha_{13}|^2)\Psi_{41}^{(1)}
 +(\alpha_{11}\alpha_{12}+\alpha_{12}\alpha_{22}+\alpha_{13}\bar{\alpha}_{23})\Psi_{42}^{(1)} \right.\vspace{0.1in}\\
&\d \left. +(\alpha_{11}\alpha_{13}+\alpha_{12}\alpha_{23}+\alpha_{13}\alpha_{33})\Psi_{43}^{(1)}\right] +\frac{1}{k^2}\left[(\alpha_{11}^2+|\alpha_{12}|^2+|\alpha_{13}|^2)\Psi_{41}^{(2)} \right.\vspace{0.1in}\\
& +(\alpha_{11}\alpha_{12}+\alpha_{12}\alpha_{22}+\alpha_{13}\bar{\alpha}_{23})\Psi_{42}^{(2)}
 +(\alpha_{11}\alpha_{13}+\alpha_{12}\alpha_{23}+\alpha_{13}\alpha_{33})\Psi_{43}^{(2)} \vspace{0.1in}\\
& \d   +\Psi_{41}^{(1)}\left[\alpha_{11}\left(\alpha_{11}\bar{\phi}_{11}^{(1)}+\bar{\alpha}_{12}\bar{\phi}_{12}^{(1)}
          +\bar{\alpha}_{13}\bar{\phi}_{13}^{(1)}\right)
     +\alpha_{12}\left(\alpha_{11}\bar{\phi}_{21}^{(1)}+\bar{\alpha}_{12}\bar{\phi}_{22}^{(1)}+\bar{\alpha}_{13}\bar{\phi}_{23}^{(1)}\right)
     \right.\vspace{0.1in}\\
 & \left.+\alpha_{13}\left(\alpha_{11}\bar{\phi}_{31}^{(1)}+\bar{\alpha}_{12}\bar{\phi}_{32}^{(1)}+\bar{\alpha}_{13}\bar{\phi}_{33}^{(1)}\right)\right]
   +\Psi_{42}^{(1)}\left[\alpha_{11}\left(\alpha_{12}\bar{\phi}_{11}^{(1)}+\alpha_{22}\bar{\phi}_{12}^{(1)}
    +\bar{\alpha}_{23}\bar{\phi}_{13}^{(1)}\right) \right.\vspace{0.1in}\\
\end{array}
\ene

\bee
\label{c42} \begin{array}{rl}
c_{42}(t,k)=&\d -\left\{\frac{\bar{\alpha}_{12}\bar{\phi}_{14}^{(1)}+\alpha_{22}\bar{\phi}_{24}^{(1)}+\alpha_{23}\bar{\phi}_{34}^{(1)}}{k}
+\frac{1}{k^2}\left[\bar{\alpha}_{12}\bar{\phi}_{14}^{(2)}+\alpha_{22}\bar{\phi}_{24}^{(2)}+\alpha_{23}\bar{\phi}_{34}^{(2)} \right.\right. \vspace{0.08in}\\
&\d \left.\left.+\Psi_{44}^{(1)}\left(\bar{\alpha}_{12}\bar{\phi}_{14}^{(1)}+\alpha_{22}\bar{\phi}_{24}^{(1)}
   +\alpha_{23}\bar{\phi}_{34}^{(1)}\right)\right]+O\left(\frac{1}{k^3}\right)\right\}e^{-2ikL} \vspace{0.08in}\\
&\d+\frac{1}{k}\left\{(|\alpha_{12}|^2+\alpha_{22}^2+|\alpha_{23}|^2)\Psi_{42}^{(1)}
 +(\alpha_{11}\bar{\alpha}_{12}+\bar{\alpha}_{12}\bar{\alpha}_{22}+\bar{\alpha}_{13}\alpha_{23})\Psi_{41}^{(1)} \right.\vspace{0.1in}\\
&\d +(\bar{\alpha}_{12}\alpha_{13}+\alpha_{22}\alpha_{23}+\alpha_{23}\alpha_{33})\Psi_{43}^{(1)} +\frac{1}{k^2}\left[(|\alpha_{12}|^2+\alpha_{22}^2+|\alpha_{23}|^2)\Psi_{42}^{(2)} \right.\vspace{0.1in}\\
& +(\alpha_{11}\bar{\alpha}_{12}+\bar{\alpha}_{12}\bar{\alpha}_{22}+\bar{\alpha}_{13}\alpha_{23})\Psi_{41}^{(2)}
 +(\bar{\alpha}_{12}\alpha_{13}+\alpha_{22}\alpha_{23}+\alpha_{23}\alpha_{33})\Psi_{43}^{(2)}  \vspace{0.1in}\\
& \d   +\Psi_{41}^{(1)}\left[\bar{\alpha}_{12}\left(\alpha_{11}\bar{\phi}_{11}^{(1)}+\bar{\alpha}_{12}\bar{\phi}_{12}^{(1)}
      +\bar{\alpha}_{13}\bar{\phi}_{13}^{(1)}\right)
 +\alpha_{22}\left(\alpha_{11}\bar{\phi}_{21}^{(1)}+\bar{\alpha}_{12}\bar{\phi}_{22}^{(1)}+\bar{\alpha}_{13}\bar{\phi}_{23}^{(1)}\right)
     \right.\vspace{0.1in}\\
& \left.+\alpha_{23}\left(\alpha_{11}\bar{\phi}_{31}^{(1)}+\bar{\alpha}_{12}\bar{\phi}_{32}^{(1)}
    +\bar{\alpha}_{13}\bar{\phi}_{33}^{(1)}\right)\right]
   +\Psi_{42}^{(1)}\left[\bar{\alpha}_{12}\left(\alpha_{12}\bar{\phi}_{11}^{(1)}+\alpha_{22}\bar{\phi}_{12}^{(1)}
    +\bar{\alpha}_{23}\bar{\phi}_{13}^{(1)}\right) \right.\vspace{0.1in}\\
\end{array}
\ene
\bee
\label{c43} \begin{array}{rl}
c_{43}(t,k)=&\d -\left\{\frac{\bar{\alpha}_{13}\bar{\phi}_{14}^{(1)}+\bar{\alpha}_{23}\bar{\phi}_{24}^{(1)}+\alpha_{33}\bar{\phi}_{34}^{(1)}}{k}
+\frac{1}{k^2}\left[\bar{\alpha}_{13}\bar{\phi}_{14}^{(2)}+\bar{\alpha}_{23}\bar{\phi}_{24}^{(2)}+\alpha_{33}\bar{\phi}_{34}^{(2)} \right.\right. \vspace{0.08in}\\
&\d \left.\left.+\Psi_{44}^{(1)}\left(\bar{\alpha}_{13}\bar{\phi}_{14}^{(1)}+\bar{\alpha}_{23}\bar{\phi}_{24}^{(1)}+\alpha_{33}\bar{\phi}_{34}^{(1)}\right)\right]+O\left(\frac{1}{k^3}\right)\right\}e^{-2ikL} \vspace{0.08in}\\
&\d+\frac{1}{k}\left\{(|\alpha_{13}|^2+|\alpha_{23}|^2+\alpha_{33}^2)\Psi_{43}^{(1)}
 +(\alpha_{11}\bar{\alpha}_{13}+\bar{\alpha}_{12}\bar{\alpha}_{23}+\bar{\alpha}_{13}\alpha_{33})\Psi_{41}^{(1)} \right.\vspace{0.1in}\\
&\d +(\bar{\alpha}_{13}\alpha_{12}+\alpha_{22}\bar{\alpha}_{23}+\alpha_{13}\bar{\alpha}_{23})\Psi_{42}^{(1)} +\frac{1}{k^2}\left[(|\alpha_{13}|^2+|\alpha_{23}|^2+\alpha_{33}^2)\Psi_{43}^{(2)} \right.\vspace{0.1in}\\
& +(\alpha_{11}\bar{\alpha}_{12}+\bar{\alpha}_{12}\bar{\alpha}_{22}+\bar{\alpha}_{13}\alpha_{23})\Psi_{41}^{(2)}
 +(\bar{\alpha}_{12}\alpha_{13}+\alpha_{22}\alpha_{23}+\alpha_{23}\alpha_{33})\Psi_{43}^{(2)}  \vspace{0.1in}\\
& \d  +\Psi_{41}^{(1)}\left[\bar{\alpha}_{13}\left(\alpha_{11}\bar{\phi}_{11}^{(1)}+\bar{\alpha}_{12}\bar{\phi}_{12}^{(1)}
      +\bar{\alpha}_{13}\bar{\phi}_{13}^{(1)}\right)
 +\bar{\alpha}_{23}\left(\alpha_{11}\bar{\phi}_{21}^{(1)}+\bar{\alpha}_{12}\bar{\phi}_{22}^{(1)}+\bar{\alpha}_{13}\bar{\phi}_{23}^{(1)}\right)
     \right.\vspace{0.1in}\\
& \left.+\alpha_{33}\left(\alpha_{11}\bar{\phi}_{31}^{(1)}+\bar{\alpha}_{12}\bar{\phi}_{32}^{(1)}
    +\bar{\alpha}_{13}\bar{\phi}_{33}^{(1)}\right)\right]
   +\Psi_{42}^{(1)}\left[\bar{\alpha}_{13}\left(\alpha_{12}\bar{\phi}_{11}^{(1)}+\alpha_{22}\bar{\phi}_{12}^{(1)}
    +\bar{\alpha}_{23}\bar{\phi}_{13}^{(1)}\right) \right.\vspace{0.1in}\\
\end{array}
\ene

\vspace{0.1in}
\noindent {\bf Proof.}  The global relation (\ref{gr}) under the vanishing initial data can be simplified as
\bes\bee
& \label{ca}
\begin{array}{rl}
c_{14}(t,k)\!=&\!\!\! \Psi_{14}(t,k)\bar{\phi}_{44}(t,\bar{k})-e^{2ikL}\big[\Psi_{11}(t,k)(\alpha_{11}\bar{\phi}_{41}(t,\bar{k})
 +\bar{\alpha}_{12}\bar{\phi}_{42}(t,\bar{k}) +\bar{\alpha}_{13}\bar{\phi}_{43}(t,\bar{k})) \vspace{0.1in}\\
 &\!\!\! + \Psi_{12}(t,k)(\alpha_{12}\bar{\phi}_{41}(t,\bar{k})
 +\alpha_{22}\bar{\phi}_{42}(t,\bar{k})+\bar{\alpha}_{23}\bar{\phi}_{43}(t,\bar{k})) \vspace{0.1in}\\
 &\!\!\! + \Psi_{13}(t,k)(\alpha_{13}\bar{\phi}_{41}(t,\bar{k})
 +\alpha_{23}\bar{\phi}_{42}(t,\bar{k})+\alpha_{33}\bar{\phi}_{43}(t,\bar{k}))\big],
 \end{array}
 \vspace{0.1in}\\
&\begin{array}{rl}
c_{24}(t,k)\!=&\!\!\! \Psi_{24}(t,k)\bar{\phi}_{44}(t,\bar{k})-e^{2ikL}\big[\Psi_{21}(t,k)(\alpha_{11}\bar{\phi}_{41}(t,\bar{k})
 +\bar{\alpha}_{12}\bar{\phi}_{42}(t,\bar{k}) +\bar{\alpha}_{13}\bar{\phi}_{43}(t,\bar{k})) \vspace{0.1in}\\
 &\!\!\! + \Psi_{22}(t,k)(\alpha_{12}\bar{\phi}_{41}(t,\bar{k})
 +\alpha_{22}\bar{\phi}_{42}(t,\bar{k})+\bar{\alpha}_{23}\bar{\phi}_{43}(t,\bar{k})) \vspace{0.1in}\\
 &\!\!\! + \Psi_{23}(t,k)(\alpha_{13}\bar{\phi}_{41}(t,\bar{k})
 +\alpha_{23}\bar{\phi}_{42}(t,\bar{k})+\alpha_{33}\bar{\phi}_{43}(t,\bar{k}))\big],
 \end{array} \vspace{0.1in}\\
& \begin{array}{rl}
c_{34}(t,k)\!=&\!\!\! \Psi_{34}(t,k)\bar{\phi}_{44}(t,\bar{k})-e^{2ikL}\big[\Psi_{31}(t,k)(\alpha_{11}\bar{\phi}_{41}(t,\bar{k})
 +\bar{\alpha}_{12}\bar{\phi}_{42}(t,\bar{k}) +\bar{\alpha}_{13}\bar{\phi}_{43}(t,\bar{k})) \vspace{0.1in}\\
 &\!\!\! + \Psi_{32}(t,k)(\alpha_{12}\bar{\phi}_{41}(t,\bar{k})
 +\alpha_{22}\bar{\phi}_{42}(t,\bar{k})+\bar{\alpha}_{23}\bar{\phi}_{43}(t,\bar{k})) \vspace{0.1in}\\
 &\!\!\! + \Psi_{33}(t,k)(\alpha_{13}\bar{\phi}_{41}(t,\bar{k})
 +\alpha_{23}\bar{\phi}_{42}(t,\bar{k})+\alpha_{33}\bar{\phi}_{43}(t,\bar{k}))\big],
 \end{array}
\ene\ees
where $\bar{\phi}_{ij}(t,\bar{k})=\overline{\phi_{ij}(t,\bar{k})}$.

Recalling the time-part of the Lax pair (\ref{mulax})
\bee\label{tlax}
 \mu_t+2ik^2[\sigma_4, \mu]=V(x,t,k)\mu,
 \ene
It follows from the first column of Eq.~(\ref{tlax}) with $\mu=\mu_2$ that we have
\bee\label{psit11}
\left\{ \begin{array}{rl}
 \Psi_{11,t}(t,k)=&\!\!\! (2ku_{01}+iu_{11})\Psi_{41}
   -i\Psi_{11}(\alpha_{11}|u_{01}|^2+\bar{\alpha}_{12}u_{01}\bar{u}_{02}+\bar{\alpha}_{13}u_{01}\bar{u}_{03})\vspace{0.08in}\\
   &\!\!\! -i\Psi_{21}(\alpha_{12}|u_{01}|^2+\alpha_{22}u_{01}\bar{u}_{02}+\bar{\alpha}_{23}u_{01}\bar{u}_{03}) \vspace{0.08in}\\
  &\!\!\! -i\Psi_{31}(\alpha_{13}|u_{01}|^2+\alpha_{23}u_{01}\bar{u}_{02}+\alpha_{33}u_{01}\bar{u}_{03}),  \vspace{0.08in}\\
 \Psi_{21,t}(t,k)=&\!\!\! (2ku_{02}+iu_{12})\Psi_{41}
   -i\Psi_{11}(\alpha_{11}u_{02}\bar{u}_{01}+\bar{\alpha}_{12}|u_{02}|^2+\bar{\alpha}_{13}u_{02}\bar{u}_{03})\vspace{0.08in}\\
   &\!\!\! -i\Psi_{21}(\alpha_{12}u_{02}\bar{u}_{01}+\alpha_{22}|u_{02}|^2+\bar{\alpha}_{23}u_{02}\bar{u}_{03}) \vspace{0.08in}\\
   &\!\!\! -i\Psi_{31}(\alpha_{13}u_{02}\bar{u}_{01}+\alpha_{23}|u_{02}|^2+\alpha_{33}u_{02}\bar{u}_{03}),  \vspace{0.08in}\\
 \Psi_{31,t}(t,k)=& (2ku_{03}+iu_{13})\Psi_{41}   -i\Psi_{11}(\alpha_{11}u_{03}\bar{u}_{01}+\bar{\alpha}_{12}u_{03}\bar{u}_{02}+\bar{\alpha}_{13}|u_{03}|^2)\vspace{0.08in}\\
   &\!\!\! -i\Psi_{21}(\alpha_{12}u_{03}\bar{u}_{01}+\alpha_{22}u_{03}\bar{u}_{02}+\bar{\alpha}_{23}|u_{03}|^2) \vspace{0.08in}\\
\end{array}\right.
\ene

The second column of Eq.~(\ref{tlax}) with $\mu=\mu_2$ yields
\bee\label{psit12}
\left\{ \begin{array}{rl}
 \Psi_{12,t}(t,k)=&\!\!\! (2ku_{01}+iu_{11})\Psi_{42}
   -i\Psi_{12}(\alpha_{11}|u_{01}|^2+\bar{\alpha}_{12}u_{01}\bar{u}_{02}+\bar{\alpha}_{13}u_{01}\bar{u}_{03})\vspace{0.08in}\\
   &\!\!\! -i\Psi_{22}(\alpha_{12}|u_{01}|^2+\alpha_{22}u_{01}\bar{u}_{02}+\bar{\alpha}_{23}u_{01}\bar{u}_{03}) \vspace{0.08in}\\
  &\!\!\! -i\Psi_{32}(\alpha_{13}|u_{01}|^2+\alpha_{23}u_{01}\bar{u}_{02}+\alpha_{33}u_{01}\bar{u}_{03}),  \vspace{0.08in}\\
 \Psi_{22,t}(t,k)=&\!\!\! (2ku_{02}+iu_{12})\Psi_{42}
   -i\Psi_{12}(\alpha_{11}u_{02}\bar{u}_{01}+\bar{\alpha}_{12}|u_{02}|^2+\bar{\alpha}_{13}u_{02}\bar{u}_{03})\vspace{0.08in}\\
   &\!\!\! -i\Psi_{22}(\alpha_{12}u_{02}\bar{u}_{01}+\alpha_{22}|u_{02}|^2+\bar{\alpha}_{23}u_{02}\bar{u}_{03}) \vspace{0.08in}\\
   &\!\!\! -i\Psi_{32}(\alpha_{13}u_{02}\bar{u}_{01}+\alpha_{23}|u_{02}|^2+\alpha_{33}u_{02}\bar{u}_{03}),  \vspace{0.08in}\\
 \Psi_{32,t}(t,k)=& (2ku_{03}+iu_{13})\Psi_{42}   -i\Psi_{12}(\alpha_{11}u_{03}\bar{u}_{01}+\bar{\alpha}_{12}u_{03}\bar{u}_{02}+\bar{\alpha}_{13}|u_{03}|^2)\vspace{0.08in}\\
   &\!\!\! -i\Psi_{22}(\alpha_{12}u_{03}\bar{u}_{01}+\alpha_{22}u_{03}\bar{u}_{02}+\bar{\alpha}_{23}|u_{03}|^2) \vspace{0.08in}\\
\end{array}\right.
\ene
The third column of Eq.~(\ref{tlax}) with $\mu=\mu_2$ yields
\bee\label{psit13}
\left\{ \begin{array}{rl}
 \Psi_{13,t}(t,k)=&\!\!\! (2ku_{01}+iu_{11})\Psi_{43}
   -i\Psi_{13}(\alpha_{11}|u_{01}|^2+\bar{\alpha}_{12}u_{01}\bar{u}_{02}+\bar{\alpha}_{13}u_{01}\bar{u}_{03})\vspace{0.08in}\\
   &\!\!\! -i\Psi_{23}(\alpha_{12}|u_{01}|^2+\alpha_{22}u_{01}\bar{u}_{02}+\bar{\alpha}_{23}u_{01}\bar{u}_{03}) \vspace{0.08in}\\
  &\!\!\! -i\Psi_{33}(\alpha_{13}|u_{01}|^2+\alpha_{23}u_{01}\bar{u}_{02}+\alpha_{33}u_{01}\bar{u}_{03}),  \vspace{0.08in}\\
 \Psi_{23,t}(t,k)=&\!\!\! (2ku_{02}+iu_{12})\Psi_{43}
   -i\Psi_{13}(\alpha_{11}u_{02}\bar{u}_{01}+\bar{\alpha}_{12}|u_{02}|^2+\bar{\alpha}_{13}u_{02}\bar{u}_{03})\vspace{0.08in}\\
   &\!\!\! -i\Psi_{23}(\alpha_{12}u_{02}\bar{u}_{01}+\alpha_{22}|u_{02}|^2+\bar{\alpha}_{23}u_{02}\bar{u}_{03}) \vspace{0.08in}\\
   &\!\!\! -i\Psi_{33}(\alpha_{13}u_{02}\bar{u}_{01}+\alpha_{23}|u_{02}|^2+\alpha_{33}u_{02}\bar{u}_{03}),  \vspace{0.08in}\\
 \Psi_{33,t}(t,k)=& (2ku_{03}+iu_{13})\Psi_{43}   -i\Psi_{13}(\alpha_{11}u_{03}\bar{u}_{01}+\bar{\alpha}_{12}u_{03}\bar{u}_{02}+\bar{\alpha}_{13}|u_{03}|^2)\vspace{0.08in}\\
   &\!\!\! -i\Psi_{23}(\alpha_{12}u_{03}\bar{u}_{01}+\alpha_{22}u_{03}\bar{u}_{02}+\bar{\alpha}_{23}|u_{03}|^2) \vspace{0.08in}\\
\end{array}\right.
\ene

The fourth column of Eq.~(\ref{tlax}) with $\mu=\mu_2$ yields
\bee\label{psit14}
\left\{ \begin{array}{rl}
 \Psi_{14,t}(t,k)=&\!\!\! (2ku_{01}+iu_{11})\Psi_{44}
   -i\Psi_{14}(4k^2+\alpha_{11}|u_{01}|^2+\bar{\alpha}_{12}u_{01}\bar{u}_{02}+\bar{\alpha}_{13}u_{01}\bar{u}_{03})\vspace{0.08in}\\
   &\!\!\! -i\Psi_{24}(\alpha_{12}|u_{01}|^2+\alpha_{22}u_{01}\bar{u}_{02}+\bar{\alpha}_{23}u_{01}\bar{u}_{03}) \vspace{0.08in}\\
  &\!\!\! -i\Psi_{34}(\alpha_{13}|u_{01}|^2+\alpha_{23}u_{01}\bar{u}_{02}+\alpha_{33}u_{01}\bar{u}_{03}),  \vspace{0.08in}\\
 \Psi_{24,t}(t,k)=&\!\!\! (2ku_{02}+iu_{12})\Psi_{44}
   -i\Psi_{14}(\alpha_{11}u_{02}\bar{u}_{01}+\bar{\alpha}_{12}|u_{02}|^2+\bar{\alpha}_{13}u_{02}\bar{u}_{03})\vspace{0.08in}\\
   &\!\!\! -i\Psi_{24}(4k^2+\alpha_{12}u_{02}\bar{u}_{01}+\alpha_{22}|u_{02}|^2+\bar{\alpha}_{23}u_{02}\bar{u}_{03}) \vspace{0.08in}\\
   &\!\!\! -i\Psi_{34}(\alpha_{13}u_{02}\bar{u}_{01}+\alpha_{23}|u_{02}|^2+\alpha_{33}u_{02}\bar{u}_{03}),  \vspace{0.08in}\\
 \Psi_{34,t}(t,k)=& (2ku_{03}+iu_{13})\Psi_{44}   -i\Psi_{14}(\alpha_{11}u_{03}\bar{u}_{01}+\bar{\alpha}_{12}u_{03}\bar{u}_{02}+\bar{\alpha}_{13}|u_{03}|^2)\vspace{0.08in}\\
   &\!\!\! -i\Psi_{24}(\alpha_{12}u_{03}\bar{u}_{01}+\alpha_{22}u_{03}\bar{u}_{02}+\bar{\alpha}_{23}|u_{03}|^2) \vspace{0.08in}\\
\end{array}\right.
\ene

Suppose that $\Psi_{j1}$'s,\, $j=1,2,3,4$ are of the form
\bee\label{psi1}
\left(\begin{array}{c} \Psi_{11} \\ \Psi_{21} \\ \Psi_{31} \\ \Psi_{41} \end{array}\right)
=\left(a_{10}(t)+\frac{a_{11}(t)}{k}+\frac{a_{12}(t)}{k^2}+\cdots\right)+\left(b_{10}(t)+\frac{b_{11}(t)}{k}+\frac{b_{12}(t)}{k^2}+\cdots\right)e^{4ik^2t},
\ene
where the $4\times 1$ column vector functions $a_{1j}(t),b_{1j}(t)\, (j=0,1,...,)$ are independent of $k$.

By substituting Eq.~(\ref{psi1}) into Eq.(\ref{psit11}) and using the initial conditions
\bee\nonumber
a_{10}(0)+b_{10}(0)=(1, 0,0,0)^T, \quad a_{11}(0)+b_{11}(0)=(0, 0, 0, 0)^T,
\ene
we have
\bee\no
\left(\begin{array}{c} \Psi_{11} \\ \Psi_{21} \\ \Psi_{31} \\ \Psi_{41} \end{array}\right)
=\left(\begin{array}{c} 1 \\ 0 \\ 0 \\ 0 \end{array}\right)
+\sum_{s=1}^2\frac{1}{k^s}\left(\begin{array}{c} \Psi_{11}^{(s)} \vspace{0.05in}\\ \Psi_{21}^{(s)} \vspace{0.05in}\\ \Psi_{31}^{(s)} \vspace{0.05in}\\ \Psi_{41}^{(s)} \end{array}\right)
+O\left(\frac{1}{k^3}\right)\qquad\qquad\qquad\qquad \vspace{0.05in}\\
+\left[\frac{1}{2ik}\left(\begin{array}{c} 0 \\ 0 \vspace{0.05in}\\ 0 \vspace{0.05in}\\
 \alpha_{11}\bar{u}_{01}(0)+\bar{\alpha}_{12}\bar{u}_{02}(0)+\bar{\alpha}_{13}\bar{u}_{03}(0) \end{array}\right)+O\left(\frac{1}{k^2}\right)  \right]e^{4ik^2t},
\ene

Similarly,  it follows from Eqs.~(\ref{psit12})-(\ref{psit14}) that we have the asymptotic formulae for $\Psi_{ij},\, i=1,2,3,4; j=2,3,4$ in the form
\bee\no
\left(\begin{array}{c} \Psi_{12} \\ \Psi_{22} \\ \Psi_{32} \\ \Psi_{42} \end{array}\right)
=\left(\begin{array}{c} 1 \\ 0 \\ 0 \\ 0 \end{array}\right)
+\sum_{s=1}^2\frac{1}{k^s}\left(\begin{array}{c} \Psi_{12}^{(s)} \vspace{0.05in}\\ \Psi_{22}^{(s)} \vspace{0.05in}\\ \Psi_{32}^{(s)} \vspace{0.05in}\\ \Psi_{42}^{(s)} \end{array}\right)
+O\left(\frac{1}{k^3}\right)\qquad\qquad\qquad\qquad \vspace{0.05in}\\
+\left[\frac{1}{2ik}\left(\begin{array}{c} 0 \\ 0 \vspace{0.05in}\\ 0 \vspace{0.05in}\\
 \alpha_{12}\bar{u}_{01}(0)+\alpha_{22}\bar{u}_{02}(0)+\bar{\alpha}_{23}\bar{u}_{03}(0) \end{array}\right)+O\left(\frac{1}{k^2}\right)\right]e^{4ik^2t},
\ene
\bee\no
\left(\begin{array}{c} \Psi_{13} \\ \Psi_{23} \\ \Psi_{33} \\ \Psi_{43} \end{array}\right)
=\left(\begin{array}{c} 1 \\ 0 \\ 0 \\ 0 \end{array}\right)
+\sum_{s=1}^2\frac{1}{k^s}\left(\begin{array}{c} \Psi_{13}^{(s)} \vspace{0.05in}\\ \Psi_{23}^{(s)} \vspace{0.05in}\\ \Psi_{33}^{(s)} \vspace{0.05in}\\ \Psi_{43}^{(s)} \end{array}\right)
+O\left(\frac{1}{k^3}\right)\qquad\qquad\qquad\qquad \vspace{0.05in}\\
+\left[\frac{1}{2ik}\left(\begin{array}{c} 0 \vspace{0.05in}\\  0 \vspace{0.05in}\\ 0 \vspace{0.05in}\\
\alpha_{13}\bar{u}_{01}(0)+\alpha_{23}\bar{u}_{02}(0)+\alpha_{33}\bar{u}_{03}(0)
\end{array}\right)+O\left(\frac{1}{k^2}\right)  \right]e^{4ik^2t},
\ene
and
\bee
\left(\begin{array}{c} \Psi_{14} \\ \Psi_{24} \\ \Psi_{34} \\ \Psi_{44} \end{array}\right)
=\left(\begin{array}{c} 1 \\ 0 \\ 0 \\ 0 \end{array}\right)
+\sum_{s=1}^2\frac{1}{k^s}\left(\begin{array}{c} \Psi_{14}^{(s)} \vspace{0.05in}\\ \Psi_{24}^{(s)} \vspace{0.05in}\\ \Psi_{34}^{(s)} \vspace{0.05in}\\ \Psi_{44}^{(s)} \end{array}\right)
+O\left(\frac{1}{k^3}\right)
+\left[\frac{i}{2k}\left(\begin{array}{c} u_{01}(0) \vspace{0.05in}\\  u_{02}(0) \vspace{0.05in}\\ u_{03}(0) \vspace{0.05in}\\ 0\end{array}\right)+O\left(\frac{1}{k^2}\right) \right]e^{-4ik^2t},
\ene

Similar to Eqs.~(\ref{psit11})-(\ref{psit14}) for $\mu_2(0,t,k)$, we also know that the function $\mu(x,t,k)=\mu_3(L, t,k)$ at $x=L$ satisfy the $t$-part of Lax pair (\ref{tlax}).

The first column of Eq.~(\ref{tlax}) with $\mu=\mu_3$ yields
\bee
\left\{ \begin{array}{rl}
 \phi_{11,t}(t,k)=&\!\!\! (2kv_{01}+iv_{11})\phi_{41}
   -i\phi_{11}(\alpha_{11}|v_{01}|^2+\bar{\alpha}_{12}v_{01}\bar{v}_{02}+\bar{\alpha}_{13}v_{01}\bar{v}_{03})\vspace{0.08in}\\
   &\!\!\! -i\phi_{21}(\alpha_{12}|v_{01}|^2+\alpha_{22}v_{01}\bar{v}_{02}+\bar{\alpha}_{23}v_{01}\bar{v}_{03}) \vspace{0.08in}\\
  &\!\!\! -i\phi_{31}(\alpha_{13}|v_{01}|^2+\alpha_{23}v_{01}\bar{v}_{02}+\alpha_{33}v_{01}\bar{v}_{03}),  \vspace{0.08in}\\
 \phi_{21,t}(t,k)=&\!\!\! (2kv_{02}+iv_{12})\phi_{41}
   -i\phi_{11}(\alpha_{11}v_{02}\bar{v}_{01}+\bar{\alpha}_{12}|v_{02}|^2+\bar{\alpha}_{13}v_{02}\bar{v}_{03})\vspace{0.08in}\\
   &\!\!\! -i\phi_{21}(\alpha_{12}v_{02}\bar{v}_{01}+\alpha_{22}|v_{02}|^2+\bar{\alpha}_{23}v_{02}\bar{v}_{03}) \vspace{0.08in}\\
   &\!\!\! -i\phi_{31}(\alpha_{13}v_{02}\bar{v}_{01}+\alpha_{23}|v_{02}|^2+\alpha_{33}v_{02}\bar{v}_{03}),  \vspace{0.08in}\\
 \phi_{31,t}(t,k)=& (2kv_{03}+iv_{13})\phi_{41}   -i\phi_{11}(\alpha_{11}v_{03}\bar{v}_{01}+\bar{\alpha}_{12}v_{03}\bar{v}_{02}+\bar{\alpha}_{13}|v_{03}|^2)\vspace{0.08in}\\
   &\!\!\! -i\phi_{21}(\alpha_{12}v_{03}\bar{v}_{01}+\alpha_{22}v_{03}\bar{v}_{02}+\bar{\alpha}_{23}|v_{03}|^2) \vspace{0.08in}\\
\end{array}\right.
\ene
The second column of Eq.~(\ref{tlax}) with $\mu=\mu_3$  yields
\bee
\left\{ \begin{array}{rl}
 \phi_{12,t}(t,k)=&\!\!\! (2kv_{01}+iv_{11})\phi_{42}
   -i\phi_{12}(\alpha_{11}|v_{01}|^2+\bar{\alpha}_{12}v_{01}\bar{v}_{02}+\bar{\alpha}_{13}v_{01}\bar{v}_{03})\vspace{0.08in}\\
   &\!\!\! -i\phi_{22}(\alpha_{12}|v_{01}|^2+\alpha_{22}v_{01}\bar{v}_{02}+\bar{\alpha}_{23}v_{01}\bar{v}_{03}) \vspace{0.08in}\\
  &\!\!\! -i\phi_{32}(\alpha_{13}|v_{01}|^2+\alpha_{23}v_{01}\bar{v}_{02}+\alpha_{33}v_{01}\bar{v}_{03}),  \vspace{0.08in}\\
 \phi_{22,t}(t,k)=&\!\!\! (2kv_{02}+iv_{12})\phi_{42}
   -i\phi_{12}(\alpha_{11}v_{02}\bar{v}_{01}+\bar{\alpha}_{12}|v_{02}|^2+\bar{\alpha}_{13}v_{02}\bar{v}_{03})\vspace{0.08in}\\
   &\!\!\! -i\phi_{22}(\alpha_{12}v_{02}\bar{v}_{01}+\alpha_{22}|v_{02}|^2+\bar{\alpha}_{23}v_{02}\bar{v}_{03}) \vspace{0.08in}\\
   &\!\!\! -i\phi_{32}(\alpha_{13}v_{02}\bar{v}_{01}+\alpha_{23}|v_{02}|^2+\alpha_{33}v_{02}\bar{v}_{03}),  \vspace{0.08in}\\
 \phi_{32,t}(t,k)=& (2kv_{03}+iv_{13})\phi_{42}   -i\phi_{12}(\alpha_{11}v_{03}\bar{v}_{01}+\bar{\alpha}_{12}v_{03}\bar{v}_{02}+\bar{\alpha}_{13}|v_{03}|^2)\vspace{0.08in}\\
   &\!\!\! -i\phi_{22}(\alpha_{12}v_{03}\bar{v}_{01}+\alpha_{22}v_{03}\bar{v}_{02}+\bar{\alpha}_{23}|v_{03}|^2) \vspace{0.08in}\\
\end{array}\right.
\ene
The third column of Eq.~(\ref{tlax}) with $\mu=\mu_3$  yields
\bee
\left\{ \begin{array}{rl}
 \phi_{13,t}(t,k)=&\!\!\! (2kv_{01}+iv_{11})\phi_{43}
   -i\phi_{13}(\alpha_{11}|v_{01}|^2+\bar{\alpha}_{12}v_{01}\bar{v}_{02}+\bar{\alpha}_{13}v_{01}\bar{v}_{03})\vspace{0.08in}\\
  &\!\!\! -i\phi_{23}(\alpha_{12}|v_{01}|^2+\alpha_{22}v_{01}\bar{v}_{02}+\bar{\alpha}_{23}v_{01}\bar{v}_{03}) \vspace{0.08in}\\
  &\!\!\! -i\phi_{33}(\alpha_{13}|v_{01}|^2+\alpha_{23}v_{01}\bar{v}_{02}+\alpha_{33}v_{01}\bar{v}_{03}),  \vspace{0.08in}\\
 \phi_{23,t}(t,k)=&\!\!\! (2kv_{02}+iv_{12})\phi_{43}
   -i\phi_{13}(\alpha_{11}v_{02}\bar{v}_{01}+\bar{\alpha}_{12}|v_{02}|^2+\bar{\alpha}_{13}v_{02}\bar{v}_{03})\vspace{0.08in}\\
   &\!\!\! -i\phi_{23}(\alpha_{12}v_{02}\bar{v}_{01}+\alpha_{22}|v_{02}|^2+\bar{\alpha}_{23}v_{02}\bar{v}_{03}) \vspace{0.08in}\\
   &\!\!\! -i\phi_{33}(\alpha_{13}v_{02}\bar{v}_{01}+\alpha_{23}|v_{02}|^2+\alpha_{33}v_{02}\bar{v}_{03}),  \vspace{0.08in}\\
 \phi_{33,t}(t,k)=& (2kv_{03}+iv_{13})\phi_{43}   -i\phi_{13}(\alpha_{11}v_{03}\bar{v}_{01}+\bar{\alpha}_{12}v_{03}\bar{v}_{02}+\bar{\alpha}_{13}|v_{03}|^2)\vspace{0.08in}\\
   &\!\!\! -i\phi_{23}(\alpha_{12}v_{03}\bar{v}_{01}+\alpha_{22}v_{03}\bar{v}_{02}+\bar{\alpha}_{23}|v_{03}|^2) \vspace{0.08in}\\
\end{array}\right.
\ene
The fourth column of Eq.~(\ref{tlax}) with $\mu=\mu_3$  yields
\bee
\left\{ \begin{array}{rl}
 \phi_{14,t}(t,k)=&\!\!\! (2kv_{01}+iv_{11})\phi_{44}
   -i\phi_{14}(4k^2+\alpha_{11}|v_{01}|^2+\bar{\alpha}_{12}v_{01}\bar{v}_{02}+\bar{\alpha}_{13}v_{01}\bar{v}_{03})\vspace{0.08in}\\
   &\!\!\! -i\phi_{24}(\alpha_{12}|v_{01}|^2+\alpha_{22}v_{01}\bar{v}_{02}+\bar{\alpha}_{23}v_{01}\bar{v}_{03}) \vspace{0.08in}\\
  &\!\!\! -i\phi_{34}(\alpha_{13}|v_{01}|^2+\alpha_{23}v_{01}\bar{v}_{02}+\alpha_{33}v_{01}\bar{v}_{03}),  \vspace{0.08in}\\
 \phi_{24,t}(t,k)=&\!\!\! (2kv_{02}+iv_{12})\phi_{44}
   -i\phi_{14}(\alpha_{11}v_{02}\bar{v}_{01}+\bar{\alpha}_{12}|v_{02}|^2+\bar{\alpha}_{13}v_{02}\bar{v}_{03})\vspace{0.08in}\\
   &\!\!\! -i\phi_{24}(4k^2+\alpha_{12}v_{02}\bar{v}_{01}+\alpha_{22}|v_{02}|^2+\bar{\alpha}_{23}v_{02}\bar{v}_{03}) \vspace{0.08in}\\
   &\!\!\! -i\phi_{34}(\alpha_{13}v_{02}\bar{v}_{01}+\alpha_{23}|v_{02}|^2+\alpha_{33}v_{02}\bar{v}_{03}),  \vspace{0.08in}\\
 \phi_{34,t}(t,k)=& (2kv_{03}+iv_{13})\phi_{44}   -i\phi_{14}(\alpha_{11}v_{03}\bar{v}_{01}+\bar{\alpha}_{12}v_{03}\bar{v}_{02}+\bar{\alpha}_{13}|v_{03}|^2)\vspace{0.08in}\\
   &\!\!\! -i\phi_{24}(\alpha_{12}v_{03}\bar{v}_{01}+\alpha_{22}v_{03}\bar{v}_{02}+\bar{\alpha}_{23}|v_{03}|^2) \vspace{0.08in}\\
\end{array}\right.
\ene

Similarly, we can also obtain the asymptotic formulae for $\phi_{ij},\, i,j=1,2,3,4$. The substitution of these formulae into Eq.~(\ref{ca}) and using the assumption that the initial and boundary data are compatible at $x=0$ and $x=L$, we find the asymptotic result (\ref{c14})
 of $c_{14}(t,k)$ for $k\to \infty$. Similarly we can also show Eqs.~(\ref{c24}) and (\ref{c34}) for $c_{24}(t,k)$ and $c_{34}(t,k)$ as $k\to \infty$.

 Similarly, we have the global relation (\ref{gr}) under the vanishing initial data as
\bee\label{c41g}
\begin{array}{rl}
c_{41}(t,k)=&\!\! -\Psi_{44}(t,k)(\alpha_{11}\bar{\phi}_{14}+\alpha_{12}\bar{\phi}_{24}+\alpha_{13}\bar{\phi}_{34})e^{-2ikL} \vspace{0.1in} \\
 &  +\Psi_{41}(t,k)\big[\alpha_{11}(\alpha_{11}\bar{\phi}_{11}+\bar{\alpha}_{12}\bar{\phi}_{12}+\bar{\alpha}_{13}\bar{\phi}_{13})
     +\alpha_{12}(\alpha_{11}\bar{\phi}_{21}+\bar{\alpha}_{12}\bar{\phi}_{22}+\bar{\alpha}_{13}\bar{\phi}_{23}) \vspace{0.1in}\\
 & +\alpha_{13}(\alpha_{11}\bar{\phi}_{31}+\bar{\alpha}_{12}\bar{\phi}_{32}+\bar{\alpha}_{13}\bar{\phi}_{33})\big]
   +\Psi_{42}(t,k)\big[\alpha_{11}(\alpha_{12}\bar{\phi}_{11}+\alpha_{22}\bar{\phi}_{12}+\bar{\alpha}_{23}\bar{\phi}_{13}) \vspace{0.1in}\\
 &  +\alpha_{12}(\alpha_{12}\bar{\phi}_{21}+\alpha_{22}\bar{\phi}_{22}+\bar{\alpha}_{23}\bar{\phi}_{23})
   +\alpha_{13}(\alpha_{12}\bar{\phi}_{31}+\alpha_{22}\bar{\phi}_{32}+\bar{\alpha}_{23}\bar{\phi}_{33})\big] \vspace{0.1in}\\
 &+\Psi_{43}(t,k)\big[\alpha_{11}(\alpha_{13}\bar{\phi}_{11}+\alpha_{23}\bar{\phi}_{12}+\alpha_{33}\bar{\phi}_{13})
   +\alpha_{12}(\alpha_{13}\bar{\phi}_{21}+\alpha_{23}\bar{\phi}_{22}+\alpha_{33}\bar{\phi}_{23}) \vspace{0.1in}\\
 &  +\alpha_{13}(\alpha_{13}\bar{\phi}_{31}+\alpha_{23}\bar{\phi}_{32}+\alpha_{33}\bar{\phi}_{33})\big],
\end{array}
\ene
\bee\label{c42g}\begin{array}{rl}
c_{42}(t,k)=&\!\! -\Psi_{44}(t,k)(\bar{\alpha}_{12}\bar{\phi}_{14}+\alpha_{22}\bar{\phi}_{24}+\alpha_{23}\bar{\phi}_{34})e^{-2ikL} \vspace{0.1in} \\
 &  +\Psi_{41}(t,k)\big[\bar{\alpha}_{12}(\alpha_{11}\bar{\phi}_{11}+\bar{\alpha}_{12}\bar{\phi}_{12}+\bar{\alpha}_{13}\bar{\phi}_{13})
     +\alpha_{22}(\alpha_{11}\bar{\phi}_{21}+\bar{\alpha}_{12}\bar{\phi}_{22}+\bar{\alpha}_{13}\bar{\phi}_{23}) \vspace{0.1in}\\
 & +\alpha_{23}(\alpha_{11}\bar{\phi}_{31}+\bar{\alpha}_{12}\bar{\phi}_{32}+\bar{\alpha}_{13}\bar{\phi}_{33})\big]
   +\Psi_{42}(t,k)\big[\bar{\alpha}_{12}(\alpha_{12}\bar{\phi}_{11}+\alpha_{22}\bar{\phi}_{12}+\bar{\alpha}_{23}\bar{\phi}_{13}) \vspace{0.1in}\\
 &  +\alpha_{22}(\alpha_{12}\bar{\phi}_{21}+\alpha_{22}\bar{\phi}_{22}+\bar{\alpha}_{23}\bar{\phi}_{23})
   +\alpha_{23}(\alpha_{12}\bar{\phi}_{31}+\alpha_{22}\bar{\phi}_{32}+\bar{\alpha}_{23}\bar{\phi}_{33})\big] \vspace{0.1in}\\
 &+\Psi_{43}(t,k)\big[\bar{\alpha}_{12}(\alpha_{13}\bar{\phi}_{11}+\alpha_{23}\bar{\phi}_{12}+\alpha_{33}\bar{\phi}_{13})
   +\alpha_{22}(\alpha_{13}\bar{\phi}_{21}+\alpha_{23}\bar{\phi}_{22}+\alpha_{33}\bar{\phi}_{23}) \vspace{0.1in}\\
 &  +\alpha_{23}(\alpha_{13}\bar{\phi}_{31}+\alpha_{23}\bar{\phi}_{32}+\alpha_{33}\bar{\phi}_{33})\big],
\end{array}
\ene
\bee\label{c43g}\begin{array}{rl}
c_{43}(t,k)=&\!\! -\Psi_{44}(t,k)(\bar{\alpha}_{13}\bar{\phi}_{14}+\bar{\alpha}_{23}\bar{\phi}_{24}+\alpha_{33}\bar{\phi}_{34})e^{-2ikL} \vspace{0.1in} \\
 &  +\Psi_{41}(t,k)\big[\bar{\alpha}_{13}(\alpha_{11}\bar{\phi}_{11}+\bar{\alpha}_{12}\bar{\phi}_{12}+\bar{\alpha}_{13}\bar{\phi}_{13})
     +\bar{\alpha}_{23}(\alpha_{11}\bar{\phi}_{21}+\bar{\alpha}_{12}\bar{\phi}_{22}+\bar{\alpha}_{13}\bar{\phi}_{23}) \vspace{0.1in}\\
 & +\alpha_{33}(\alpha_{11}\bar{\phi}_{31}+\bar{\alpha}_{12}\bar{\phi}_{32}+\bar{\alpha}_{13}\bar{\phi}_{33})\big]
   +\Psi_{42}(t,k)\big[\bar{\alpha}_{13}(\alpha_{12}\bar{\phi}_{11}+\alpha_{22}\bar{\phi}_{12}+\bar{\alpha}_{23}\bar{\phi}_{13}) \vspace{0.1in}\\
 &  +\bar{\alpha}_{23}(\alpha_{12}\bar{\phi}_{21}+\alpha_{22}\bar{\phi}_{22}+\bar{\alpha}_{23}\bar{\phi}_{23})
   +\alpha_{13}(\alpha_{12}\bar{\phi}_{31}+\alpha_{22}\bar{\phi}_{32}+\bar{\alpha}_{23}\bar{\phi}_{33})\big] \vspace{0.1in}\\
 &+\Psi_{43}(t,k)\big[\bar{\alpha}_{13}(\alpha_{13}\bar{\phi}_{11}+\alpha_{23}\bar{\phi}_{12}+\alpha_{33}\bar{\phi}_{13})
   +\bar{\alpha}_{23}(\alpha_{13}\bar{\phi}_{21}+\alpha_{23}\bar{\phi}_{22}+\alpha_{33}\bar{\phi}_{23}) \vspace{0.1in}\\
 &  +\alpha_{33}(\alpha_{13}\bar{\phi}_{31}+\alpha_{23}\bar{\phi}_{32}+\alpha_{33}\bar{\phi}_{33})\big],
\end{array}
\ene
where $\bar{\phi}_{ij}=\bar{\phi}_{ij}(t,\bar{k})=\overline{\phi_{ij}(t,\bar{k})}$, such that we can show Eqs.~(\ref{c41})-(\ref{c43}) for $c_{4j}(t,k),\, j=1,2,3$ as $k\to \infty$.  $\square$

\subsection*{\it 4.3.\, The relation between Dirichlet and Neumann boundary value problems}

\quad In what follows we show that the spectral functions $S(k)$ and $S_L(k)$ can be expressed in terms of the prescribed Dirichlet and Neumann boundary data and the initial data using the solution of a system of integral equations.
Introduce the new notations as
\bee
 F_{\pm} (t,k)=F(t,k)\pm F(t, -k), \quad \Sigma_{\pm}(k)=e^{2ikL}\pm e^{-2ikL}.
\ene
The sign $\partial D_j,\ j=1,2,3,4$ stands for the boundary of the $j$th quadrant $D_j$, oriented so that $D_j$ lies to the left of $\partial D_j$.
$\partial D_3^0$ denotes the boundary contour which has not contain the zeros of $\Sigma_-(k)$ and $\partial D_3^0=-\partial D_1^0$.

\vspace{0.1in}
\noindent{\bf Theorem 4.2.} {\it Let $q_{0j}(x)=q_j(x,t=0)=0,\, j=1,2,3$ be the initial data of Eq.~(\ref{pnls}) on the interval $x\in [0, L]$ and  $T<\infty$. (i) For the Dirichlet problem, the boundary data $u_{0j}(t)$ and $v_{0j}(t)\, (j=1,2,3)$ on the interval $t\in [0, T)$ are sufficiently smooth and compatible with the initial data $q_{0j}(x),\, (j=1,2,3)$ at points $(x_2, t_2)=(0, 0)$ and $(x_3, t_3)=(L, 0)$, respectively, i.e., $u_{0j}(0)=q_{0j}(0),\, v_{0j}(0)=q_{0j}(L), \, j=1,2,3$; (ii) For the Neumann problem, the boundary data $u_{1j}(t)$ and $v_{0j}(t)\, (j=1,2,3)$ on the interval $t\in [0, T)$ are sufficiently smooth and compatible with the initial data $q_{0j}(x),\, (j=1,2,3)$ at the origin $(x_2, t_2)=(0, 0)$ and $(x_3, t_3)=(L, 0)$, respectively.}

{\it For simplicity, let $n_{33,44}(\mathbb{S})(k)$ have no zeros in the domain $D_1$. Then the spectral functions $S(k)$ and $S_L(k)$ are defined by
\bee\label{skm}
S(k)\!\!=\!\!e^{2ik^2T\hat{\sigma}_4}\left[P\left(
\begin{array}{cccc}
\overline{\Psi_{11}(T,\bar{k})} & \overline{\Psi_{21}(T,\bar{k})} & \overline{\Psi_{31}(T,\bar{k})} &
 \overline{\Psi_{41}(T,\bar{k})} \vspace{0.1in}\\
\overline{\Psi_{12}(T,\bar{k})} & \overline{\Psi_{22}(T,\bar{k})} & \overline{\Psi_{32}(T,\bar{k})} & \overline{\Psi_{42}(T,\bar{k})} \vspace{0.1in}\\
\overline{\Psi_{13}(T,\bar{k})} & \overline{\Psi_{23}(T,\bar{k})} & \overline{\Psi_{33}(T,\bar{k})} &
\overline{\Psi_{43}(T,\bar{k})} \vspace{0.1in}\\
\overline{\Psi_{14}(T,\bar{k})} & \overline{\Psi_{24}(T,\bar{k})} & \overline{\Psi_{34}(T,\bar{k})} & \overline{\Psi_{44}(T,\bar{k})}
 \end{array}\right)P\right],\,\,
\ene
\bee\label{slm}
S_L(k)\!=\!e^{2ik^2T\hat{\sigma}_4}\left[P\left(\begin{array}{cccc}
 \overline{\phi_{11}(T,\bar{k})} & \overline{\phi_{21}(T,\bar{k})} & \overline{\phi_{31}(T,\bar{k})} & \overline{\phi_{41}(T,\bar{k})} \vspace{0.1in}\\
\overline{\phi_{12}(T,\bar{k})} & \overline{\phi_{22}(T,\bar{k})} & \overline{\phi_{32}(T,\bar{k})} & \overline{\phi_{42}(T,\bar{k})} \vspace{0.1in}\\
\overline{\phi_{13}(T,\bar{k})} & \overline{\phi_{23}(T,\bar{k})} & \overline{\phi_{33}(T,\bar{k})} & \overline{\phi_{43}(T,\bar{k})} \vspace{0.1in}\\
\overline{\phi_{14}(T,\bar{k})} & \overline{\phi_{24}(T,\bar{k})} & \overline{\phi_{34}(T,\bar{k})} & \overline{\phi_{44}(T,\bar{k})}
 \end{array}\right)P\right],\,\,
\ene
where the matrix $P$ is given by Eq.~(\ref{mp}), and the complex-valued functions $\{\Psi_{ij}(t,k)\}_{i,j=1}^4$ have the following system of integral equations
\bee\label{psit10}
\left\{\begin{array}{l}
 \Psi_{11,t}(t,k)=\!\! \d 1+\int_0^t\big[(2ku_{01}+iu_{11})\Psi_{41}
   -i\Psi_{11}(\alpha_{11}|u_{01}|^2+\bar{\alpha}_{12}u_{01}\bar{u}_{02}+\bar{\alpha}_{13}u_{01}\bar{u}_{03})\vspace{0.08in}\\
  \quad  -i\Psi_{21}(\alpha_{12}|u_{01}|^2\!+\!\alpha_{22}u_{01}\bar{u}_{02}\!+\!\bar{\alpha}_{23}u_{01}\bar{u}_{03}) \!-\!i\Psi_{31}(\alpha_{13}|u_{01}|^2\!+\!\alpha_{23}u_{01}\bar{u}_{02}\!+\!\alpha_{33}u_{01}\bar{u}_{03})\big](t',k)dt',  \vspace{0.08in}\\
 \Psi_{21,t}(t,k)=\!\!\! \d \int_0^t\big[ (2ku_{02}+iu_{12})\Psi_{41}
   -i\Psi_{11}(\alpha_{11}u_{02}\bar{u}_{01}+\bar{\alpha}_{12}|u_{02}|^2+\bar{\alpha}_{13}u_{02}\bar{u}_{03})\vspace{0.08in}\\
 \quad -i\Psi_{21}(\alpha_{12}u_{02}\bar{u}_{01}\!+\!\alpha_{22}|u_{02}|^2\!+\!\bar{\alpha}_{23}u_{02}\bar{u}_{03})  \!-\!i\Psi_{31}(\alpha_{13}u_{02}\bar{u}_{01}\!+\!\alpha_{23}|u_{02}|^2\!+\!\alpha_{33}u_{02}\bar{u}_{03})\big](t',k)dt',  \vspace{0.08in}\\
 \Psi_{31,t}(t,k)=\!\!\! \d \int_0^t\big[ (2ku_{03}+iu_{13})\Psi_{41}   -i\Psi_{11}(\alpha_{11}u_{03}\bar{u}_{01}+\bar{\alpha}_{12}u_{03}\bar{u}_{02}+\bar{\alpha}_{13}|u_{03}|^2)\vspace{0.08in}\\
  \quad -i\Psi_{21}(\alpha_{12}u_{03}\bar{u}_{01}\!+\!\alpha_{22}u_{03}\bar{u}_{02}\!+\!\bar{\alpha}_{23}|u_{03}|^2)  \!-\!i\Psi_{31}(\alpha_{13}u_{03}\bar{u}_{01}\!+\!\alpha_{23}u_{03}\bar{u}_{02}\!+\!\alpha_{33}|u_{03}|^2)\big](t',k)dt',   \vspace{0.08in}\\
 \Psi_{41,t}(t,k)=\!\!\! \d \int_0^te^{4ik^2(t-t')}\big\{\Psi_{11}[\alpha_{11}(2k\bar{u}_{01}-i\bar{u}_{11})+\bar{\alpha}_{12}(2k\bar{u}_{02}-i\bar{u}_{12})
                     +\bar{\alpha}_{13}(2k\bar{u}_{03}-i\bar{u}_{13})] \vspace{0.08in}\\
                \quad+\Psi_{21}[\alpha_{12}(2k\bar{u}_{01}-i\bar{u}_{11})+\alpha_{22}(2k\bar{u}_{02}-i\bar{u}_{12})
                      +\bar{\alpha}_{23}(2k\bar{u}_{03}-i\bar{u}_{13})] \vspace{0.08in}\\
\end{array}\right.
\ene

\bee\label{psit20}
\left\{\begin{array}{l}
 \Psi_{12,t}(t,k)=\!\! \d \int_0^t\big[(2ku_{01}+iu_{11})\Psi_{42}
    -i\Psi_{12}(\alpha_{11}|u_{01}|^2+\bar{\alpha}_{12}u_{01}\bar{u}_{02}+\bar{\alpha}_{13}u_{01}\bar{u}_{03})\vspace{0.08in}\\
  \quad -i\Psi_{22}(\alpha_{12}|u_{01}|^2+\alpha_{22}u_{01}\bar{u}_{02}+\bar{\alpha}_{23}u_{01}\bar{u}_{03})
    -i\Psi_{32}(\alpha_{13}|u_{01}|^2+\alpha_{23}u_{01}\bar{u}_{02}+\alpha_{33}u_{01}\bar{u}_{03})\big](t',k)dt',  \vspace{0.08in}\\
 \Psi_{22,t}(t,k)=\!\! \d 1+\int_0^t\big[ (2ku_{02}+iu_{12})\Psi_{42}
   -i\Psi_{12}(\alpha_{11}u_{02}\bar{u}_{01}+\bar{\alpha}_{12}|u_{02}|^2+\bar{\alpha}_{13}u_{02}\bar{u}_{03})\vspace{0.08in}\\
  \quad -i\Psi_{22}(\alpha_{12}u_{02}\bar{u}_{01}+\alpha_{22}|u_{02}|^2+\bar{\alpha}_{23}u_{02}\bar{u}_{03}) -i\Psi_{32}(\alpha_{13}u_{02}\bar{u}_{01}+\alpha_{23}|u_{02}|^2+\alpha_{33}u_{02}\bar{u}_{03})\big](t',k)dt',  \vspace{0.08in}\\
 \Psi_{32,t}(t,k)=\!\! \d \int_0^t\big[ (2ku_{03}+iu_{13})\Psi_{42}   -i\Psi_{12}(\alpha_{11}u_{03}\bar{u}_{01}+\bar{\alpha}_{12}u_{03}\bar{u}_{02}+\bar{\alpha}_{13}|u_{03}|^2)\vspace{0.08in}\\
  \quad -i\Psi_{22}(\alpha_{12}u_{03}\bar{u}_{01}+\alpha_{22}u_{03}\bar{u}_{02}+\bar{\alpha}_{23}|u_{03}|^2) -i\Psi_{32}(\alpha_{13}u_{03}\bar{u}_{01}+\alpha_{23}u_{03}\bar{u}_{02}+\alpha_{33}|u_{03}|^2)\big](t',k)dt',   \vspace{0.08in}\\
 \Psi_{42,t}(t,k)=\!\!\! \d \int_0^te^{4ik^2(t-t')}\big\{\Psi_{12}[\alpha_{11}(2k\bar{u}_{01}-i\bar{u}_{11})+\bar{\alpha}_{12}(2k\bar{u}_{02}-i\bar{u}_{12})
                     +\bar{\alpha}_{13}(2k\bar{u}_{03}-i\bar{u}_{13})] \vspace{0.08in}\\
               \quad+\Psi_{22}[\alpha_{12}(2k\bar{u}_{01}-i\bar{u}_{11})+\alpha_{22}(2k\bar{u}_{02}-i\bar{u}_{12})
                      +\bar{\alpha}_{23}(2k\bar{u}_{03}-i\bar{u}_{13})] \vspace{0.08in}\\
\end{array}\right.
\ene

\bee\label{psit30}
\left\{ \begin{array}{l}
 \Psi_{13,t}(t,k)=\!\! \d \int_0^t\big[(2ku_{01}+iu_{11})\Psi_{43}
   -i\Psi_{13}(\alpha_{11}|u_{01}|^2+\bar{\alpha}_{12}u_{01}\bar{u}_{02}+\bar{\alpha}_{13}u_{01}\bar{u}_{03})\vspace{0.08in}\\
   \quad -i\Psi_{22}(\alpha_{12}|u_{01}|^2+\alpha_{22}u_{01}\bar{u}_{02}+\bar{\alpha}_{23}u_{01}\bar{u}_{03}) -i\Psi_{32}(\alpha_{13}|u_{01}|^2+\alpha_{23}u_{01}\bar{u}_{02}+\alpha_{33}u_{01}\bar{u}_{03})\big](t',k)dt',  \vspace{0.08in}\\
 \Psi_{23,t}(t,k)=\!\! \d \int_0^t\big[ (2ku_{02}+iu_{12})\Psi_{42}
   -i\Psi_{13}(\alpha_{11}u_{02}\bar{u}_{01}+\bar{\alpha}_{12}|u_{02}|^2+\bar{\alpha}_{13}u_{02}\bar{u}_{03})\vspace{0.08in}\\
   \quad -i\Psi_{23}(\alpha_{12}u_{02}\bar{u}_{01}+\alpha_{22}|u_{02}|^2+\bar{\alpha}_{23}u_{02}\bar{u}_{03})  -i\Psi_{33}(\alpha_{13}u_{02}\bar{u}_{01}+\alpha_{23}|u_{02}|^2+\alpha_{33}u_{02}\bar{u}_{03})\big](t',k)dt',  \vspace{0.08in}\\
 \Psi_{33,t}(t,k)=\!\!\! \d 1+\int_0^t\big[ (2ku_{03}+iu_{13})\Psi_{43}   -i\Psi_{13}(\alpha_{11}u_{03}\bar{u}_{01}+\bar{\alpha}_{12}u_{03}\bar{u}_{02}+\bar{\alpha}_{13}|u_{03}|^2)\vspace{0.08in}\\
   \quad -i\Psi_{23}(\alpha_{12}u_{03}\bar{u}_{01}+\alpha_{22}u_{03}\bar{u}_{02}+\bar{\alpha}_{23}|u_{03}|^2)  -i\Psi_{33}(\alpha_{13}u_{03}\bar{u}_{01}+\alpha_{23}u_{03}\bar{u}_{02}+\alpha_{33}|u_{03}|^2)\big](t',k)dt',   \vspace{0.08in}\\
 \Psi_{43,t}(t,k)=\!\! \d \int_0^te^{4ik^2(t-t')}\big\{\Psi_{13}[\alpha_{11}(2k\bar{u}_{01}-i\bar{u}_{11})+\bar{\alpha}_{12}(2k\bar{u}_{02}-i\bar{u}_{12})
                     +\bar{\alpha}_{13}(2k\bar{u}_{03}-i\bar{u}_{13})] \vspace{0.08in}\\
            \quad    +\Psi_{23}[\alpha_{12}(2k\bar{u}_{01}-i\bar{u}_{11})+\alpha_{22}(2k\bar{u}_{02}-i\bar{u}_{12})
                      +\bar{\alpha}_{23}(2k\bar{u}_{03}-i\bar{u}_{13})] \vspace{0.08in}\\
\end{array}\right.
\ene
and
\bee\label{psit40}
\left\{ \begin{array}{l}
 \Psi_{14,t}(t,k)=\!\! \d \int_0^te^{-4ik^2(t-t')}\big[(2ku_{01}+iu_{11})\Psi_{44}
   -i\Psi_{14}(\alpha_{11}|u_{01}|^2+\bar{\alpha}_{12}u_{01}\bar{u}_{02}+\bar{\alpha}_{13}u_{01}\bar{u}_{03})\vspace{0.08in}\\
   \quad -i\Psi_{24}(\alpha_{12}|u_{01}|^2+\alpha_{22}u_{01}\bar{u}_{02}+\bar{\alpha}_{23}u_{01}\bar{u}_{03})  -i\Psi_{34}(\alpha_{13}|u_{01}|^2+\alpha_{23}u_{01}\bar{u}_{02}+\alpha_{33}u_{01}\bar{u}_{03})\big](t',k)dt',  \vspace{0.08in}\\
 \Psi_{24,t}(t,k)=\!\! \d \int_0^te^{-4ik^2(t-t')}\big[ (2ku_{02}+iu_{12})\Psi_{44}
   -i\Psi_{14}(\alpha_{11}u_{02}\bar{u}_{01}+\bar{\alpha}_{12}|u_{02}|^2+\bar{\alpha}_{13}u_{02}\bar{u}_{03})\vspace{0.08in}\\
   \quad -i\Psi_{24}(\alpha_{12}u_{02}\bar{u}_{01}+\alpha_{22}|u_{02}|^2+\bar{\alpha}_{23}u_{02}\bar{u}_{03})
   -i\Psi_{34}(\alpha_{13}u_{02}\bar{u}_{01}+\alpha_{23}|u_{02}|^2+\alpha_{33}u_{02}\bar{u}_{03})\big](t',k)dt',  \vspace{0.08in}\\
 \Psi_{34,t}(t,k)=\!\! \d \int_0^te^{-4ik^2(t-t')}\big[ (2ku_{03}+iu_{13})\Psi_{44}   -i\Psi_{14}(\alpha_{11}u_{03}\bar{u}_{01}+\bar{\alpha}_{12}u_{03}\bar{u}_{02}+\bar{\alpha}_{13}|u_{03}|^2)\vspace{0.08in}\\
 \quad -i\Psi_{24}(\alpha_{12}u_{03}\bar{u}_{01}+\alpha_{22}u_{03}\bar{u}_{02}+\bar{\alpha}_{23}|u_{03}|^2)  -i\Psi_{34}(\alpha_{13}u_{03}\bar{u}_{01}+\alpha_{23}u_{03}\bar{u}_{02}+\alpha_{33}|u_{03}|^2)\big](t',k)dt',   \vspace{0.08in}\\
 \Psi_{44,t}(t,k)=\!\! \d 1+\int_0^t\big\{\Psi_{14}[\alpha_{11}(2k\bar{u}_{01}-i\bar{u}_{11})+\bar{\alpha}_{12}(2k\bar{u}_{02}-i\bar{u}_{12})
                     +\bar{\alpha}_{13}(2k\bar{u}_{03}-i\bar{u}_{13})] \vspace{0.08in}\\
               \quad  +\Psi_{24}[\alpha_{12}(2k\bar{u}_{01}-i\bar{u}_{11})+\alpha_{22}(2k\bar{u}_{02}-i\bar{u}_{12})
                      +\bar{\alpha}_{23}(2k\bar{u}_{03}-i\bar{u}_{13})] \vspace{0.08in}\\
\end{array}\right.
\ene }

{\it The functions $\{\phi_{ij}(t,k)\}_{i,j=1}^4$ are of the same integral equations (\ref{psit10})-(\ref{psit40}) by replacing
the functions $\{u_{0j},\, u_{1j}\}$ with $\{v_{0j},\, v_{1j}\},\, (j=1,2,3)$, that is,
$\phi_{ij}(t,k)=\Phi_{ij}(t,k)\big|_{\{u_{0l}(t)=v_{0l}(t),\, u_{1l}(t)=v_{1l}(t)\}},\, (i,j=1,2,3,4; l=1,2,3)$

(i) For the given Dirichlet problem, the unknown Neumann boundary data $\{u_{1j}(t)\}_{j=1}^3$ and $\{v_{1j}(t)\}_{j=1}^3$,\, $0<t<T$ can be given by
\bee\label{u11}
\begin{array}{rl}
 u_{11}(t)=&\!\!\!\!\d\int_{\partial D_3^0}\left[\frac{2\Sigma_+}{i\pi\Sigma_-}(k\Psi_{14-}+iu_{01})
 +u_{01}(2\Psi_{44-} -\bar{\phi}_{44-})\right]dk \vspace{0.08in}\\
&\!\!\!\! +\d\int_{\partial D_3^0}\frac{4}{\pi\Sigma_-}\left[\alpha_{11}
\left(-ik\bar{\phi}_{41-}+\alpha_{11}v_{01}+\alpha_{12}v_{02}+\alpha_{13}v_{03}\right)\right.\vspace{0.08in}\\
&\quad \d+\bar{\alpha}_{12}\left(-ik\bar{\phi}_{42-}+\bar{\alpha}_{12}v_{01}+\alpha_{22}v_{02}
+\alpha_{23}v_{03}\right)  \vspace{0.08in}\\
 &\quad \d\left.+\bar{\alpha}_{13}\left(-ik\bar{\phi}_{43-}+\bar{\alpha}_{13}v_{01}
    +\bar{\alpha}_{23}v_{02}+\alpha_{33}v_{03}\right)\right]dk \vspace{0.08in}\\
&\!\!\!\! +\d\int_{\partial D_3^0}\frac{4k}{i\pi\Sigma_-}
\left[\Psi_{14}(\bar{\phi}_{44}-1)e^{-2ikL}-(\Psi_{11}\!-\!1)(\alpha_{11}\bar{\phi}_{41}+\bar{\alpha}_{12}\bar{\phi}_{42} +\bar{\alpha}_{13}\bar{\phi}_{43}) \right.\vspace{0.08in}\\
&\quad\left.- \Psi_{12}(\alpha_{12}\bar{\phi}_{41}+\alpha_{22}\bar{\phi}_{42}+\bar{\alpha}_{23}\bar{\phi}_{43})
 \!- \!\Psi_{13}(\alpha_{13}\bar{\phi}_{41} +\alpha_{23}\bar{\phi}_{42}+\alpha_{33}\bar{\phi}_{43})\right]_-dk,\qquad
\end{array}
\ene
\bee\label{u12}
\begin{array}{rl}
 u_{12}(t)=&\!\!\!\!\d\int_{\partial D_3^0}\left[\frac{2\Sigma_+}{i\pi\Sigma_-}(k\Psi_{24-}+iu_{02})
  +u_{02}(2\Psi_{44-} -\bar{\phi}_{44-}) \right]dk \vspace{0.08in}\\
&\!\!\!\! +\d\int_{\partial D_3^0}\frac{4}{\pi\Sigma_-}
 \left[\alpha_{12}\left(-ik\bar{\phi}_{41-}+\alpha_{11}v_{01}+\alpha_{12}v_{02}+\alpha_{13}v_{03}\right)\right.\vspace{0.08in}\\
&\quad \d+\alpha_{22}\left(-ik\bar{\phi}_{42-}+\bar{\alpha}_{12}v_{01}+\alpha_{22}v_{02}
+\alpha_{23}v_{03}\right)  \vspace{0.08in}\\
 &\quad \d\left.+\bar{\alpha}_{23}\left(-ik\bar{\phi}_{43-}+\bar{\alpha}_{13}v_{01}
    +\bar{\alpha}_{23}v_{02}+\alpha_{33}v_{03}\right)\right]dk \vspace{0.08in}\\
&\!\!\!\! +\d\int_{\partial D_3^0}\frac{4k}{i\pi\Sigma_-}
\left[\Psi_{24}(\bar{\phi}_{44}-1)e^{-2ikL}-\Psi_{21}(\alpha_{11}\bar{\phi}_{41}+\bar{\alpha}_{12}\bar{\phi}_{42} +\bar{\alpha}_{13}\bar{\phi}_{43}) \right.\vspace{0.08in}\\
&\quad\left.- (\Psi_{22}\!-\!1)(\alpha_{12}\bar{\phi}_{41}+\alpha_{22}\bar{\phi}_{42}+\bar{\alpha}_{23}\bar{\phi}_{43})
 \!-\! \Psi_{23}(\alpha_{13}\bar{\phi}_{41} +\alpha_{23}\bar{\phi}_{42}+\alpha_{33}\bar{\phi}_{43})\right]_-dk,
\end{array}  \ene
\bee\label{u13}
\begin{array}{rl}
 u_{13}(t)=&\!\!\!\!\d\int_{\partial D_3^0}\left[\frac{2\Sigma_+}{i\pi\Sigma_-}(k\Psi_{34-}+iu_{03})
 +u_{03}(2\Psi_{44-} -\bar{\phi}_{44-}) \right]dk \vspace{0.08in}\\
&\!\!\!\! +\d\int_{\partial D_3^0}\frac{4}{\pi\Sigma_-}
 \left[\alpha_{13}\left(-ik\bar{\phi}_{41-}+\alpha_{11}v_{01}+\alpha_{12}v_{02}+\alpha_{13}v_{03}\right)\right.\vspace{0.08in}\\
&\quad \d+\alpha_{23}\left(-ik\bar{\phi}_{42-}+\bar{\alpha}_{12}v_{01}+\alpha_{22}v_{02}
+\alpha_{23}v_{03}\right)  \vspace{0.08in}\\
 &\quad \d\left.+\alpha_{33}\left(-ik\bar{\phi}_{43-}+\bar{\alpha}_{13}v_{01}
    +\bar{\alpha}_{23}v_{02}+\alpha_{33}v_{03}\right)\right]dk \vspace{0.08in}\\
&\!\!\!\! +\d\int_{\partial D_3^0}\frac{4k}{i\pi\Sigma_-}
\left[\Psi_{34}(\bar{\phi}_{44}-1)e^{-2ikL}-\Psi_{31}(\alpha_{11}\bar{\phi}_{41}+\bar{\alpha}_{12}\bar{\phi}_{42} +\bar{\alpha}_{13}\bar{\phi}_{43}) \right.\vspace{0.08in}\\
&\quad\left.- \Psi_{32}(\alpha_{12}\bar{\phi}_{41}+\alpha_{22}\bar{\phi}_{42}+\bar{\alpha}_{23}\bar{\phi}_{43})
 \!-\! (\Psi_{33}\!-\!1)(\alpha_{13}\bar{\phi}_{41}+\alpha_{23}\bar{\phi}_{42}+\alpha_{33}\bar{\phi}_{43})\right]_-dk,
\end{array}%
\ene

(ii) For the known Neumann  problem, the unknown Dirichlet boundary data $\{u_{0j}(t)\}_{j=1}^3$ and $\{v_{0j}(t)\}_{j=1}^3,\, 0<t<T$ can be given by
\bes\bee \label{u01}
&\begin{array}{rl}
u_{01}(t)=&\!\!\!\!\!\! \d \int_{\partial D_3^0}\frac{1}{\pi\Sigma_-}\left[\Sigma_+\Psi_{14+}-2
(\alpha_{11}\bar{\phi}_{41}+\bar{\alpha}_{12}\bar{\phi}_{42} +\bar{\alpha}_{13}\bar{\phi}_{43})_+\right]dk \vspace{0.08in}\\
&\!\!\!\!\!\! +\d\int_{\partial D_3^0}\frac{2}{\pi\Sigma_-}\left\{\Psi_{14}(\bar{\phi}_{44}-1)e^{-2ikL}-\big[(\Psi_{11}-1)(\alpha_{11}\bar{\phi}_{41}+\bar{\alpha}_{12}\bar{\phi}_{42} +\bar{\alpha}_{13}\bar{\phi}_{43}) \right.\vspace{0.08in}\\
&\left. + \Psi_{12}(\alpha_{12}\bar{\phi}_{41}+\alpha_{22}\bar{\phi}_{42}+\bar{\alpha}_{23}\bar{\phi}_{43})
 + \Psi_{13}(\alpha_{13}\bar{\phi}_{41}
 +\alpha_{23}\bar{\phi}_{42}+\alpha_{33}\bar{\phi}_{43})\big]\right\}_+dk,\qquad\qquad
\end{array}\vspace{0.08in}\\
& \label{u02} \begin{array}{rl}
u_{02}(t)=&\!\!\!\!\!\! \d \int_{\partial D_3^0}\frac{1}{\pi\Sigma_-}\left[\Sigma_+\Psi_{24+}-2
(\alpha_{12}\bar{\phi}_{41}+\alpha_{22}\bar{\phi}_{42} +\bar{\alpha}_{23}\bar{\phi}_{43})_+\right]dk \vspace{0.08in}\\
&\!\!\!\!\!\! +\d\int_{\partial D_3^0}\frac{2}{\pi\Sigma_-}\left\{\Psi_{24}(\bar{\phi}_{44}-1)e^{-2ikL}-\big[\Psi_{21}(\alpha_{11}\bar{\phi}_{41}+\bar{\alpha}_{12}\bar{\phi}_{42} +\bar{\alpha}_{13}\bar{\phi}_{43}) \right.\vspace{0.08in}\\
&\left. + (\Psi_{22}-1)(\alpha_{12}\bar{\phi}_{41}+\alpha_{22}\bar{\phi}_{42}+\bar{\alpha}_{23}\bar{\phi}_{43})
 + \Psi_{23}(\alpha_{13}\bar{\phi}_{41}
 +\alpha_{23}\bar{\phi}_{42}+\alpha_{33}\bar{\phi}_{43})\big]\right\}_+dk,\quad
\end{array}\vspace{0.08in}\\
& \label{u03} \begin{array}{rl}
u_{03}(t)=&\!\!\!\!\!\! \d \int_{\partial D_3^0}\frac{1}{\pi\Sigma_-}\left[\Sigma_+\Psi_{34+}-2
(\alpha_{13}\bar{\phi}_{41}+\alpha_{23}\bar{\phi}_{42} +\alpha_{33}\bar{\phi}_{43})_+\right]dk \vspace{0.08in}\\
&\!\!\!\!\!\! +\d\int_{\partial D_3^0}\frac{2}{\pi\Sigma_-}\left\{\Psi_{34}(\bar{\phi}_{44}-1)e^{-2ikL}-\big[\Psi_{31}(\alpha_{11}\bar{\phi}_{41}+\bar{\alpha}_{12}\bar{\phi}_{42} +\bar{\alpha}_{13}\bar{\phi}_{43}) \right.\vspace{0.08in}\\
&\left. + \Psi_{32}(\alpha_{12}\bar{\phi}_{41}+\alpha_{22}\bar{\phi}_{42}+\bar{\alpha}_{23}\bar{\phi}_{43})
 + (\Psi_{33}-1)(\alpha_{13}\bar{\phi}_{41}
 +\alpha_{23}\bar{\phi}_{42}+\alpha_{33}\bar{\phi}_{43})\big]\right\}_+dk,\quad
\end{array}%
\ene
where $\Psi_{14}=\Psi_{14}(t,k),\, \bar{\phi}_{44}=\overline{\phi_{44}(t, \bar{k})}=\bar{\phi}_{44}(t, \bar{k})$ and other functions have the similar expressions.}

\vspace{0.1in}

\noindent {\bf Proof.} We can show that Eqs.~(\ref{skm}) and (\ref{slm}) hold by means of Eqs.~(\ref{sss}) and (\ref{sl}) with replacing $T$ by $t$, that is,  $S(k)=e^{-2ik^2t\hat{\sigma}_4}\mu_2^{-1}(0,t,k)$ and $S_L(k)=e^{-2ik^2t\hat{\sigma}_4}\mu_3^{-1}(L,t,k)$ and the symmetry relation (\ref{symmetry}).
Moreover, Eqs.~(\ref{psit10})-(\ref{psit40}) for $\Psi_{ij}(t,k),\, i,j=1,2,3,4$ can be obtained by using the Volteral integral equations of $\mu_2(0,t,k)$. Similarly, the expressions of $\phi_{ij}(t,k),\, (i,j=1,2,3,4)$ can be found by means of the Volteral integral equations of $\mu_3(L,t,k)$.

In what follows we show Eqs.~(\ref{u11})-(\ref{v013}), that is the map between Dirichlet and Neumann boundary conditions.

(i) The Cauchy's theorem is employed to study Eq.~(\ref{mu2asy}) to  generate
\bee
\label{psi33}
\begin{array}{rl}
i\pi \Psi_{44}^{(1)}(t)=& \d-\left(\int_{\partial D_2}+\int_{\partial D_4}\right)[\Psi_{44}(t,k)-1]dk
= \d \left(\int_{\partial D_1}+\int_{\partial D_3}\right)[\Psi_{44}(t,k)-1]dk \vspace{0.08in}\\
=&  \d\int_{\partial D_3}[\Psi_{44}(t,k)-1]dk-\int_{\partial D_3}[\Psi_{44}(t,-k)-1]dk =\int_{\partial D_3}\Psi_{44-}(t,k)dk,
\end{array}
\ene
and
\bee\label{psi13}
\begin{array}{rl}
i\pi \Psi_{14}^{(2)}(t)=&\!\!\! \d \left(\int_{\partial D_1}+\int_{\partial D_3}\right)\left[k\Psi_{14}(t,k)+\frac{i}{2}u_{01}(t)\right]dk, =\d\int_{\partial D_3}\left[k\Psi_{14}(t,k)+\frac{i}{2}u_{01}(t)\right]_-dk, \vspace{0.08in}\\
=&\!\!\! \d\int_{\partial D_3^0}\left\{k\Psi_{14}(t,k)+\frac{i}{2}u_{01}(t)+\frac{2e^{-2ikL}}{\Sigma_-(k)}\left[k\Psi_{14}(t,k)+\frac{i}{2}u_{01}(t)\right]\right\}_-dk+C_1(t), \vspace{0.08in}\\
=&\!\!\! \d\int_{\partial D_3^0}\frac{\Sigma_+}{\Sigma_-}(k\Psi_{14-}+iu_{01})dk+C_1(t),
\end{array}\ene
where we have introduced the function $C_1(t)$ as
\bee\no
C_1(t)=-\d\int_{\partial D_3^0}\left\{\frac{2e^{-2ikL}}{\Sigma_-}\left[k\Psi_{14}(t,k)+\frac{i}{2}u_{01}(t)\right]\right\}_-dk,
\ene

We use the global relation (\ref{ca}) to further reduce $C_1(t)$ in the form
\bee \label{it} \begin{array}{rl}
C_1(t)=&\!\!\!-\d\int_{\partial D_3^0}\left\{\frac{2e^{-2ikL}}{\Sigma_-}\left[k\Psi_{14}(t,k)+\frac{i}{2}u_{01}(t)\right]\right\}_-dk \vspace{0.08in}\\
=&\!\!\! \d\int_{\partial D_3^0}\left\{\frac{2e^{-2ikL}}{\Sigma_-}\left[-kc_{14}+\Psi_{14}^{(1)}+
\frac{\Psi_{14}^{(1)}\bar{\phi}_{44}^{(1)}}{k}-(\alpha_{11}\bar{\phi}_{41}^{(1)}+\bar{\alpha}_{12}\bar{\phi}_{42}^{(1)}
+\bar{\alpha}_{13}\bar{\phi}_{43}^{(1)})e^{2ikL}\right]\right\}_-dk \vspace{0.08in}\\
&\!\!\!-\d\int_{\partial D_3^0}\left\{\frac{2e^{-2ikL}}{\Sigma_-}\left[
\frac{\Psi_{14}^{(1)}\bar{\phi}_{44}^{(1)}}{k}+
\Big(\alpha_{11}(k\bar{\phi}_{41}-\bar{\phi}_{41}^{(1)})
+\bar{\alpha}_{12}(k\bar{\phi}_{42}-\bar{\phi}_{42}^{(1)}) \right.\right.\vspace{0.08in}\\
&\left.\left.\qquad\quad +\bar{\alpha}_{13}(k\bar{\phi}_{43}-\bar{\phi}_{43}^{(1)})\Big)
  e^{2ikL}\right]\right\}_-dk \vspace{0.08in}\\
&\!\!\! +\d\int_{\partial D_3^0}\left\{\frac{2ke^{-2ikL}}{\Sigma_-}\left[
\Psi_{14}(\bar{\phi}_{44}-1)-\big[(\Psi_{11}-1)(\alpha_{11}\bar{\phi}_{41}+\bar{\alpha}_{12}\bar{\phi}_{42} +\bar{\alpha}_{13}\bar{\phi}_{43}) \right.\right.\vspace{0.08in}\\
&\left.\left. + \Psi_{12}(\alpha_{12}\bar{\phi}_{41}+\alpha_{22}\bar{\phi}_{42}+\bar{\alpha}_{23}\bar{\phi}_{43})
 + \Psi_{13}(\alpha_{13}\bar{\phi}_{41}
 +\alpha_{23}\bar{\phi}_{42}+\alpha_{33}\bar{\phi}_{43})\big]e^{2ikL}\right]\right\}_-dk,
\end{array}
\ene

By applying the Cauchy's theorem and asymptotic (\ref{c14}) to Eq.~(\ref{it}), we find that $C_1(t)$ can reduce to
\bee\label{it2}
\begin{array}{rl}
C_1(t)=&-i\pi\Psi_{14}^{(2)}-\d\int_{\partial D_3^0}\left\{\frac{i}{2}u_{01}\bar{\phi}_{44-}
  +\frac{2i}{\Sigma_-}\left[\alpha_{11}\left(-ik\bar{\phi}_{41-}+\alpha_{11}v_{01}
  +\alpha_{12}v_{02}+\alpha_{13}v_{03}\right)\right.\right.\vspace{0.08in}\\
&\d+\bar{\alpha}_{12}\left(-ik\bar{\phi}_{42-}+\bar{\alpha}_{12}v_{01}+\alpha_{22}v_{02}+\alpha_{23}v_{03}
\right) \left.\left.+\bar{\alpha}_{13}\left(-ik\bar{\phi}_{43-}+\bar{\alpha}_{13}v_{01}+\bar{\alpha}_{23}v_{02}+\alpha_{33}v_{03}
\right)
 \right]\right\}dk \vspace{0.08in}\\
&+\d\int_{\partial D_3^0}\frac{2k}{\Sigma_-}
\left[\Psi_{14}(\bar{\phi}_{44}-1)e^{-2ikL}-(\Psi_{11}-1)(\alpha_{11}\bar{\phi}_{41}+\bar{\alpha}_{12}\bar{\phi}_{42} +\bar{\alpha}_{13}\bar{\phi}_{43}) \right.\vspace{0.08in}\\
&\left.- \Psi_{12}(\alpha_{12}\bar{\phi}_{41}+\alpha_{22}\bar{\phi}_{42}+\bar{\alpha}_{23}\bar{\phi}_{43})
 - \Psi_{13}(\alpha_{13}\bar{\phi}_{41}
 +\alpha_{23}\bar{\phi}_{42}+\alpha_{33}\bar{\phi}_{43})\right]_-dk,
\end{array}
\ene

It follows from Eqs.~(\ref{psi13}) and (\ref{it2}) that we have
\bee\label{psi13g}
\begin{array}{rl}
2i\pi \Psi_{13}^{(2)}(t)=&\!\!\!\!\! \d\int_{\partial D_3^0}\!\!
\left[\frac{\Sigma_+}{\Sigma_-}(k\Psi_{14-}\!+iu_{01})-\frac{i}{2}u_{01}\bar{\phi}_{44-}
  \right]dk \vspace{0.08in}\\
&+\d\int_{\partial D_3^0}\frac{2i}{\Sigma_-}\left[\alpha_{11}\left(-ik\bar{\phi}_{41-}+\alpha_{11}v_{01}
  +\bar{\alpha}_{12}v_{02}+\bar{\alpha}_{13}v_{03}\right)\right.\vspace{0.08in}\\
&\d+\bar{\alpha}_{12}\left(-ik\bar{\phi}_{42-}+\alpha_{12}v_{01}+\alpha_{22}v_{02}+\bar{\alpha}_{23}v_{03}
\right)  \vspace{0.08in}\\ &\d\left.\left.+\bar{\alpha}_{13}\left(-ik\bar{\phi}_{43-}+\alpha_{13}v_{01}+\alpha_{23}v_{02}+\alpha_{33}v_{03}
\right)
 \right]\right\}dk  \vspace{0.08in}\\
&+\d\int_{\partial D_3^0}\frac{2k}{\Sigma_-}
\left[\Psi_{14}(\bar{\phi}_{44}-1)e^{-2ikL}-(\Psi_{11}-1)(\alpha_{11}\bar{\phi}_{41}+\bar{\alpha}_{12}\bar{\phi}_{42} +\bar{\alpha}_{13}\bar{\phi}_{43}) \right.\vspace{0.08in}\\
&\left.- \Psi_{12}(\alpha_{12}\bar{\phi}_{41}+\alpha_{22}\bar{\phi}_{42}+\bar{\alpha}_{23}\bar{\phi}_{43})
 - \Psi_{13}(\alpha_{13}\bar{\phi}_{41}
 +\alpha_{23}\bar{\phi}_{42}+\alpha_{33}\bar{\phi}_{43})\right]_-dk,
\end{array}\ene
Thus substituting Eqs.~(\ref{psi33}) and (\ref{psi13g}) into the third one of system (\ref{ud}), we can get Eq.~(\ref{u11}).
 Similarly, we can also show  Eqs.~(\ref{u12}) and (\ref{u13}).

 To use Eq.~(\ref{vd}) to show Eq.~(\ref{v11}) for $v_{11}(t)$ we need to find these functions $\phi_{44}^{(1)}(t,k)$ and $\phi_{14}^{(2)}(t,k)$.  Applying the Cauchy's theorem to Eq.~(\ref{mu3asy}), we have
\bee\label{phi13}
\begin{array}{rl}
 i\pi [\alpha_{11}\phi_{14}^{(2)}&\!\!\!+\bar{\alpha}_{12}\phi_{24}^{(2)}+\bar{\alpha}_{13}\phi_{34}^{(2)}]\qquad \qquad\vspace{0.08in}\\
=&\!\!\! \d\int_{\partial D_3}\left[\alpha_{11}(k\phi_{14}(t,k)-\phi_{14}^{(1)})+\bar{\alpha}_{12}(k\phi_{24}(t,k)-\phi_{24}^{(1)})
+\bar{\alpha}_{13}(k\phi_{34}(t,k)-\phi_{34}^{(1)})\right]_-dk, \vspace{0.08in}\\
=&\!\!\! \d\int_{\partial D_3^0}\left\{\alpha_{11}\left[k\phi_{14}(t,k)-\phi_{14}^{(1)}-\frac{2e^{2ikL}}{\Sigma_-}(k\phi_{14}-\phi_{14}^{(1)})\right] \right.\vspace{0.08in}\\
&\d \left.+\bar{\alpha}_{12}\left[k\phi_{24}(t,k)-\phi_{24}^{(1)}-\frac{2e^{2ikL}}{\Sigma_-}(k\phi_{24}-\phi_{24}^{(1)})\right]\right. \vspace{0.08in}\\
&\d \left.+\bar{\alpha}_{13}\left[k\phi_{34}(t,k)-\phi_{34}^{(1)}-\frac{2e^{2ikL}}{\Sigma_-}(k\phi_{34}-\phi_{34}^{(1)})\right]
 \right\}_-dk+C_2(t), \vspace{0.08in}\\
=&\!\!\! \d\int_{\partial D_3^0}-\frac{\Sigma_+}{\Sigma_-}\left[\alpha_{11}(k\phi_{14-}-2\phi_{14}^{(1)})+\bar{\alpha}_{12}(k\phi_{24-}-2\phi_{24}^{(1)})
+\bar{\alpha}_{13}(k\phi_{34-}-2\phi_{34}^{(1)})\right]dk+C_2(t),
\end{array}\ene
where we have introduced the function $C_2(t)$ as
\bee\no
C_2(t)=\d\int_{\partial D_3^0}\left\{\frac{2e^{2ikL}}{\Sigma_-}\left[\alpha_{11}(k\phi_{14}-\phi_{14}^{(1)})+\bar{\alpha}_{12}(k\phi_{24}-\phi_{24}^{(1)})
+\bar{\alpha}_{13}(k\phi_{34}-\phi_{34}^{(1)})\right]\right\}_-dk,
\ene

We use the global relation (\ref{c41g}) to further reduce $C_2(t)$ in the form
\bee \label{jt} \begin{array}{rl}
C_2(t)=&\!\!\!\d\int_{\partial D_3^0}\left\{\frac{2}{\Sigma_-}
\left[-k\bar{c}_{41}(t,\bar{k})-\left(\alpha_{11}\phi_{14}^{(1)}+\bar{\alpha}_{12}\phi_{24}^{(1)}+\bar{\alpha}_{13}\phi_{34}^{(1)}\right)e^{2ikL} \right.\right. \\
&\d -\bar{\Psi}_{44}^{(1)}\left(\alpha_{11}\phi_{14}^{(1)}+\bar{\alpha}_{12}\phi_{24}^{(1)}
 +\bar{\alpha}_{13}\phi_{34}^{(1)}\right)\frac{e^{2ikL}}{k}+(\alpha_{11}^2+|\alpha_{12}|^2+|\alpha_{13}|^2)\bar{\Psi}_{41}^{(1)}\\
& +(\alpha_{11}\bar{\alpha}_{12}+\bar{\alpha}_{12}\alpha_{22}+\bar{\alpha}_{13}\alpha_{23})\bar{\Psi}_{42}^{(1)}  \left.\left.+(\alpha_{11}\bar{\alpha}_{13}+\bar{\alpha}_{12}\bar{\alpha}_{23}+\bar{\alpha}_{13}\alpha_{33})\Psi_{43}^{(1)}
\right]\right\}_-dk \vspace{0.08in}\\
&\!\!\!+\d\int_{\partial D_3^0}\left\{\frac{2}{\Sigma_-}\left[\bar{\Psi}_{44}^{(1)}\left(\alpha_{11}\phi_{14}^{(1)}+\bar{\alpha}_{12}\phi_{24}^{(1)}
 +\bar{\alpha}_{13}\phi_{34}^{(1)}\right)\frac{e^{2ikL}}{k}\right.\right. \\
& +(\alpha_{11}^2+|\alpha_{12}|^2+|\alpha_{13}|^2)(k\bar{\Psi}_{41}-\bar{\Psi}_{41}^{(1)})
 +(\alpha_{11}\bar{\alpha}_{12}+\bar{\alpha}_{12}\alpha_{22}+\bar{\alpha}_{13}\alpha_{23})(k\bar{\Psi}_{42}-\bar{\Psi}_{42}^{(1)}) \vspace{0.1in}\\
&\d \left.\left.+(\alpha_{11}\bar{\alpha}_{13}+\bar{\alpha}_{12}\bar{\alpha}_{23}+\bar{\alpha}_{13}\alpha_{33})(k\bar{\Psi}_{43}-\bar{\Psi}_{43}^{(1)})\right]\right\}_-dk \vspace{0.08in}\\
&\!\!\!+\d\int_{\partial D_3^0}\frac{2k}{\Sigma_-}\left\{(1-\bar{\Psi}_{44})(\alpha_{11}\phi_{14}+\bar{\alpha}_{12}\phi_{24}+\bar{\alpha}_{13}\phi_{34})e^{2ikL}
\right.\vspace{0.1in} \\
 &  +\bar{\Psi}_{41}\big[\alpha_{11}(\alpha_{11}(\phi_{11}-1)+\alpha_{12}\phi_{12}+\alpha_{13}\phi_{13})
     +\bar{\alpha}_{12}(\alpha_{11}\phi_{21}+\alpha_{12}(\phi_{22}-1)+\alpha_{13}\phi_{23}) \vspace{0.1in}\\
 & +\bar{\alpha}_{13}(\alpha_{11}\phi_{31}+\alpha_{12}\phi_{32}+\alpha_{13}(\phi_{33}-1))\big]
  +\bar{\Psi}_{42}\big[\alpha_{11}(\bar{\alpha}_{12}(\phi_{11}-1)+\alpha_{22}\phi_{12}+\alpha_{23}\phi_{13}) \vspace{0.1in}\\
 &  +\bar{\alpha}_{12}(\bar{\alpha}_{12}\phi_{21}+\alpha_{22}(\phi_{22}-1)+\alpha_{23}\phi_{23})
   +\bar{\alpha}_{13}(\bar{\alpha}_{12}\phi_{31}+\alpha_{22}\phi_{32}+\alpha_{23}(\phi_{33}-1))\big] \vspace{0.1in}\\
 &+\bar{\Psi}_{43}\big[\alpha_{11}(\bar{\alpha}_{13}(\phi_{11}-1)+\bar{\alpha}_{23}\phi_{12}+\bar{\alpha}_{33}\phi_{13})
   +\bar{\alpha}_{12}(\bar{\alpha}_{13}\phi_{21}+\bar{\alpha}_{23}(\phi_{22}-1)+\alpha_{33}\phi_{23}) \vspace{0.1in}\\
 & \left. +\bar{\alpha}_{13}(\bar{\alpha}_{13}\phi_{31}+\bar{\alpha}_{23}\phi_{32}+\alpha_{33}(\phi_{33}-1))\big]\right\}_-dk,\\
\end{array}
\ene
We need to further reduce $C_2(t)$ by using the asymptotic (\ref{c41}) and the Cauchy's theorem
such that we have $C_2(t)$ in the form
\bee\label{jtg}
C_2(t)=\d -i\pi [\alpha_{11}\phi_{14}^{(2)}+\bar{\alpha}_{12}\phi_{24}^{(2)}+\bar{\alpha}_{13}\phi_{34}^{(2)}] -\d\int_{\partial D_3^0}\frac{i}{2}\bar{\Psi}_{44-}^{(1)}(\alpha_{11}v_{01}+\bar{\alpha}_{12}v_{02}+\bar{\alpha}_{13}v_{03})dk
+I_1(t),
\ene

It follows from Eqs.~(\ref{phi13}) and (\ref{jtg})  that we have
\bee\begin{array}{rl}
 2i\pi [\alpha_{11}\phi_{14}^{(2)}&\!\!+\bar{\alpha}_{12}\phi_{24}^{(2)}+\bar{\alpha}_{13}\phi_{34}^{(2)}] \qquad \vspace{0.08in}\\
=&\!\!\! -\d\int_{\partial D_3^0}\frac{\Sigma_+(k)}{\Sigma_-(k)}\left[\alpha_{11}(k\phi_{14-}+iv_{01})+\bar{\alpha}_{12}(k\phi_{24-}
 +iv_{02}+\bar{\alpha}_{13}(k\phi_{34-}+iv_{03})\right]dk \vspace{0.08in}\\
& -\d\int_{\partial D_3^0}\frac{i}{2}\bar{\Psi}_{44-}^{(1)}(\alpha_{11}v_{01}+\bar{\alpha}_{12}v_{02}+\bar{\alpha}_{13}v_{03})dk
+I_1(t),
\end{array}
\ene
where $I(t)$ is given by Eq.~(\ref{i1}). Similarly, we can also the expressions of $i\pi [\alpha_{12}\phi_{14}^{(2)}+\alpha_{22}\phi_{24}^{(2)}+\bar{\alpha}_{23}\phi_{34}^{(2)}]$
and $2i\pi [\alpha_{13}\phi_{14}^{(2)}+\alpha_{23}\phi_{24}^{(2)}+\alpha_{33}\phi_{34}^{(2)}]$ such that we can show that Eq.~(\ref{v11}) holds.

(ii) We now deduce the Dirichlet boundary value problems (\ref{u01})-(\ref{u03}) at $x=0$ from the Neumann boundary value problems. It follows from the first one of Eq.~(\ref{ud}) that $u_{01}(t)$ can be expressed by means of $\Psi_{14}^{(1)}$.
Applying the Cauchy's theorem to Eq.~(\ref{mu2asy}) yields
\bee\label{psi13g2}
\begin{array}{rl}
i\pi \Psi_{14}^{(1)}(t)=&\!\!\! \d \left(\int_{\partial D_1}+\int_{\partial D_3}\right)\Psi_{14}(t,k)dk =\int_{\partial D_3}\Psi_{14-}(t,k)dk \vspace{0.08in}\\
=&\!\!\! \d\int_{\partial D_3^0}\left[\Psi_{14-}(t,k)+\frac{2}{\Sigma_-(k)}(e^{-2ikL}\Psi_{14})_+\right]dk+C_3(t) \vspace{0.08in}\\
=&\!\!\! \d \int_{\partial D_3^0}\frac{\Sigma_+(k)}{\Sigma_-(k)}\Psi_{14+}dk+C_3(t),
\end{array}\ene
where $C_3(t)$ is defined by
\bee\label{k1}
C_3(t)=-\d\int_{\partial D_3^0}\frac{2}{\Sigma_-(k)}(e^{-2ikL}\Psi_{14})_+dk,
\ene

By applying the global relation (\ref{ca}), the Cauchy's theorem and asymptotics (\ref{c14}) to Eq.~(\ref{k1}), we find
\bee \label{k1g} \begin{array}{rl}
C_3(t)=&-\d\int_{\partial D_3^0}\frac{2}{\Sigma_-}(e^{-2ikL}\Psi_{14})_+dk \vspace{0.08in}\\
=&\d\int_{\partial D_3^0}\frac{2}{\Sigma_-}\left[-c_{14}e^{-2ikL}-(\alpha_{11}\bar{\phi}_{41}+\bar{\alpha}_{12}\bar{\phi}_{42} +\bar{\alpha}_{13}\bar{\phi}_{43})\right]_+dk \vspace{0.08in}\\
&\!\!\! +\d\int_{\partial D_3^0}\frac{2}{\Sigma_-}\left[
\Psi_{14}(\bar{\phi}_{44}-1)e^{-2ikL}-\big[(\Psi_{11}-1)(\alpha_{11}\bar{\phi}_{41}+\bar{\alpha}_{12}\bar{\phi}_{42} +\bar{\alpha}_{13}\bar{\phi}_{43}) \right.\vspace{0.08in}\\
&\left. + \Psi_{12}(\alpha_{12}\bar{\phi}_{41}+\alpha_{22}\bar{\phi}_{42}+\bar{\alpha}_{23}\bar{\phi}_{43})
 + \Psi_{13}(\alpha_{13}\bar{\phi}_{41}
 +\alpha_{23}\bar{\phi}_{42}+\alpha_{33}\bar{\phi}_{43})\big]\right]_+dk, \\
=& -i\pi\Psi_{14}^{(1)}-\d\int_{\partial D_3^0}\frac{2}{\Sigma_-}(\alpha_{11}\bar{\phi}_{41}+\bar{\alpha}_{12}\bar{\phi}_{42} +\bar{\alpha}_{13}\bar{\phi}_{43})_+dk \vspace{0.08in}\\
&\!\!\! +\d\int_{\partial D_3^0}\frac{2}{\Sigma_-}\left\{
\Psi_{14}(\bar{\phi}_{44}-1)e^{-2ikL}-\big[(\Psi_{11}-1)(\alpha_{11}\bar{\phi}_{41}+\bar{\alpha}_{12}\bar{\phi}_{42} +\bar{\alpha}_{13}\bar{\phi}_{43}) \right.\vspace{0.08in}\\
&\left. + \Psi_{12}(\alpha_{12}\bar{\phi}_{41}+\alpha_{22}\bar{\phi}_{42}+\bar{\alpha}_{23}\bar{\phi}_{43})
 + \Psi_{13}(\alpha_{13}\bar{\phi}_{41}
 +\alpha_{23}\bar{\phi}_{42}+\alpha_{33}\bar{\phi}_{43})\big]\right\}_+dk, \\
\end{array}
\ene

Eqs.~(\ref{psi13g2}) and (\ref{k1g}) imply that
\bee \label{psi13g2g}
\begin{array}{rl}
2i\pi \Psi_{14}^{(1)}(t)=&\!\!\!\!\!\! \d \int_{\partial D_3^0}\left[\frac{\Sigma_+(k)}{\Sigma_-(k)}\Psi_{14+}-\frac{2}{\Sigma_-}(\alpha_{11}\bar{\phi}_{41}+\bar{\alpha}_{12}\bar{\phi}_{42} +\bar{\alpha}_{13}\bar{\phi}_{43})_+\right]dk \vspace{0.08in}\\
&\!\!\! +\d\int_{\partial D_3^0}\frac{2}{\Sigma_-}\left\{
\Psi_{14}(\bar{\phi}_{44}-1)e^{-2ikL}-\big[(\Psi_{11}-1)(\alpha_{11}\bar{\phi}_{41}+\bar{\alpha}_{12}\bar{\phi}_{42} +\bar{\alpha}_{13}\bar{\phi}_{43}) \right.\vspace{0.08in}\\
&\left. + \Psi_{12}(\alpha_{12}\bar{\phi}_{41}+\alpha_{22}\bar{\phi}_{42}+\bar{\alpha}_{23}\bar{\phi}_{43})
 + \Psi_{13}(\alpha_{13}\bar{\phi}_{41}
 +\alpha_{23}\bar{\phi}_{42}+\alpha_{33}\bar{\phi}_{43})\big]\right\}_+dk,
\end{array}
\ene
Thus, substituting Eq.~(\ref{psi13g2g}) into the first one of Eq.~(\ref{ud}) yields Eq.~(\ref{u01}). Similarly, by applying the expressions of $\Psi_{24}^{(1)}(t)$ and $\Psi_{34}^{(1)}(t)$  to the second one of Eq.~(\ref{ud}), we can find Eqs.~(\ref{u02}) and (\ref{u03}).

Similarly we also can show that the Dirichlet boundary value problems (\ref{v013}) at $x=L$ hold from the Neumann boundary value problems.  $\square$

\subsection*{\it 4.3.\, The effective characterizations }

\quad Substituting the perturbated expressions for eigenfunctions and initial boundary conditions
\bee\label{solue}
\left\{\begin{array}{l}
\Psi_{ij}(t,k)=\Psi_{ij}^{[0]}(t,k)+\epsilon\Psi_{ij}^{[1]}(t,k)+\epsilon^2\Psi_{ij}^{[2]}(t,k)+\cdots, \quad  i,j=1,2,3,4, \vspace{0.08in}\\
\phi_{ij}(t,k)=\phi_{ij}^{[0]}(t,k)+\epsilon\phi_{ij}^{[1]}(t,k)+\epsilon^2\phi_{ij}^{[2]}(t,k)+\cdots, \quad  i,j=1,2,3,4, \vspace{0.08in}\\
u_{sj}(t)=\epsilon u_{sj}^{[1]}(t)+\epsilon^2 u_{sj}^{[2]}(t)+\cdots, \quad  s=0,1; j=1,2,3, \vspace{0.08in}\\
v_{sj}(t)=\epsilon v_{sj}^{[1]}(t)+\epsilon^2 v_{sj}^{[2]}(t)+\cdots, \quad  s=0,1; j=1,2,3,
\end{array}\right.
\ene
into Eqs.~(\ref{psit11})-(\ref{psit14}), where $\epsilon>0$ is a small parameter, we have these terms of $O(1)$,
and $O(\epsilon)$  as
\bee
O(1): \left\{\begin{array}{l}
 \Psi_{jj}^{[0]}=1,\quad j=1,2,3,4, \vspace{0.1in}\\
 \Psi_{ij}^{[0]}=0,\quad i,j=1,2,3,4,\, i\not=j,
\end{array}\right. \qquad\qquad
\ene
\bee \label{epsilon}
O(\epsilon): \left\{\begin{array}{l}
 \Psi_{js}^{[1]}=\!\Psi_{44}^{[1]}=0, \quad j,s,=1,2,3, \vspace{0.1in}\\
 \Psi_{j4}^{[1]}=\!\d\int_0^t\!e^{-4ik^2(t-t')}\left(2ku_{0j}^{[1]}+iu_{1j}^{[1]}\right)(t')dt',\quad j=1,2,3,\vspace{0.1in}\\
  \Psi_{41}^{[1]}\!=\!\!\!\d\int_0^t\!\!\!e^{4ik^2(t-t')}\!\left[\alpha_{11}\!\left(2k\bar{u}_{01}^{[1]}\!-\!i\bar{u}_{11}^{[1]}\right)
   \!+\!\bar{\alpha}_{12}\!\left(2k\bar{u}_{02}^{[1]}\!-\!i\bar{u}_{12}^{[1]}\right)
   \!+\!\bar{\alpha}_{13}\!\left(2k\bar{u}_{03}^{[1]}\!-\!i\bar{u}_{13}^{[1]}\right)\!\right]\! (t')dt',\vspace{0.1in}\\
 \Psi_{42}^{[1]}\!=\!\!\!\d\int_0^t\!\!\!e^{4ik^2(t-t')}\!\left[\alpha_{12}\!\left(2k\bar{u}_{01}^{[1]}\!-\!i\bar{u}_{11}^{[1]}\right)
   \!+\!\alpha_{22}\!\left(2k\bar{u}_{02}^{[1]}\!-\!i\bar{u}_{12}^{[1]}\right)
   \!+\!\bar{\alpha}_{23}\!\left(2k\bar{u}_{03}^{[1]}\!-\!i\bar{u}_{13}^{[1]}\right)\!\right]\! (t')dt',\vspace{0.1in}\\
   \Psi_{43}^{[1]}\!=\!\!\!\d\int_0^t\!\!\!e^{4ik^2(t-t')}\!\left[\alpha_{13}\!\left(2k\bar{u}_{01}^{[1]}\!-\!i\bar{u}_{11}^{[1]}\right)
   \!+\!\alpha_{23}\!\left(2k\bar{u}_{02}^{[1]}\!-\!i\bar{u}_{12}^{[1]}\right)
   \!+\!\alpha_{33}\!\left(2k\bar{u}_{03}^{[1]}\!-\!i\bar{u}_{13}^{[1]}\right)\!\right]\! (t')dt',\vspace{0.1in}\\
 \end{array}\right.
\ene

Similarly, we can also obtain the analogous expressions for $\{\phi_{ij}^{[l]}\}_{i,j=1}^4, l=0,1$
by means of the boundary values at $x=L$, that is, $\{v_{ij}^{[l]}\}, i=0,1; j=1,2,3; l=0,1$.

If we assume that $m_{44}(\mathbb{S})$ has no zeros, then we expand Eqs.~(\ref{u11})-(\ref{v11}) to have
\bes\bee\label{u11a}
&\begin{array}{rl}
 u_{11}^{[n]}(t)=&\!\!\!\!\d\int_{\partial D_3^0}\left\{\frac{2\Sigma_+}{i\pi\Sigma_-}(k\Psi_{14-}^{[n]}+iu_{01}^{[n]})
dk+\frac{4}{\pi\Sigma_-}\left[\alpha_{11}\!\!\left(-ik\bar{\phi}_{41-}^{[n]}
   +\alpha_{11}v_{01}^{[n]}+\alpha_{12}v_{02}^{[n]}+\alpha_{13}v_{03}^{[n]}\right)\right.\right.\vspace{0.08in}\\
& \d+\bar{\alpha}_{12}\left(-ik\bar{\phi}_{42-}^{[n]}+\bar{\alpha}_{12}v_{01}^{[n]}+\alpha_{22}v_{02}^{[n]}
+\alpha_{23}v_{03}^{[n]}\right)  \vspace{0.08in}\\
 & \d\left.\left.+\bar{\alpha}_{13}\left(-ik\bar{\phi}_{43-}^{[n]}+\bar{\alpha}_{13}v_{01}^{[n]}
    +\bar{\alpha}_{23}v_{02}^{[n]}+\alpha_{33}v_{03}^{[n]}\right)\right]\right\}dk+ {\rm lower \,\, order \,\, terms},  \vspace{0.08in}\\
 \end{array} \\
\label{u12a}
&\begin{array}{rl}
 u_{12}^{[n]}(t)=&\!\!\!\!\d\int_{\partial D_3^0}\left\{\frac{2\Sigma_+}{i\pi\Sigma_-}(k\Psi_{24-}^{[n]}+iu_{02}^{[n]})
 dk+\frac{4}{\pi\Sigma_-}\left[\alpha_{12}\!\!
\left(-ik\bar{\phi}_{41-}^{[n]}+\alpha_{11}v_{01}^{[n]}+\alpha_{12}v_{02}^{[n]}+\alpha_{13}v_{03}^{[n]}\right)\right.\right.\vspace{0.08in}\\
& \d+\alpha_{22}\left(-ik\bar{\phi}_{42-}^{[n]}+\bar{\alpha}_{12}v_{01}^{[n]}+\alpha_{22}v_{02}^{[n]}+\alpha_{23}v_{03}^{[n]}\right)  \vspace{0.08in}\\
 & \d\left.\left.+\bar{\alpha}_{23}\left(-ik\bar{\phi}_{43-}^{[n]}+\bar{\alpha}_{13}v_{01}^{[n]}
    +\bar{\alpha}_{23}v_{02}^{[n]}+\alpha_{33}v_{03}^{[n]}\right)\right]\right\}dk+ {\rm lower \,\, order \,\, terms},  \vspace{0.08in}\\
 \end{array} \\
\label{u13a}
&\begin{array}{rl}
 u_{13}^{[n]}(t)=&\!\!\!\!\d\int_{\partial D_3^0}\left\{\frac{2\Sigma_+}{i\pi\Sigma_-}(k\Psi_{34-}^{[n]}+iu_{03}^{[n]})
 dk+\frac{4}{\pi\Sigma_-}\left[\alpha_{13}\!\!\left(-ik\bar{\phi}_{41-}^{[n]}+\alpha_{11}v_{01}^{[n]}+\alpha_{12}v_{02}^{[n]}
 +\alpha_{13}v_{03}^{[n]}\right)\right.\right.\vspace{0.08in}\\
& \d+\alpha_{23}\left(-ik\bar{\phi}_{42-}^{[n]}+\bar{\alpha}_{12}v_{01}^{[n]}+\alpha_{22}v_{02}^{[n]}+\alpha_{23}v_{03}^{[n]}\right)  \vspace{0.08in}\\
 & \d\left.\left.+\alpha_{33}\left(-ik\bar{\phi}_{43-}^{[1]}+\bar{\alpha}_{13}v_{01}^{[n]}
    +\bar{\alpha}_{23}v_{02}^{[n]}+\alpha_{33}v_{03}^{[n]}\right)\right]\right\}dk+ {\rm lower \,\, order \,\, terms},  \vspace{0.08in}\\
 \end{array}
\ene\ees
where `lower order terms' stands for the result involving known terms of lower order.

The terms of $O(\epsilon^n)$ in Eqs.~(\ref{psit10})-(\ref{psit40}) and the similar equations for $\phi_{ij}$ yield
\bee\label{pp0}
\left\{\begin{array}{rl}
\Psi_{j4}^{[n]}(t,k)=&\!\!\! \displaystyle\int_0^t e^{-4ik^2(t-t')}\left(2ku_{0j}^{[n]}+iu_{1j}^{[n]}\right)(t')dt'
   + {\rm lower \,\, order \,\, terms},  \quad j=1,2,3,  \vspace{0.1in}\\
\bar{\phi}_{41}^{[n]}(t,\bar{k})=&\!\!\!\d \int_0^te^{-4ik^2(t-t')}\left[\alpha_{11}(2kv_{01}^{[n]}+iv_{11}^{[n]})+\alpha_{12}(2kv_{02}^{[n]}+iv_{12}^{[n]})
                     +\alpha_{13}(2kv_{03}^{[n]}+iv_{13}^{[n]})\right](t')dt' \vspace{0.1in}\\
      &\!\!\!+ {\rm lower \,\, order \,\, terms},  \vspace{0.1in}\\
\bar{\phi}_{42}^{[n]}(t,\bar{k})=&\!\!\!\d \int_0^te^{-4ik^2(t-t')}\left[\bar{\alpha}_{12}(2kv_{01}^{[n]}+iv_{11}^{[n]})+\alpha_{22}(2kv_{02}^{[n]}+iv_{12}^{[n]})
                     +\alpha_{23}(2kv_{03}^{[n]}+iv_{13}^{[n]})\right](t')dt' \vspace{0.1in}\\
                     &\!\!\!+ {\rm lower \,\, order \,\, terms}, \vspace{0.1in}\\
\bar{\phi}_{43}^{[n]}(t,\bar{k})=&\!\!\!\d \int_0^te^{-4ik^2(t-t')}\left[\bar{\alpha}_{13}(2kv_{01}^{[n]}+iv_{11}^{[n]})+\bar{\alpha}_{23}(2kv_{02}^{[n]}+iv_{12}^{[n]})
                     +\alpha_{33}(2kv_{03}^{[n]}+iv_{13}^{[n]})\right](t')dt' \vspace{0.1in}\\
                     &\!\!\!+ {\rm lower \,\, order \,\, terms},
\end{array}\right.
\ene
which leads to
\bee\label{pp-}
\begin{array}{rl}
\Psi_{j4-}^{[n]}(t,k)=&\!\!\! \displaystyle 4k\int_0^t e^{-4ik^2(t-t')}u_{0j}^{[n]}(t')dt'
   + {\rm lower \,\, order \,\, terms},  \quad j=1,2,3,   \vspace{0.1in}\\
\bar{\phi}_{41-}^{[n]}(t,\bar{k})=&\!\!\!\d 4k\int_0^te^{-4ik^2(t-t')}\left(\alpha_{11}v_{01}^{[n]}+\alpha_{12}v_{02}^{[n]}                    +\alpha_{13}v_{03}^{[n]}\right)(t')dt'+ {\rm lower \,\, order \,\, terms},  \vspace{0.1in}\\
\bar{\phi}_{42-}^{[n]}(t,\bar{k})=&\!\!\!\d 4k\int_0^te^{-4ik^2(t-t')}\left(\bar{\alpha}_{12}v_{01}^{[n]}+\alpha_{22}v_{02}^{[n]}
                     +\alpha_{23}v_{03}^{[n]}\right)(t')dt'+ {\rm lower \,\, order \,\, terms}, \vspace{0.1in}\\
\bar{\phi}_{43-}^{[n]}(t,\bar{k})=&\!\!\!\d 4k \int_0^te^{-4ik^2(t-t')}\left(\bar{\alpha}_{13}v_{01}^{[n]}+\bar{\alpha}_{23}v_{02}^{[n]}
                     +\alpha_{33}v_{03}^{[n]}\right)(t')dt'+ {\rm lower \,\, order \,\, terms},
\end{array}
\ene
It follows from system (\ref{pp-}) that $\Psi_{1j-}^{[n]}$ and $\phi_{4j-}^{[n]},\, j=1,2,3$ can be generated at each step from the known Dirichlet
boundary data $u_{0j}^{[n]}$ and $v_{0j}^{[n]}$ such that we know that the Neumann boundary data $u_{1j}^{[n]}$ can then be given
by Eqs.~(\ref{u11a})-(\ref{u13a}). Similarly, we also show that the Neumann boundary data $v_{1j}^{[n]}$ can then be determined
by the known Dirichlet boundary data $u_{0j}^{[n]}$ and $v_{0j}^{[n]}$ .

Similarly, the substitution of Eq.~(\ref{solue}) into Eqs.~(\ref{u01}) and (\ref{u02}) yields the terms of $O(\epsilon^n)$ as
\bes\bee \label{u01e}
& u_{01}^{[n]}(t)=\d \int_{\partial D_3^0}\left[\frac{\Sigma_+}{\pi\Sigma_-}\Psi_{14+}^{[n]}-\frac{2}{\pi\Sigma_-}\left(\alpha_{11}\bar{\phi}_{41}^{[n]}
+\bar{\alpha}_{12}\bar{\phi}_{42}^{[n]} +\bar{\alpha}_{13}\bar{\phi}_{43}^{[n]}\right)\right]dk+ {\rm lower \,\, order \,\, terms},\qquad \\
& \label{u02e}
u_{02}^{[n]}(t)=\d \int_{\partial D_3^0}\left[\frac{\Sigma_+}{\pi\Sigma_-}\Psi_{24+}^{[n]}
  -\frac{2}{\pi\Sigma_-}\left(\alpha_{12}\bar{\phi}_{41}^{[n]}
   +\alpha_{22}\bar{\phi}_{42}^{[n]} +\bar{\alpha}_{23}\bar{\phi}_{43}^{[n]}\right)\right]dk+ {\rm lower \,\, order \,\, terms},\qquad \\
& \label{u03e}
u_{03}^{[n]}(t)=\d \int_{\partial D_3^0}\left[\frac{\Sigma_+}{\pi\Sigma_-}\Psi_{34+}^{[n]}-\frac{2}{\pi\Sigma_-}\left(\alpha_{13}\bar{\phi}_{41}^{[n]}
+\alpha_{23}\bar{\phi}_{42}^{[n]} +\alpha_{33}\bar{\phi}_{43}^{[n]}\right)\right]dk+ {\rm lower \,\, order \,\, terms},\qquad
\ene\ees

Eq.~(\ref{pp0}) implies that
\bee\label{pp+}
\begin{array}{rl}
\Psi_{j4+}^{[n]}(t,k)=&\!\!\! \d 2i\int_0^t e^{-4ik^2(t-t')}u_{1j}^{[n]}(t')dt'+ {\rm lower \,\, order \,\, terms},  \quad j=1,2,3,  \vspace{0.1in}\\
\bar{\phi}_{41+}^{[n]}(t,\bar{k})=&\!\!\!\d \int_0^te^{-4ik^2(t-t')}\left(\alpha_{11}v_{11}^{[n]}+\alpha_{12}v_{12}^{[n]}                    +\alpha_{13}v_{13}^{[n]}\right)(t')dt' + {\rm lower \,\, order \,\, terms},  \vspace{0.1in}\\
\bar{\phi}_{42+}^{[n]}(t,\bar{k})=&\!\!\!\d \int_0^te^{-4ik^2(t-t')}\left(\bar{\alpha}_{12}v_{11}^{[n]}+\alpha_{22}v_{12}^{[n]}
                     +\alpha_{23}v_{13}^{[n]}\right)(t')dt' + {\rm lower \,\, order \,\, terms}, \vspace{0.1in}\\
\bar{\phi}_{43+}^{[n]}(t,\bar{k})=&\!\!\!\d \int_0^te^{-4ik^2(t-t')}\left(\bar{\alpha}_{13}v_{11}^{[n]})+\bar{\alpha}_{23}v_{12}^{[n]}
                     +\alpha_{33}v_{13}^{[n]}\right)(t')dt' + {\rm lower \,\, order \,\, terms},
\end{array}
\ene

It follows from system (\ref{pp+}) that $\Psi_{j4+}^{[n]}$ and $\phi_{4j+}^{[n]},\, j=1,2,3$ can be generated at each step from the known Neumann
boundary data $u_{1j}^{[n]}$ and $v_{1j}^{[n]}$ such that we know that the Dirichlet boundary data $u_{0j}^{[n]}$ can then be given
by Eqs.~(\ref{u01e})-(\ref{u03e}). Similarly, we also show that the Dirichlet boundary data $v_{0j}^{[n]}$ can then be determined
by the known Neumann boundary data $u_{1j}^{[n]}$ and $v_{1j}^{[n]}$.

\subsection*{\it 4.4. \, The large $L$ limit from the interval to the half-line}

\quad The formulae for the initial and boundary value conditions $u_{0j}(t)$ and $u_{1j}(t),\, j=1,2,2$ of Theorem 4.2 in the limit $L\to \infty$ can reduce to the corresponding
ones on the half-line. Since when $L\to \infty$,
\bee \label{limit}
\begin{array}{l} v_{0j}\to 0,\quad v_{1j}\to 0,\quad j=1,2,3, \quad
 \phi_{ij}\to \delta_{ij}, \quad \d\frac{\Sigma_+(k)}{\Sigma_-(k)}\to 1 \quad {\rm as} \quad k\to \infty \quad {\rm in} \quad D_3,
\end{array}
\ene

Thus, according to Eq.~(\ref{limit}), the $L\to \infty$ limits of Eqs.~(\ref{u11}), (\ref{u12}), (\ref{u01}), and (\ref{u02}) yield the unknown Neumann boundary data
\bee
 u_{1j}(t)=\d\frac{2}{\pi}\int_{\partial D_3^0}\left[u_{0j}(\Psi_{44-}+1)-ik\Psi_{j4-}\right]dk,\quad j=1,2,3,
\ene
for the given Dirichlet boundary problem, and the unknown Dirichlet boundary data
\bee
 u_{0j}(t)=\d \frac{1}{\pi}\int_{\partial D_3^0}\Psi_{j4+}dk,\quad j=1,2,3,
\ene
for the given Neumann boundary problem.

\section{The GLM representation and equivalence}

\quad In this section we deduce the eigenfunctions $\Psi(t,k)$ and $\phi(t,k)$ in terms of the Gel'fand-Levitan-Marchenko (GLM) approach~\cite{glm1,glm2,glm3,glm4}.
Moreover, the global relation can be used to find the unknown Neumann (Dirichlet) boundary values from the given
Dirichlet (Neumann) boundary values by means of the GLM representations. Moreover, the GLM representations are shown to be equivalent to the
ones obtained in Sec. 4. Finally, the linearizable boundary conditions are presented for  the  GLM representations.

\vspace{0.1in}
\noindent \subsection*{\it 5.1. The GLM representation}
\vspace{0.1in}

\noindent {\bf Proposition 5.1.} {\it The eigenfunctions $\Psi(t,k)$ and $\phi(t,k)$ possess the GLM representation
\bes\bee\label{psiglm}
&\Psi(t,k)=\mathbb{I}+\d\int_{-t}^t\left[L(t,s)+\left(k+\frac{i}{2}U^{(0)}\sigma_4\right)G(t,s)\right]e^{-2ik^2(s-t)\sigma_4}ds,\\
\label{phiglm}
&\phi(t,k)=\mathbb{I}+\d\int_{-t}^t\left[\mathcal{L}(t,s)+\left(k+\frac{i}{2}\mathcal{U}^{(L)}\sigma_4\right)\mathcal{G}(t,s)\right]e^{-2ik^2(s-t)\sigma_4}ds,
\ene\ees
where the $4\times 4$ matrix-valued functions $L(t, s)=(L_{ij})_{4\times 4}$ and $G(t, s)=(G_{ij})_{4\times 4},\, -t\leq s\leq t$ satisfy a
Goursat system
\bee\label{lge}
\left\{
\begin{array}{l}
\d L_t(t, s)\!+\!\sigma_4L_s(t,s)\sigma_4= i\sigma_4U_x^{(0)}L(t,s) \!-\!\frac{1}{2}\left[(U^{(0)})^3\!+\!i\dot{U}^{(0)}\sigma_4+[U_x^{(0)}, U^{(0)}]\right]G(t,s), \vspace{0.1in}\\
G_t(t, s)+\sigma_4G_s(t,s)\sigma_4=2U^{(0)}L(t,s)+i\sigma_4U_x^{(0)} G(t,s),
\end{array}\right.
\ene
with the initial conditions
\bee\label{lgi}
\left\{
\begin{array}{rl}
L_{lj}(t,-t)=&\!\!\!L_{44}(t,-t)=0, \quad l,j=1,2,3, \vspace{0.1in}\\
G_{lj}(t,-t)=&\!\!\! G_{44}(t,-t)=0, \quad l,j=1,2,3, \vspace{0.1in}\\
G_{14}(t,t)=&\!\!\! u_{01}(t),\,\, G_{24}(t,t)=u_{02}(t),\,\, G_{34}(t,t)=u_{03}(t), \vspace{0.1in}\\
G_{41}(t,t)=&\!\!\! \alpha_{11}\bar{u}_{01}(t)+\bar{\alpha}_{12}\bar{u}_{02}(t)+\bar{\alpha}_{13}\bar{u}_{03}(t), \vspace{0.1in}\\
G_{42}(t,t)=&\!\!\!\alpha_{12}\bar{u}_{01}(t)+\alpha_{22}\bar{u}_{02}(t)+\bar{\alpha}_{23}\bar{u}_{03}(t), \vspace{0.1in}\\
G_{43}(t,t)=&\!\!\!\alpha_{13}\bar{u}_{01}(t)+\alpha_{23}\bar{u}_{02}(t)+\alpha_{33}\bar{u}_{03}(t), \vspace{0.1in}\\
L_{14}(t,t)=&\!\!\! \frac{i}{2}u_{11}(t),\,\,L_{24}(t,t)=\frac{i}{2}u_{12}(t), \,\,L_{34}(t,t)=\frac{i}{2}u_{13}(t), \vspace{0.1in}\\
L_{41}(t,t)=&\!\!\! -\frac{i}{2}(\alpha_{11}\bar{u}_{11}(t)+\bar{\alpha}_{12}\bar{u}_{12}(t)+\bar{\alpha}_{13}\bar{u}_{13}(t)),\vspace{0.1in}\\
L_{42}(t,t)=&\!\!\! -\frac{i}{2}(\alpha_{12}\bar{u}_{11}(t)+\alpha_{22}\bar{u}_{12}(t)+\bar{\alpha}_{23}\bar{u}_{13}(t)),\vspace{0.1in}\\
L_{43}(t,t)=&\!\!\! -\frac{i}{2}(\alpha_{13}\bar{u}_{11}(t)+\alpha_{23}\bar{u}_{12}(t)+\alpha_{33}\bar{u}_{13}(t)),
\end{array}\right.
\ene
\bee \label{U0}
U^{(0)}=\left(\begin{array}{cccc}
            0 & 0 & 0 & u_{01}(t) \\
            0&  0 & 0 & u_{02}(t) \\
            0 & 0 & 0 & u_{03}(t) \\
            p_{01}(t) & p_{02}(t) & p_{03}(t) & 0
            \end{array}\right), \qquad
U_x^{(0)}=\left(\begin{array}{cccc}
            0 & 0 & 0 & u_{11}(t) \\
            0&  0 & 0 & u_{12}(t) \\
            0 & 0 & 0 & u_{13}(t) \\
            p_{11}(t)& p_{12}(t) & p_{13}(t) & 0
            \end{array}\right),
            \ene
with
\bee \no
 \begin{array}{l} p_{01}=\alpha_{11}\bar{u}_{01}+\bar{\alpha}_{12}\bar{u}_{02}+\bar{\alpha}_{13}\bar{u}_{03},\,
 p_{02}=\alpha_{12}\bar{u}_{01}+\alpha_{22}\bar{u}_{02}+\bar{\alpha}_{23}\bar{u}_{03},\,
 p_{03}=\alpha_{13}\bar{u}_{01}+\alpha_{23}\bar{u}_{02}+\alpha_{33}\bar{u}_{03},\vspace{0.1in}\\
p_{11}=\alpha_{11}\bar{u}_{11}+\bar{\alpha}_{12}\bar{u}_{12}+\bar{\alpha}_{13}\bar{u}_{13},\,
 p_{12}=\alpha_{12}\bar{u}_{11}+\alpha_{22}\bar{u}_{12}+\bar{\alpha}_{23}\bar{u}_{13},\,
 p_{13}=\alpha_{13}\bar{u}_{11}+\alpha_{23}\bar{u}_{12}+\alpha_{33}\bar{u}_{13},
 \end{array}
 \ene

Similarly, $\mathcal{L}(t,s),\, \mathcal{G}(t,s)$ satisfy the similar Eqs.~(\ref{lge}) and (\ref{lgi}) with $u_{0j}\to v_{0j},\, u_{1j}\to v_{1j},\,
 U^{(0)}\to \mathcal{U}^{(L)}=U^{(0)}\big|_{u_{0j}\to v_{0j}},\, U_x^{(0)}\to \mathcal{U}_x^{(L)}=U_x^{(0)}\big|_{u_{1j}\to v_{1j}}$.}

\vspace{0.1in}
\noindent {\bf Proof.} We assume that  the function
\bee \label{psi}
\psi(t,k)=e^{-2ik^2t\sigma_4}+\d\int_{-t}^t[L_0(t,s)+kG(t,s)]e^{-2ik^2s\sigma_4}ds,
\ene
satisfies the time-part of Lax pair (\ref{lax}) with the boundary data $\psi(0,k)=\mathbb{I}$ at $x=0$, where $L_0(t,s)$ and $G(t,s)$ are the unknown $4\times 4$ matrix-valued functions. We substitute Eq.~(\ref{psi}) into the time-part of Lax pair (\ref{lax}) with the boundary data (\ref{ibv}) and use the identity
\bee
\d\int_{-t}^tF(t,s)e^{-2ik^2s\sigma_4}ds=\frac{i}{2k^2}\left[F(t,t)e^{-2ik^2t\sigma_4}-F(t,-t)e^{2ik^2t\sigma_4}-
  \d\int_{-t}^tF_s(t,s)e^{-2ik^2s\sigma_4}ds\right]\sigma_4,
\ene
where the function $F(t,s)$ is a $4\times 4$ matrix-valued function. As a consequence, we find
\bee\label{lg}
\left\{
\begin{array}{l}
L_0(t, -t)+\sigma_4L_0(t,-t)\sigma_4=-iU^{(0)}G(t,-t)\sigma_4, \vspace{0.1in}\\
G(t, -t)+\sigma_4G(t,-t)\sigma_4=0, \vspace{0.1in}\\
L_0(t, t)-\sigma_4L_0(t,t)\sigma_4=iU^{(0)}G(t,t)\sigma_4+V_0^{(0)}, \vspace{0.1in}\\
G(t, t)-\sigma_4G(t,t)\sigma_4=2U^{(0)}, \vspace{0.1in}\\
L_{0t}(t, s)+\sigma_4L_{0s}(t,s)\sigma_4=-iU^{(0)}G_s(t,s)\sigma_4+V_0^{(0)}L_0(t,s), \vspace{0.1in}\\
G_{t}(t, s)+\sigma_4G_s(t,s)\sigma_4=2U^{(0)}L_0(t,s)+V_0^{(0)}G(t,s),
\end{array}\right.
\ene
where $U^{(0)}$ is given by Eq.~(\ref{U0}) and
\bee \no V_0^{(0)}=-i(U_{x}^{(0)}+U^{(0)2})\sigma_4=-i\left(\begin{array}{cccc}
           u_{01}p_{01} & u_{01}p_{02} & u_{01}p_{03}  & -u_{11} \vspace{0.1in}\\
           u_{02}p_{01} & u_{02}p_{02} & u_{02}p_{03} & -u_{21} \vspace{0.1in}\\
           u_{03}p_{01} & u_{03}p_{02} & u_{03}p_{03} & -u_{31} \vspace{0.1in}\\
         p_{11} & p_{21} & p_{31} & -(u_{01}p_{01}+u_{02}p_{02}+u_{03}p_{03})
            \end{array}\right),
                     \ene

To reduce system (\ref{lg}) we further introduce the new matrix $L(t,s)$ by
\bee
L(t,s)=L_0(t,s)-\frac{i}{2}U^{(0)}\sigma_4G(t,s),
\ene
such that the first four equations of system (\ref{lg}) become
\bee\no
\left\{
\begin{array}{l}
L(t, -t)+\sigma_4L(t,-t)\sigma_4=0, \vspace{0.1in}\\
G(t, -t)+\sigma_4G(t,-t)\sigma_4=0, \vspace{0.1in}\\
L(t, t)-\sigma_4L(t,t)\sigma_4=V_0^{(0)}, \vspace{0.1in}\\
G(t, t)-\sigma_4G(t,t)\sigma_4=2U^{(0)},
\end{array}\right.
\ene
which leads to Eq.~(\ref{lgi}), and from the last two equations of system (\ref{lg}) we have Eq.~(\ref{lge}). By means of  transformation
(\ref{mud}), that is, $\mu_2(0, t, k)=\Psi(t,k)=\psi(t,k)e^{2ik^2t\sigma_4}$, we know that $\Psi(t,k)$ is given by Eq.~(\ref{psiglm}).
Similarly, we can also show that Eq.~(\ref{phiglm}) holds. $\square$

\vspace{0.1in}

For convenience, we rewrite a $4\times 4$ matrix $C=(C_{ij})_{4\times 4}$ as
\bee\no
C=(C_{ij})_{4\times 4}=\left(\!\!\!\begin{array}{cc}
 \tilde{C}_{3\times 3} & \tilde{C}_{j4} \vspace{0.05in}\cr
 \tilde{C}_{4j} & C_{44}
\end{array}\!\!\!\right),\,
\tilde{C}_{3\times 3}=(C_{ij})_{3\times 3},\,
\tilde{C}_{j4}=(C_{14}, C_{24}, C_{34})^T, \,
\tilde{C}_{4j}=(C_{41}, C_{42}, C_{43}),
\ene

The Dirichlet and Neumann boundary values at $x=0, L$ are simply written as
\bee\begin{array}{l}
 u_j(t)=(u_{j1}(t), u_{j2}(t), u_{j3}(t)),\quad
 v_j(t)=(v_{j1}(t), v_{j2}(t), v_{j3}(t)), \, j=1,2,3, \vspace{0.1in}\\
w_{j0}(t)=(p_{j1}(t), p_{j2}(t), p_{j3}(t)),\quad
w_{jL}(t)=(p_{j1}(t), p_{j2}(t), p_{j3}(t))|_{u_{sj}\to v_{sj}},\, s=0,1; j=1,2,3,
  \end{array}
\ene

For a matrix-valued function $F(t,s)$, we introduce the $\hat{F}(t,k)$ by
\bee\no
 \hat{F}(t,k)=\d\int_{-t}^{t}F(t,s)e^{2ik^2(s-t)}ds,
\ene
Thus, the GLM expressions (\ref{psiglm}) and (\ref{phiglm}) of $\{\Psi_{ij},\, \phi_{ij}\}$ can be rewritten as
\bes
\bee \label{psiglmg}
&\left\{\begin{array}{l}
\d\tilde{\Psi}_{3\times 3}(t,k)=\mathbb{I}+\hat{\tilde{L}}_{3\times 3}-\frac{i}{2}u_0^T(t)\hat{\tilde{G}}_{4j}+k\hat{\tilde{G}}_{3\times 3},  \vspace{0.1in}\\
\d\tilde{\Psi}_{j4}(t,k)=\hat{\tilde{L}}_{j4}-\frac{i}{2}u_0^T(t)\hat{\tilde{G}}_{44}+k\hat{\tilde{G}}_{j4}, \quad j=1,2,3, \vspace{0.1in}\\
\d\tilde{\Psi}_{4j}(t,k)=\hat{\tilde{L}}_{4j}+\frac{i}{2}\bar{u}_0(t)\mathcal{M}\hat{\tilde{G}}_{3\times 3}+k\hat{\tilde{G}}_{4j}, \quad j=1,2,3,  \vspace{0.1in}\\
\d\tilde{\Psi}_{44}(t,k)=1+\hat{\tilde{L}}_{44}+\frac{i}{2}\bar{u}_0(t)\mathcal{M}\hat{\tilde{G}}_{j4}+k\hat{\tilde{G}}_{44},
\end{array}\right.\\
\label{phiglmg}
&\left\{\begin{array}{l}
\d\tilde{\phi}_{3\times 3}(t,k)
=\mathbb{I}+\hat{\tilde{\mathcal{L}}}_{3\times 3}-\frac{i}{2}v_0^T(t)\hat{\tilde{\mathcal{G}}}_{4j}+k\hat{\tilde{\mathcal{G}}}_{3\times 3},  \vspace{0.1in}\\
\d\tilde{\phi}_{j4}(t,k)=\hat{\tilde{\mathcal{L}}}_{j4}-\frac{i}{2}v_0^T(t)\hat{\tilde{\mathcal{G}}}_{44}+k\hat{\tilde{\mathcal{G}}}_{j4},  \quad j=1,2,3, \vspace{0.1in}\\
\d\tilde{\phi}_{4j}(t,k)=\hat{\tilde{\mathcal{L}}}_{4j}+\frac{i}{2}\bar{v}_0(t)\mathcal{M}\hat{\tilde{\mathcal{G}}}_{3\times 3}+k\hat{\tilde{\mathcal{G}}}_{4j}, \quad j=1,2,3,  \vspace{0.1in}\\
\d\tilde{\phi}_{44}(t,k)=1+\hat{\tilde{\mathcal{L}}}_{44}+\frac{i}{2}\bar{v}_0(t)\mathcal{M}\hat{\tilde{\mathcal{G}}}_{j4}
+k\hat{\tilde{\mathcal{G}}}_{44},
\end{array}\right.
\ene\ees

For the given Eqs.~(\ref{psiglmg}) and (\ref{phiglmg}) we have the following proposition.

\vspace{0.1in}
\noindent {\bf Proposition 5.2.}
\bes\bee
& \label{f12e} \d \lim_{t'\to t}\d\int_{\partial D_1^0}\!\!\frac{ke^{4ik^2(t-t')}}{\Sigma_-}\!\left(\tilde{F}_{j4}e^{-2ikL}\right)_-dk\!=\!\!
\d\int_{\partial D_1^0}\!\!\left[\frac{i k}{2}u_0^T\left(\hat{\tilde{G}}_{44}\!-\!\bar{\hat{\tilde{\mathcal{G}}}}_{44}\right)\!+\!\frac{k}{\Sigma_-}
\left(\tilde{F}_{j4}e^{-2ikL}\right)_-\right]dk,\qquad \vspace{0.1in}\\
&\label{f21e} \d\lim_{t'\to t}\d\int_{\partial D_1^0}\frac{ke^{4ik^2(t-t')}}{\Sigma_-}\tilde{F}_{4j-}dk=
\d\int_{\partial D_1^0}\left[\frac{ik}{2}\mathcal{M}^T v_0^T\left(\hat{\tilde{\mathcal{G}}}_{44}-\bar{\hat{\tilde{G}}}_{44}\right)+\frac{k}{\Sigma_-}\tilde{F}_{4j-}\right]dk, \vspace{0.1in}\\
&\label{f12e+}
\d \lim_{t'\to t}\d\int_{\partial D_1^0}\frac{e^{4ik^2(t-t')}}{\Sigma_-}\left(\tilde{F}_{j4}e^{-2ikL}\right)_+dk=
\d\int_{\partial D_1^0}\frac{1}{\Sigma_-}\left(\tilde{F}_{j4}e^{-2ikL}\right)_+dk, \vspace{0.1in}\\
&\label{f21e+}
\d\lim_{t'\to t}\d\int_{\partial D_1^0}\frac{e^{4ik^2(t-t')}}{\Sigma_-}\tilde{F}_{4j+}dk=
\d\int_{\partial D_1^0}\frac{1}{\Sigma_-}\tilde{F}_{4j+}dk, \qquad\qquad\qquad\qquad
\ene
\ees
{\it where the vector-valued functions $\tilde{F}_{j4}(t,k)$ and $\tilde{F}_{4j}(t,k)\, (j=1,2,3)$ are defined by}
\bee
\label{f12eq}
\begin{array}{rl}
\tilde{F}_{j4}(t,k)= &\!\!\!\!  \d-\frac{i}{2}u_0^T\hat{\tilde{G}}_{44}+\frac{i}{2}\mathcal{M}^T\bar{\hat{\tilde{\mathcal{G}}}}_{3\times 3}^T\mathcal{M}v_0^Te^{2ikL}  \vspace{0.1in}\\
 &\!\!\!\! \d+ \left(\hat{\tilde{L}}_{j4}-\frac{i}{2}u_0^T\hat{\tilde{G}}_{44}+k\hat{\tilde{G}}_{j4}\right)
   \left(\bar{\hat{\tilde{\mathcal{L}}}}_{44}-\frac{i}{2}\bar{\hat{\tilde{\mathcal{G}}}}_{j4}^T\mathcal{M}v_0^T
   +k\bar{\hat{\tilde{\mathcal{G}}}}_{44}\right)  \vspace{0.1in}\\
  &\!\!\!\!  \d -\left(\hat{\tilde{L}}_{3\times 3}-\frac{i}{2}u_0^T\hat{\tilde{G}}_{4j}+k\hat{\tilde{G}}_{3\times 3}\right)\mathcal{M}^T
  \left(\bar{\hat{\tilde{\mathcal{L}}}}_{4j}^T-\frac{i}{2}\bar{\hat{\tilde{\mathcal{G}}}}_{3\times 3}^T\mathcal{M}v_0^T
  +k\bar{\hat{\tilde{\mathcal{G}}}}_{4j}^T\right)e^{2ikL},
\end{array}
\ene
\bee
\label{f21eq}
\begin{array}{rl}
\tilde{F}_{4j}(t,k)= &\!\!\!\!  \d -\frac{i}{2}\bar{\hat{\tilde{G}}}^T_{3\times 3}\mathcal{M}u_0^T+\frac{i}{2}\mathcal{M}^Tv_0^T\hat{\tilde{\mathcal{G}}}_{44}e^{2ikL} \vspace{0.1in}\\
 &\!\!\!\! \d+\mathcal{M}^T\left(\hat{\tilde{\mathcal{L}}}_{3\times 3}-\frac{i}{2}v_0^T\hat{\tilde{\mathcal{G}}}_{4j}+k\hat{\tilde{\mathcal{G}}}_{3\times 3}\right)\mathcal{M}^T
  \left(\bar{\hat{\tilde{L}}}^T_{4j}-\frac{i}{2}\bar{\hat{\tilde{G}}}^T_{3\times 3}\mathcal{M}u_0^T+k\bar{\hat{\tilde{G}}}^T_{4j}\right) \qquad  \vspace{0.1in}\\
  &\!\!\!\! \d -\mathcal{M}^T \left(\hat{\tilde{\mathcal{L}}}_{j4}-\frac{i}{2}v_0^T\hat{\tilde{\mathcal{G}}}_{44}+k\hat{\tilde{\mathcal{G}}}_{j4}\right)
   \left(\bar{\hat{\tilde{L}}}_{44}-\frac{i}{2}\bar{\hat{\tilde{G}}}^T_{j4}\mathcal{M}u_0^T+k\bar{\hat{\tilde{G}}}_{44}\right) e^{2ikL},  \end{array}
\ene

\vspace{0.1in}
\noindent{\bf Proof.} Similar to the proof of Lemma 4.3 in Ref.~\cite{m3}, we here show Eq.~(\ref{f12e}) in detail. We multiply Eq.~(\ref{f12eq}) by $\frac{k}{\Sigma_-}e^{4ik^2(t-t')}$ with $0<t'<t$ and integrate along along $\partial D_1^0$ with respect to $dk$ to yield
\bee\label{f12ana}
\begin{array}{l}
\d\int_{\partial D_1^0}\frac{k}{\Sigma_-}e^{4ik^2(t-t')}(\tilde{F}_{j4}e^{-2ikL})_-dk=
\d\int_{\partial D_1^0}\frac{i k}{2}e^{4ik^2(t-t')}u_0^T\hat{\tilde{G}}_{44}dk
-\d\int_{\partial D_1^0} k^3e^{4ik^2(t-t')}\hat{\tilde{G}}_{j4}\bar{\hat{\tilde{\mathcal{G}}}}_{44}dk  \vspace{0.1in}\\
\qquad-\d\int_{\partial D_1^0} k e^{4ik^2(t-t')}\left(\hat{\tilde{L}}_{j4}-\frac{i}{2}u_0^T\hat{\tilde{G}}_{44}\right)
\left(\bar{\hat{\tilde{\mathcal{L}}}}_{44}-\frac{i}{2}\bar{\hat{\tilde{\mathcal{G}}}}_{j4}^T\mathcal{M}v_0^T\right)dk \vspace{0.1in}\\
\qquad\d +\d\int_{\partial D_1^0} \frac{k^2\Sigma_+}{\Sigma_-}e^{4ik^2(t-t')}\left[\left(\hat{\tilde{L}}_{j4}
-\frac{i}{2}u_0^T\hat{\tilde{G}}_{44}\right)\bar{\hat{\tilde{\mathcal{G}}}}_{44}
+\hat{\tilde{G}}_{j4}\left(\bar{\hat{\tilde{\mathcal{L}}}}_{44}-\frac{i}{2}\bar{\hat{\tilde{\mathcal{G}}}}_{j4}\mathcal{M}v_0^T\right)\right]dk
    \vspace{0.1in}\\
\qquad -\d\int_{\partial D_1^0}\frac{2 k^2}{\Sigma_-}e^{4ik^2(t-t')}\left[\left(\hat{\tilde{L}}_{3\times 3}-\frac{i}{2}u_0^T\hat{\tilde{G}}_{4j}\right)\mathcal{M}^T\bar{\hat{\tilde{\mathcal{G}}}}_{4j}^T+
\hat{\tilde{G}}_{3\times 3}\mathcal{M}^T\left(\bar{\hat{\tilde{\mathcal{L}}}}_{4j}^T-\frac{i}{2}\bar{\hat{\tilde{\mathcal{G}}}}_{3\times 3}^T\mathcal{M}v_0^T\right)\right]dk,
\end{array}
\ene

To further analyse the above equation, the following identities are introduced
\bee\label{con1}
\d\int_{\partial D_1}ke^{4ik^2(t-t')}\hat{F}(t,k)dk
 =\left\{\begin{array}{l}\d \frac{\pi}{2} F(t, 2t'-t),\quad 0<t'<t, \vspace{0.1in}\\
                        \d\frac{\pi}{4}F(t, t),\quad  0<t'=t,
\end{array}\right.
\ene
and
\bee \label{con2}
\d\int_{\partial D_1^0}\frac{k^2}{\Sigma_-}e^{4ik^2(t-t')}\hat{F}(t,k)dk=
2\d\int_{\partial D_1^0}\frac{k^2}{\Sigma_-}\left[\int_0^{t'}e^{4ik^2(t-t')}\hat{F}(t,2\tau-t)d\tau-\frac{F(t, 2t'-t)}{4ik^2}\right]dk,
\ene
which also holds for the cases that $\frac{k^2}{\Sigma_-}$ is taken place by $\frac{k^2\Sigma_+}{\Sigma_-}$ or $k^2$.

It follows from the first integral on the RHS of Eq.~(\ref{f12ana}) and Eq.~(\ref{con1}) that we have
\bes\bee\label{inte1}
&\d\lim_{t'\to t}\d\int_{\partial D_1^0}\frac{i k}{2}e^{4ik^2(t-t')}u_0^T\hat{\tilde{G}}_{44}dk
=\lim_{t'\to t}\frac{i\pi}{2} u_0^T\tilde{G}_{22}(t, 2t'-t)
=\frac{i\pi}{4}u_0^T\tilde{G}_{44}(t, t),\\
& \label{inte2}
\d\lim_{t'\to t}\d\int_{\partial D_1^0}\frac{i k}{2}e^{4ik^2(t-t')}u_0^T\hat{\tilde{G}}_{44}dk
=\d\int_{\partial D_1^0}\frac{i k}{2}u_0^T\hat{\tilde{G}}_{44}dk
=\frac{i\pi}{8}u_0^T\tilde{G}_{44}(t,t),
\ene\ees

Therefore, we know that the first integral on the RHS of Eq.~(\ref{f12ana}) yields the following two terms
\bee
\d \lim_{t'\to t}\int_{\partial D_1^0}\frac{i k}{2}e^{4ik^2(t-t')}u_0^T\hat{\tilde{G}}_{44}dk
=\int_{\partial D_1^0}\frac{i k}{2}u_0^T\hat{\tilde{G}}_{44}dk\Big|_{(\ref{inte1})}
+\int_{\partial D_1^0}\frac{i k}{2}u_0^T\hat{\tilde{G}}_{44}dk\Big|_{(\ref{inte2})},
\ene

Nowadays we study the second integral on the RHS of Eq.~(\ref{f12ana}). It follows from the second integral on the RHS of Eq.~(\ref{f12ana}) and Eq.~(\ref{con2}) that we have
\bee \label{k3} \begin{array}{l}
-\d\int_{\partial D_1^0} k^3e^{4ik^2(t-t')}\hat{\tilde{G}}_{j4}\bar{\hat{\tilde{\mathcal{G}}}}_{44}dk
=-2\d\int_{\partial D_1^0} k^3\int_0^{t}e^{4ik^2(\tau-t')}\tilde{G}_{j4}(t, 2\tau-t)\bar{\hat{\tilde{\mathcal{G}}}}_{44}d\tau dk \\
\qquad\qquad =-2\d\int_{\partial D_1^0} k^3\left[\int_0^{t'}e^{4ik^2(\tau-t')}\tilde{G}_{j4}(t, 2\tau-t)d\tau-\frac{\tilde{G}_{j4}(t, 2t'-t)}{4ik^2}\right]\bar{\hat{\tilde{\mathcal{G}}}}_{44}dk,
\end{array}
\ene
Thus we take the limit $t'\to t$ of Eq.~(\ref{k3}) to have
\bee\no
-\lim_{t'\to t}\d\int_{\partial D_1^0} k^3e^{4ik^2(t-t')}\hat{\tilde{G}}_{j4}\bar{\hat{\tilde{\mathcal{G}}}}_{44}dk
=-\d\int_{\partial D_1^0} k^3\hat{\tilde{G}}_{j4}\bar{\hat{\tilde{\mathcal{G}}}}_{44}dk
+\d\int_{\partial D_1^0}\frac{k}{2i}u_0^T\bar{\hat{\tilde{\mathcal{G}}}}_{44}dk
\ene

Finally, following the proof in Ref.~\cite{m3} we can show the limits $t'\to t$ of the rest three integrals (i.e., the third, fourth and fifth integrals) of Eq.~(\ref{f12ana}) can be deduced by simply making the limit $t'\to t$ inside the every integral, that is, no additional terms
arise in these integrals. For example,
\bee \no
\lim_{t'\to t}\d\int_{\partial D_1^0} k e^{4ik^2(t-t')}\left(\hat{\tilde{L}}_{j4}-\frac{i}{2}u_0^T\hat{\tilde{G}}_{44}\right)
\left(\bar{\hat{\tilde{\mathcal{L}}}}_{44}-\frac{i}{2}\bar{\hat{\tilde{\mathcal{G}}}}_{j4}^T\mathcal{M}v_0^T\right)dk \\
=\d\int_{\partial D_1^0} k \left(\hat{\tilde{L}}_{j4}-\frac{i}{2}u_0^T\hat{\tilde{G}}_{44}\right)
\left(\bar{\hat{\tilde{\mathcal{L}}}}_{44}-\frac{i}{2}\bar{\hat{\tilde{\mathcal{G}}}}_{j4}^T\mathcal{M}v_0^T\right)dk.
\no\ene
Thus we complete the proof of Eq.~(\ref{f12e}). Similarly, we can show that Eqs.~(\ref{f21e}), (\ref{f12e+}) and (\ref{f21e+}) also hold.
$\square$

\vspace{0.1in}
\noindent {\bf Theorem  5.3.} {\it  \it Let $q_{0j}(x)=q_j(x,t=0)=0,\, j=1,2,3$ be the initial data of Eq.~(\ref{pnls}) on the interval $x\in [0, L]$ and  $T<\infty$. For the Dirichlet problem, the boundary data $u_{0j}(t)$ and $v_{0j}(t)\, (j=1,2,3)$ on the interval $t\in [0, T)$ are sufficiently smooth and compatible with the initial data $q_{j0}(x)\, (j=1,2,3)$ at the points $(x_2, t_2)=(0, 0)$ and  $(x_3, t_3)=(L, 0)$, respectively. For the Neumann problem, the boundary data $u_{1j}(t)$ and $v_{1j}(t)\, (j=1,2,3)$ on the interval $t\in [0, T)$ are sufficiently smooth and compatible with the initial data $q_{0j}(x)\, (j=1,2,3)$ at the points $(x_2, t_2)=(0, 0)$ and  $(x_3, t_3)=(L, 0)$, respectively. For simplicity, let $n_{33,44}(\mathbb{S})(k)$ have no zeros in the domain $D_1$. Then the spectral functions $S(k)$ and $S_L(k)$ are defined by Eqs.~(\ref{skm}) and (\ref{slm}) with $\Psi(t,k)$ and $\phi(t,k)$ given by Eq.~(\ref{psiglm}) and (\ref{phiglm}).

(i) For the given Dirichlet boundary values $u_0(t)$ and $v_0(t)$, the unknown Neumann boundary values $u_1(t)$ and $v_1(t)$ are given by
\bes
\bee
& \label{u1}
\begin{array}{rl}
u_1^T(t)=&\!\!\! \d\frac{4}{i\pi}\int_{\partial D_1^0}\left\{\frac{\Sigma_+}{\Sigma_-}\left[k^2\hat{\tilde{G}}_{j4}(t,t)+
\frac{i}{2}u_0^T(t)\right]
-\d\frac{2\mathcal{M}^T}{\Sigma_-}\left[k^2\bar{\hat{\tilde{\mathcal{G}}}}_{4j}^T(t, t)+\frac{i}{2}\mathcal{M}v_0^T(t)\right]\right. \vspace{0.1in}\\
&\left.+\d\frac{i k}{2}u_0^T\left(\hat{\tilde{G}}_{44}-\bar{\hat{\tilde{\mathcal{G}}}}_{44}\right)+\frac{k}{\Sigma_-}\left[\tilde{F}_{j4}e^{-2ikL}\right]_-\right\}dk,
\end{array} \vspace{0.1in}\\
&\label{v1}
\begin{array}{rl}
v_1^T(t)=&\!\!\! \d\frac{4}{i\pi}\int_{\partial D_1^0}\left\{-\frac{\Sigma_+}{\Sigma_-}\left[k^2\hat{\tilde{\mathcal{G}}}_{j4}(t,t)+\frac{i}{2}v_0^T(t)\right]
+\d\frac{2\mathcal{M}^T}{\Sigma_-}\left[k^2\bar{\hat{\tilde{G}}}_{4j}^T(t, t)+\frac{i}{2}\mathcal{M}u_0^T(t)\right]\right. \vspace{0.1in}\\
&\left.+\d\frac{i k}{2}v_0^T\left(\hat{\tilde{\mathcal{G}}}_{44}-\bar{\hat{\tilde{G}}}_{44}\right)
+\frac{k}{\Sigma_-}\mathcal{M}^T\tilde{F}_{4j-}\right\}dk,
\end{array}
\ene\ees

(ii) For the given Neumann boundary values $u_1(t)$ and $v_1(t)$, the unknown Dirichlet boundary values $u_0(t)$ and $v_0(t)$ are given by
\bes\bee
&\label{u0}
\begin{array}{rl}
u_0^T(t)=&\!\!\!\d\frac{2}{\pi}
\d\int_{\partial D_1^0}\left[\frac{\Sigma_+}{\Sigma_-}\hat{\tilde{L}}_{j4}-\frac{2\mathcal{M}^T}{\Sigma_-}\bar{\hat{\tilde{\mathcal{L}}}}_{4j}^T
+\frac{1}{\Sigma_-}\left(\tilde{F}_{j4}e^{-2ikL}\right)_+\right]dk,
\end{array}  \vspace{0.1in}\\
&\label{v0}
\begin{array}{rl}
v_0^T(t) = &\!\!\!\d\frac{2}{\pi}\int_{\partial D_1^0}\left[
\frac{2\mathcal{M}^T}{\Sigma_-}\bar{\hat{\tilde{L}}}^T_{j4}-\frac{1}{\Sigma_-}\hat{\tilde{\mathcal{L}}}_{4j}
+\frac{\mathcal{M}^T}{\Sigma_-}\tilde{F}_{4j+}\right]dk,\qquad\quad
\end{array}
\ene\ees
where $\tilde{F}_{j4}(t,k)$ and $\tilde{F}_{4j}(t,k)$ are defined by Eqs.~(\ref{f12eq}) and (\ref{f21eq}).}

\vspace{0.1in}
\noindent {\bf Proof.} By means of the global relation (\ref{gr}) and Proposition 5.1, we can show that the spectral functions $S(k)$ and $S_L(k)$ are defined by Eqs.~(\ref{skm}) and (\ref{slm}) with $\Psi(t,k)$ and $\phi(t,k)$ given by Eq.~(\ref{psiglm}) and (\ref{phiglm}).

 (i) we firstly consider the Dirichlet problem. It follows from the global relation (\ref{gr}) with the vanishing initial data
\bee
 c(t,k)=\mu_2(0, t,k)e^{ikL\hat{\sigma}_4}\mu_3^{-1}(L, t,k),
\ene
that we find
\bes\bee
\label{c12}
&\tilde{c}_{j4}(t,k)=-\tilde{\Psi}_{3\times 3}\mathcal{M}^T\bar{\tilde{\phi}}_{4j}^T(t,\bar{k})e^{2ikL}+\tilde{\Psi}_{j4}\bar{\tilde{\phi}}_{44}(t,\bar{k}),\qquad\quad
\vspace{0.1in}\\
\label{c21} &
\tilde{c}_{4j}(t,k)=\tilde{\Psi}_{4j}\mathcal{M}^T\bar{\tilde{\phi}}_{3\times 3}^T(t,\bar{k})\mathcal{M}^T
 -\tilde{\Psi}_{44}\bar{\tilde{\phi}}_{j4}^T(t,\bar{k})\mathcal{M}^Te^{-2ikL},
\ene\ees

Substituting Eqs.~(\ref{psiglmg}) and (\ref{phiglmg}) into Eq.~(\ref{c12}) yields
\bee\label{c12g}
\mathcal{M}^T\bar{\hat{\tilde{\mathcal{L}}}}_{4j}^Te^{2ikL}-\hat{\tilde{L}}_{j4}=k\hat{\tilde{G}}_{j4}
-k\mathcal{M}^T\bar{\hat{\tilde{\mathcal{G}}}}_{4j}^Te^{2ikL}+\tilde{F}_{j4}(t,k)-\tilde{c}_{j4}(t,k),
\ene
where $\tilde{F}_{j4}(t,k)$ is given by Eq.~(\ref{f12eq}). Eq.~(\ref{c12g}) with $k\to -k$ further yields
\bee\label{c12gk}
\mathcal{M}^T\bar{\hat{\tilde{\mathcal{L}}}}_{4j}^Te^{-2ikL}-\hat{\tilde{L}}_{j4}=-k\hat{\tilde{G}}_{j4}
+k\mathcal{M}^T\bar{\hat{\tilde{\mathcal{G}}}}_{4j}^Te^{-2ikL}+\tilde{F}_{j4}(t,-k)-\tilde{c}_{j4}(t,-k),
\ene

It follows from Eqs~(\ref{c12g}) and (\ref{c12gk}) that we  get
\bee
\label{c12g1}
\d\hat{\tilde{L}}_{j4}=\frac{k\Sigma_+}{\Sigma_-}\hat{\tilde{G}}_{j4}
-\frac{2 k}{\Sigma_-}\mathcal{M}^T\bar{\hat{\tilde{\mathcal{G}}}}_{4j}^T+\frac{1}{\Sigma_-}\left\{[\tilde{F}_{j4}(t, k)-\tilde{c}_{j4}(t, k)]e^{-2ikL}\right\}_-.
\ene

We multiply Eq.~(\ref{c12g1})  by $k e^{4ik^2(t-t')}$ with $0<t'<t$ and integrate them along $\partial D_1^0$ with respect to
$dk$, respectively to yield
\bee
\label{c12g22}
\begin{array}{rl}\d\int_{\partial D_1^0}k e^{4ik^2(t-t')}\hat{\tilde{L}}_{j4}dk=
&\!\!\!\d\int_{\partial D_1^0} e^{4ik^2(t-t')}\frac{k^2\Sigma_+}{\Sigma_-}\hat{\tilde{G}}_{j4}dk
-\d\int_{\partial D_1^0}e^{4ik^2(t-t')}\frac{2k^2}{\Sigma_-}\mathcal{M}^T\bar{\hat{\tilde{\mathcal{G}}}}_{4j}^Tdk \vspace{0.1in}\\
&\!\!\!+\d\int_{\partial D_1^0} \frac{ke^{4ik^2(t-t')}}{\Sigma_-}[\tilde{F}_{j4}(t,k)e^{-2ikL}]_-dk,
\end{array}
\ene
where we have used
\bee\no
\d\int_{\partial D_1^0} k e^{4ik^2(t-t')}\tilde{c}_{j4-}(t, k)dk
=\d\int_{\partial D_1^0}k e^{4ik^2(t-t')}(\tilde{c}_{j4}(t, k)e^{-2ikL})_-dk=0
\ene
in terms of their analytical properties in $D_1^0$.

Based on these conditions given by Eqs.~(\ref{con1}) and (\ref{con2}), Eq.~(\ref{c12g22}) can become
\bee
\label{l22}
\begin{array}{rl}
\d\frac{\pi}{2}\tilde{L}_{j4}(t, 2t'-t)=&\!\!\! 2\d\int_{\partial D_1^0}\frac{k^2\Sigma_+}{\Sigma_-}\left[\int_0^{t'}e^{4ik^2(t-t')}
\tilde{G}_{j4}(t,2\tau-t)d\tau-\frac{\tilde{G}_{j4}(t, 2t'-t)}{4ik^2}\right]dk \vspace{0.1in}\\
&-4\d\int_{\partial D_1^0}\frac{k^2\mathcal{M}^T}{\Sigma_-}\left[\int_0^{t'}e^{4ik^2(t-t')}
\bar{\tilde{\mathcal{G}}}_{4j}^T(t,2\tau-t)d\tau-\frac{\bar{\tilde{\mathcal{G}}}_{4j}^T(t, 2t'-t)}{4ik^2}\right]dk \vspace{0.1in}\\
&+\d\int_{\partial D_1^0}\frac{k}{\Sigma_-}e^{4ik^2(t-t')}[\tilde{F}_{j4}(t, k)e^{-2ikL}]_-dk,
\end{array}
\ene

We choose the limit $t'\to t$ of Eq.~(\ref{l22}) with the initial data (\ref{lgi}) and Proposition 5.2 to find
\bee\label{l22a}
\begin{array}{rl}
\d\frac{\pi}{2}\tilde{L}_{j4}(t, t)=&\!\!\! 2\d\lim_{t'\to t}\int_{\partial D_1^0}\frac{k^2\Sigma_+}{\Sigma_-}\left[\int_0^{t'}e^{4ik^2(t-t')}
\tilde{G}_{j4}(t,2\tau-t)d\tau-\frac{\tilde{G}_{j4}(t, 2t'-t)}{4ik^2}\right]dk \vspace{0.1in}\\
&-4\d\lim_{t'\to t}\int_{\partial D_1^0}\frac{k^2\mathcal{M}^T}{\Sigma_-}\left[\int_0^{t'}e^{4ik^2(t-t')}\bar{\tilde{\mathcal{G}}}_{4j}^T(t,2\tau-t)d\tau-\frac{\bar{\tilde{\mathcal{G}}}_{4j}^T(t, 2t'-t)}{4ik^2}\right]dk \vspace{0.1in}\\
&+\d\lim_{t'\to t}\int_{\partial D_1^0}\frac{k}{\Sigma_-}e^{4ik^2(t-t')}[\tilde{F}_{j4}(t, k)e^{-2ikL}]_-dk \vspace{0.1in}\\
=&\!\!\! \d\int_{\partial D_1^0}\left\{\frac{\Sigma_+}{\Sigma_-}\left[k^2\hat{\tilde{G}}_{j4}(t,t)+\frac{i}{2}\tilde{G}_{j4}(t, t)\right]
-\d\frac{2\mathcal{M}^T}{\Sigma_-}\left[k^2\bar{\hat{\tilde{\mathcal{G}}}}_{4j}^T(t, t)+\frac{i}{2}\bar{\tilde{\mathcal{G}}}_{4j}^T(t, t)\right]\right. \vspace{0.1in}\\
&\left.
+\d\frac{i k}{2}u_0^T\left(\hat{\tilde{G}}_{44}-\bar{\hat{\tilde{\mathcal{G}}}}_{44}\right)+\frac{k}{\Sigma_-}\left(\tilde{F}_{j4}e^{-2ikL}\right)_-\right\}dk,
\end{array}
\ene
Since the initial data (\ref{lgi}) are of the form
\bee \label{gl22b}
\tilde{L}_{j4}(t, t)=\frac{i}{2}u_1^T(t)=\frac{i}{2}(u_{11}(t), u_{12}(t), u_{13}(t))^T,
\ene
then we know that Eq.~(\ref{u1}) holds ny means of Eqs.~(\ref{l22a}) and (\ref{gl22b}).

To show Eq.~(\ref{v1}) we rewrite Eq.~(\ref{c21}) in the form
\bee \label{c21g}
\bar{\tilde{c}}^T_{4j}(t,\bar{k})=\mathcal{M}^T\tilde{\phi}_{3\times 3}\mathcal{M}^T\bar{\tilde{\Psi}}^T_{4j}(t,\bar{k})
 -\mathcal{M}^T\tilde{\phi}_{j4}\bar{\tilde{\Psi}}_{44}^T(t,\bar{k})e^{2ikL},
\ene

We substitute Eqs.~(\ref{psiglmg}) and (\ref{phiglmg}) into Eq.~(\ref{c21g}) to have
\bee\label{c21gg}
-\bar{\hat{\tilde{L}}}^T_{4j}+\mathcal{M}^T\hat{\tilde{\mathcal{L}}}_{j4}e^{2ikL}=k\bar{\hat{\tilde{G}}}^T_{4j}
-k\mathcal{M}^T\hat{\tilde{\mathcal{G}}}_{j4}e^{2ikL}+\tilde{F}_{4j}(t,k)-\bar{\tilde{c}}^T_{4j}(t,\bar{k}),
\ene
where $\tilde{F}_{4j}(t,k)$ is given  by Eq.~(\ref{f21eq}). Eq.~(\ref{c21gg}) with $k\to -k$ yields
\bee\label{c21gk}
-\bar{\hat{\tilde{L}}}^T_{4j}+\mathcal{M}^T\hat{\tilde{\mathcal{L}}}_{j4}e^{-2ikL}=-k\bar{\hat{\tilde{G}}}^T_{4j}
+k\mathcal{M}^T\hat{\tilde{\mathcal{G}}}_{j4}e^{-2ikL}+\tilde{F}_{4j}(t,-k)-\bar{\tilde{c}}^T_{4j}(t,-\bar{k}),
\ene

It follows from Eqs.~(\ref{c21gg}) and (\ref{c21gk}) that we have
\bee
\label{c21g1}
\d \mathcal{M}^T\hat{\tilde{\mathcal{L}}}_{j4}=\frac{2k}{\Sigma_-}\bar{\hat{\tilde{G}}}^T_{4j}
-\frac{k\Sigma_+}{\Sigma_-}\mathcal{M}^T\hat{\tilde{\mathcal{G}}}_{j4}+\frac{1}{\Sigma_-}[\tilde{F}_{4j}(t, k)-\bar{\tilde{c}}_{4j}^T(t, \bar{k})]_-
\ene

We multiply Eq.~(\ref{c21g1})  by $k e^{4ik^2(t-t')}$ with $0<t'<t$, integrate them along $\partial D_1^0$ with respect to
$dk$, and use these conditions given by Eqs.~(\ref{con1}) and (\ref{con2}) to yield
\bee
\label{2l2}
\begin{array}{rl}
\d\frac{\pi}{2}\mathcal{M}^T\tilde{\mathcal{L}}_{j4}(t, 2t'-t)=&\!\!\! -2 \d\int_{\partial D_1^0}\frac{k^2\Sigma_+}{\Sigma_-}\mathcal{M}^T\left[\int_0^{t'}e^{4ik^2(t-t')}
\tilde{\mathcal{G}}_{j4}(t,2\tau-t)d\tau-\frac{\tilde{\mathcal{G}}_{j4}(t, 2t'-t)}{4ik^2}\right]dk \vspace{0.1in}\\
&+4\d\int_{\partial D_1^0}\frac{ k^2}{\Sigma_-}\left[\int_0^{t'}e^{4ik^2(t-t')}\bar{\tilde{G}}_{4j}^T(t,2\tau-t)d\tau-\frac{\bar{\tilde{G}}_{4j}^T(t, 2t'-t)}{4ik^2}\right]dk \vspace{0.1in}\\
&+\d\int_{\partial D_1^0}\frac{k}{\Sigma_-}e^{4ik^2(t-t')}\tilde{F}_{4j-}(t, k)dk,
\end{array}
\ene
where we have used the relation
\bee\no
\d\int_{\partial D_1^0} \frac{k}{\Sigma_-}e^{4ik^2(t-t')}\bar{\tilde{c}}^T_{4j-}(t, \bar{k})dk=0
\ene
due to the analytical property of the integrand in $D_1^0$.

We consider the limit $t'\to t$ of Eq.~(\ref{2l2}) with the initial data (\ref{lgi}) and Proposition 5.2 to have
\bee\label{2l2a}
\begin{array}{rl}
\d\frac{\pi}{2}\mathcal{M}^T\tilde{\mathcal{L}}_{j4}(t, t)=&\!\!\! \d \int_{\partial D_1^0}\left\{-\frac{\Sigma_+}{\Sigma_-}\mathcal{M}^T\left[k^2\hat{\tilde{\mathcal{G}}}_{j4}(t,t)+\frac{i}{2}\tilde{\mathcal{G}}_{j4}(t, t)\right]
+\d\frac{2}{\Sigma_-}\left[k^2\bar{\hat{\tilde{G}}}_{4j}^T(t, t)+\frac{i}{2}\bar{\tilde{G}}_{4j}^T(t, t)\right]\right. \vspace{0.1in}\\
&\left.+\d\frac{ik}{2}\mathcal{M}^Tv_0^T\left(\hat{\tilde{\mathcal{G}}}_{44}-\bar{\hat{\tilde{G}}}_{44}\right)+\frac{k}{\Sigma_-}\tilde{F}_{4j-}(t,k)\right\}dk,
\end{array}
\ene

Since the initial conditions are of the form
\bee \label{2l2b}
\tilde{\mathcal{L}}_{j4}(t, t)=\frac{i}{2}v_1^T(t)=\frac{i}{2}(v_{11}(t), v_{12}(t), v_{13}(t))^T,
\ene
then we have Eq.~(\ref{v1}) by combining Eqs.~(\ref{2l2a}) and (\ref{2l2b}).

\vspace{0.1in}

 (ii) We now turn to consider the Neumann problem. It follows from Eqs~(\ref{c12g}), (\ref{c12gk}), (\ref{c21gg}) and (\ref{c21gk}) that we have
\bes\bee
&\label{c12g+}
\d\hat{\tilde{G}}_{j4}=\frac{1}{k\Sigma_-}\left\{\Sigma_+\hat{\tilde{L}}_{j4}-2\mathcal{M}^T \bar{\hat{\tilde{\mathcal{L}}}}_{4j}^T+\left[(\tilde{F}_{j4}(t, k)-\tilde{c}_{j4}(t, k))e^{-2ikL}\right]_+\right\}, \\
&\label{c21g+}
\d\hat{\tilde{\mathcal{G}}}_{j4}=\frac{1}{k\Sigma_-}
\left\{2\mathcal{M}^T\bar{\hat{\tilde{L}}}^T_{j4}-\Sigma_+\hat{\tilde{\mathcal{L}}}_{j4}+\mathcal{M}^T\left[\tilde{F}_{4j}(t, k)-\bar{\tilde{c}}^T_{44}(t, \bar{k})\right]_+\right\}.\qquad\quad
\ene\ees

We multiply Eqs.~(\ref{c12g+}) and (\ref{c21g+}) by $k e^{4ik^2(t-t')}$ with $0<t'<t$, integrate them along $\partial D_1^0$ with respect to
$dk$, and use these conditions given by Eqs.~(\ref{con1}) and (\ref{con2}) to yield
\bes\bee
&\label{g12+1}
\begin{array}{rl}
\d\frac{\pi}{2}\tilde{G}_{j4}(t, 2t'-t)=&\!\!\!
\d\int_{\partial D_1^0}\frac{2\Sigma_+}{\Sigma_-}\left[\int_0^{t'}e^{4ik^2(t-t')}
\tilde{L}_{j4}(t,2\tau-t)d\tau-\frac{\tilde{L}_{j4}(t, 2t'-t)}{4ik^2}\right]dk \vspace{0.1in}\\
&-\d\int_{\partial D_1^0}\frac{4\mathcal{M}^T}{\Sigma_-}\left[\int_0^{t'}e^{4ik^2(t-t')}\bar{\tilde{\mathcal{L}}}_{4j}^T(t,2\tau-t)d\tau
-\frac{\bar{\tilde{\mathcal{L}}}_{4j}^T(t, 2t'-t)}{4ik^2}\right]dk \vspace{0.1in}\\
&+\d\int_{\partial D_1^0}\frac{e^{4ik^2(t-t')}}{\Sigma_-}(\tilde{F}_{j4}e^{-2ikL})_+dk,
\end{array}  \\
&\label{g12+2}
\begin{array}{rl}
\d\frac{\pi}{2}\tilde{\mathcal{G}}_{j4}(t, 2t'-t)=&\!\!\!
\d\int_{\partial D_1^0}\frac{4\mathcal{M}^T}{\Sigma_-}\left[\int_0^{t'}e^{4ik^2(t-t')}
\bar{\tilde{L}}^T_{j4}(t,2\tau-t)d\tau-\frac{\bar{\tilde{L}}^T_{j4}(t, 2t'-t)}{4ik^2}\right]dk \vspace{0.1in}\\
&-\d\int_{\partial D_1^0}\frac{2 }{\Sigma_-}\left[\int_0^{t'}e^{4ik^2(t-t')}\tilde{\mathcal{L}}_{4j}(t,2\tau-t)d\tau-\frac{\tilde{\mathcal{L}}_{4j}(t, 2t'-t)}{4ik^2}\right]dk \vspace{0.1in}\\
&+\d\int_{\partial D_1^0}\frac{\mathcal{M}^T}{\Sigma_-}e^{4ik^2(t-t')}\tilde{F}_{4j+}dk,
\end{array}
\ene\ees
where we have used the analytical property of the matrix-valued functions
\bee\no
\d\int_{\partial D_1^0} \frac{1}{\Sigma_-}e^{4ik^2(t-t')}(\tilde{c}_{j4}(t, k)e^{-2ikL})_+dk=\d\int_{\partial D_1^0} \frac{1}{\Sigma_-}e^{4ik^2(t-t')}\bar{\tilde{c}}^T_{4j+}(t, \bar{k})dk=0.
\ene

We consider the limits $t'\to t$ of Eqs.~(\ref{g12+1}) and (\ref{g12+2}) with the initial data (\ref{lgi}) and Proposition 5.2 to find
\bes\bee
&\label{g12g1}
\begin{array}{rl}
\d\frac{\pi}{2}\tilde{G}_{j4}(t, t)=&\!\!\!
\d\int_{\partial D_1^0}\left[\frac{\Sigma_+}{\Sigma_-}\hat{\tilde{L}}_{j4}-\frac{2\mathcal{M}^T}{\Sigma_-}\bar{\hat{\tilde{\mathcal{L}}}}_{4j}^T
+\frac{1}{\Sigma_-}(\tilde{F}_{j4}e^{-2ikL})_+\right]dk,
\end{array}  \\
&\label{g12g2}
\begin{array}{rl}
\d\frac{\pi}{2}\tilde{\mathcal{G}}_{j4}(t, t)=&\!\!\!
\d\int_{\partial D_1^0}\left(\frac{2\mathcal{M}^T}{\Sigma_-}\bar{\hat{\tilde{L}}}^T_{j4}
-\frac{1}{\Sigma_-}\hat{\tilde{\mathcal{L}}}_{4j}+\frac{\mathcal{M}^T}{\Sigma_-}\tilde{F}_{4j+}\right)dk,
\end{array}
\ene\ees
Since the initial conditions are of the form
\bee \label{l22b}
\begin{array}{l}
\tilde{G}_{j4}(t, t)=u_0^T(t)=(u_{01}(t), u_{02}(t), u_{03}(t))^T,\quad
\tilde{\mathcal{G}}_{j4}(t, t)=v_0^T(t)=(v_{01}(t), v_{02}(t), v_{03}(t))^T,
\end{array}
\ene
then we have Eqs.~(\ref{u0}) and (\ref{v0}) by using Eqs.~(\ref{g12g1}) and (\ref{g12g2}). This completes the proof of the Theorem.
$\square$

\vspace{0.1in}
\noindent \subsection*{\it 5.2. Equivalence of the two distinct representations}
\vspace{0.1in}

\quad We now show that the above-mentioned  GLM  representation for the Dirichlet and Neumann boundary data in Theorem 5.3 is equivalent to one in Theorem 4.2.

\vspace{0.1in}
{\it Case i. From the Dirichlet boundary conditions to the Neumann boundary ones}
\vspace{0.1in}

It follows  from Eqs.~(\ref{psiglmg}) and (\ref{phiglmg}) that we obtain
\bee \label{mpsi}
\hat{\tilde{G}}_{j4}=\frac{1}{2k}\tilde{\Psi}_{j4-},\quad
\hat{\tilde{\mathcal{G}}}_{j4}=\frac{1}{2k}\tilde{\mathcal{\phi}}_{j4-},\quad
\hat{\tilde{G}}_{44}=\frac{1}{2k}\tilde{\Psi}_{44-},\quad
\hat{\tilde{\mathcal{G}}}_{44}=\frac{1}{2k}\tilde{\mathcal{\phi}}_{44-},
\ene

Substituting Eqs.~(\ref{f12eq}) and (\ref{mpsi}) into Eq.~(\ref{u1}) yields
\bee
& \label{u1x}
\begin{array}{rl}
u_1^T(t)=&\!\!\! \d\frac{4}{i\pi}\int_{\partial D_1^0}\left\{\frac{\Sigma_+}{\Sigma_-}\left[k^2\hat{\tilde{G}}_{j4}(t,t)+
\frac{i}{2}u_0^T(t)\right]
-\d\frac{2\mathcal{M}^T}{\Sigma_-}\left[k^2\bar{\hat{\tilde{\mathcal{G}}}}_{4j}^T(t, t)+\frac{i}{2}\mathcal{M}v_0^T(t)\right]\right. \vspace{0.1in}\\
&\left.+\d ik u_0^T\hat{\tilde{G}}_{44}+\frac{ k}{2i}u_0^T\bar{\hat{\tilde{\mathcal{G}}}}_{44}
 +\frac{k}{\Sigma_-}\left[\tilde{\Psi}_{j4}(\bar{\tilde{\phi}}_{44}-1)e^{-2ikL}
-(\tilde{\Psi}_{3\times 3}-\mathbb{I})\mathcal{M}^T\bar{\tilde{\phi}}_{4j}^T\right]_-\right\}dk \vspace{0.1in}\\
=&\!\!\! \d\int_{\partial D_1^0}\left\{\frac{2\Sigma_+}{i\pi\Sigma_-}\left[k\tilde{\Psi}_{j4-}+iu_0^T(t)\right]
+\d\frac{4i\mathcal{M}^T}{\pi\Sigma_-}\left[ k\bar{\tilde{\mathcal{\phi}}}^T_{4j-}+i\mathcal{M} v_0^T(t)\right]+\frac{1}{\pi}u_0^T(2\tilde{\Psi}_{44-}-\bar{\tilde{\phi}}_{44-})\right. \vspace{0.1in}\\
&\left.\d +\frac{4k}{i\pi\Sigma_-}\left[\tilde{\Psi}_{j4}(\bar{\tilde{\phi}}_{44}-1)e^{-2ikL}
-(\tilde{\Psi}_{3\times 3}-\mathbb{I})\mathcal{M}^T\bar{\tilde{\phi}}_{4j}^T\right]_- \right\}dk,
\end{array}
\ene
Since the integrand in Eq.~(\ref{u1x}) is an odd function about $k$, which makes sure that the contour $\partial D_1^0$ can be replaced by
$\partial D_3^0$, thus we can find the same Neumann boundary data $u_{1j}(t)\, (j=1,2,3)$ at $x=0$ given by Eqs.~(\ref{u11})-(\ref{u13}) from Eq.~(\ref{u1x}). Similarly, we can also find the Neumann boundary data $v_{1j}(t)\, (j=1,2,3)$ at $x=L$ given by Eq.~(\ref{v11}) from Eq.~(\ref{v1}).

\vspace{0.1in}
{\it Case ii. From the  Neumann  boundary conditions to the Dirichlet boundary ones}
\vspace{0.1in}

Eqs.~(\ref{psiglmg}) and (\ref{phiglmg}) imply that
\bee\label{ll}
\hat{\tilde{L}}_{j4}=\frac{1}{2}\tilde{\Psi}_{j4+}(t,k)+\frac{i}{2}u_0^T\hat{\tilde{G}}_{44},\quad
\bar{\hat{\tilde{\mathcal{L}}}}^T_{4j}=\frac{1}{2}\bar{\tilde{\phi}}^T_{4j+}(t,k)
 +\frac{i}{2}\bar{\hat{\tilde{\mathcal{G}}}}^T_{3\times 3}\mathcal{M}v_0^T,
\ene
The substitution of Eqs.~(\ref{ll}) and (\ref{f12eq}) into Eq.~(\ref{u0}) yields
\bee\label{u0x}
\begin{array}{rl}
u_0^T(t)=&\!\!\!\d\frac{2}{\pi}
\d\int_{\partial D_1^0}\left[\frac{\Sigma_+}{\Sigma_-}\hat{\tilde{L}}_{j4}-\frac{2\mathcal{M}^T}{\Sigma_-}\bar{\hat{\tilde{\mathcal{L}}}}_{4j}^T
+\frac{1}{\Sigma_-}\left(\tilde{F}_{j4}e^{-2ikL}\right)_+\right]dk \vspace{0.1in}\\
=&\!\!\!\!\!\!\d
\d\int_{\partial D_1^0}\!\!\left\{\frac{\Sigma_+}{\pi\Sigma_-}\tilde{\Psi}_{j4+}\!-\!\frac{2\mathcal{M}^T}{\pi\Sigma_-}\bar{\tilde{\phi}}_{4j+}^T \right. \d\left.\!+\!\frac{2}{\pi\Sigma_-}\!\left[
\tilde{\Psi}_{j4}(\bar{\tilde{\phi}}_{44}(t,\bar{k})\!-\!1)e^{-2ikL}\!-\!(\tilde{\Psi}_{3\times 3}\!-\!\mathbb{I})\mathcal{M}^T\bar{\tilde{\phi}}_{4j}^T
\right]_{\!+}\!\right\}\!dk,
\end{array}
\ene

Since the integrand in Eq.~(\ref{u0x}) is an odd function about $k$, which makes sure that the contour $\partial D_1^0$ can be replaced by
$\partial D_3^0$, thus Eq.~(\ref{u0x}) yields the Dirichlet boundary values $u_{0j}(t),\, j=1,2,3$ again.
Similarly, we can also deduce the Dirichlet boundary values $v_{0j}(t),\, j=1,2,3$ from Eq.~(\ref{v0}).

 \subsection*{\it 5.3. \, Linearizable boundary conditions for the GLM representation}

\quad In what follows we further explore the linearizable boundary conditions for the GLM representation given in Theorem 5.3.

\vspace{0.1in}
\noindent {\bf Proposition 5.4.} {\it Let $q_j(x, t=0)=q_{0j}(x),\, j=1,2,3$ be the initial conditions of the gtc-NLS equation (\ref{pnls}) on the interval $x\in [0, L]$, and one of the following boundary conditions, either

(i) the Dirichlet boundary conditions at $x=0, L$, $q_j(x=0,t)=u_{0j}(t)=0$ and $q_j(x=L,t)=v_{0j}(t)=0,\, j=1,2,3,$

or

(ii) the Robin boundary conditions $x=0, L$, $q_{jx}(x=0,t)-\chi q_j(x=0,t)=u_{1j}(t)-\chi u_{0j}(t)=0,\, j=1,2,3$ and $q_{jx}(x=L,t)-\vartheta q_j(x=L,t)=v_{1j}(t)-\vartheta v_{0j}(t)=0,\, j=1,2$, where $\chi$ and $\vartheta$ are both real parameters.

Then the eigenfunctions $\Psi(t,k)$ and $\phi(t,k)$ can be expressed as

(i) \bes\bee\label{psiglm1}
&\Psi(t,k)=\mathbb{I}+\left(\begin{array}{cc} \hat{\tilde{L}}_{3\times 3} & \hat{\tilde{L}}_{j4} \vspace{0.1in}\\
      \hat{\tilde{L}}_{4j} & \hat{\tilde{L}}_{44} \end{array}\right), \\
\label{phiglm1}
&\phi(t,k)=\mathbb{I}+\left(\begin{array}{cc} \hat{\tilde{\mathcal{L}}}_{3\times 3} & \hat{\tilde{\mathcal{L}}}_{j4} \vspace{0.1in}\\
      \hat{\tilde{\mathcal{L}}}_{4j} & \hat{\tilde{\mathcal{L}}}_{44} \end{array}\right),
\ene\ees
where the $4\times 4$ matrix-valued function $L(t, s)=(L_{ij})_{4\times 4}$  satisfies a reduced Goursat system
\bee \label{lge2}
\left\{\begin{array}{l}
\tilde{L}_{3\times 3t}+\tilde{L}_{3\times 3s}=iu_1^T(t)\tilde{L}_{4j}, \vspace{0.1in} \\
\tilde{L}_{j4t}-\tilde{L}_{j4s}=iu_1^T(t)\tilde{L}_{44}, \quad j=1,2,3, \vspace{0.1in}\\
\tilde{L}_{4jt}-\tilde{L}_{4js}=-i \bar{u}_1(t)\mathcal{M}\tilde{L}_{3\times 3}, \quad j=1,2,3,\vspace{0.1in}\\
\tilde{L}_{44t}+\tilde{L}_{44s}=-i\bar{u}_1(t)\mathcal{M}\tilde{L}_{j4},
\end{array}\right.
\ene
with the initial data (cf. Eq.~(\ref{lgi}))
\bee\label{lgi2}
\tilde{L}_{3\times 3}(t, -t)=0_{3\times 3},\quad \tilde{L}_{44}(t, -t)=0, \quad \tilde{L}_{j4}(t, t)=\frac{i}{2}u_1^T(t),\quad
\tilde{L}_{4j}(t, t)=-\frac{i}{2}\bar{u}_1(t)\mathcal{M},
\ene
Similarly, the $4\times 4$ matrix-valued function $\mathcal{L}(t, s)=(\mathcal{L}_{ij})_{4\times 4}$  satisfies
the analogous system (\ref{lge2}) with $u_1(t)$ replaced by $v_1(t)$.}

{\it (ii)
\bes\bee
\label{psiglm2}
&\Psi(t,k)=\d\mathbb{I}+\left(\begin{array}{cc} \hat{\tilde{L}}_{3\times 3} & \hat{\tilde{L}}_{j4} \vspace{0.1in}\\
      \hat{\tilde{L}}_{4j} & \hat{\tilde{L}}_{44} \end{array}\right)+\left(\begin{array}{cc} -\d\frac{i}{2}u_0^T(t)\hat{\tilde{G}}_{4j} & k\hat{\tilde{G}}_{j4} \vspace{0.1in}\\
 k\hat{\tilde{G}}_{4j} & \d\frac{i}{2}\bar{u}_0(t)\mathcal{M}\hat{\tilde{G}}_{j4}
 \end{array}
 \right),\\
\label{phiglm2}
&\phi(t,k)=\d\mathbb{I}+\left(\begin{array}{cc} \hat{\tilde{\mathcal{L}}}_{3\times 3} & \hat{\tilde{\mathcal{L}}}_{j4} \vspace{0.1in}\\
      \hat{\tilde{\mathcal{L}}}_{4j} & \hat{\tilde{\mathcal{L}}}_{44} \end{array}\right)+\left(\begin{array}{cc} -\d\frac{i}{2}v_0^T(t)\hat{\tilde{\mathcal{G}}}_{4j} & k\hat{\tilde{\mathcal{G}}}_{j4} \vspace{0.1in}\\
 k\hat{\tilde{\mathcal{G}}}_{4j} &\d \frac{i}{2}\bar{v}_0(t)\mathcal{M}\hat{\tilde{\mathcal{G}}}_{j4}
 \end{array}
 \right),
\ene\ees
where the $4\times 4$ matrix-valued functions $L(t, s)=(L_{ij})_{4\times 4}$ and $G(t, s)=(G_{ij})_{4\times 4}$  satisfy the reduced nonlinear
Goursat system
\bee \label{lge2g}
\label{linearl1}
\left\{\begin{array}{l}
\tilde{L}_{3\times 3t}+\tilde{L}_{3\times 3s}=i\chi u_0^T(t)\tilde{L}_{4j}
  +\d\frac{1}{2}\left[i\dot{u}_0^T(t)-u_0^T(t)\bar{u}_0(t)\mathcal{M}u_0^T(t)\right]\tilde{G}_{4j}, \vspace{0.1in} \\
\tilde{L}_{44t}+\tilde{L}_{44s}=-i\chi\bar{u}_0(t)\mathcal{M}\tilde{L}_{j4}
-\d\frac{1}{2}\left[i\dot{\bar{u}}_0(t)\mathcal{M}+\bar{u}_0(t)\mathcal{M}u^T_0(t)\bar{u}_0(t)\mathcal{M}\right]\tilde{G}_{j4}, \vspace{0.1in}\\
\tilde{L}_{j4t}-\tilde{L}_{j4s}=i\chi u_0^T(t)\tilde{L}_{44}, \vspace{0.1in}\\
\tilde{L}_{4jt}-\tilde{L}_{4js}=-i\chi\bar{u}_0(t)\mathcal{M}\tilde{L}_{3\times 3}, \vspace{0.1in}\\
\tilde{G}_{j4t}-\tilde{G}_{j4s}=2u_0^T(t)\tilde{L}_{44}, \vspace{0.1in}\\
\tilde{G}_{4jt}-\tilde{G}_{4js}=2 \bar{u}_0(t)\mathcal{M}\tilde{L}_{3\times 3},
\end{array}\right.
\ene
with the initial data (cf. Eq.~(\ref{lgi}))
\bee\label{lgi2g}
\left\{\begin{array}{l}
\tilde{L}_{3\times 3}(t, -t)=0_{3\times 3},\vspace{0.1in}\\ \tilde{L}_{44}(t, -t)=0, \vspace{0.1in}\\
\tilde{L}_{j4}(t, t)=\d\frac{i}{2}\chi u_0^T(t),\vspace{0.1in}\\
\tilde{L}_{4j}(t, t)=-\d\frac{i}{2}\chi \bar{u}_0(t)\mathcal{M},\vspace{0.1in}\\
\tilde{G}_{j4}(t, t)=u_0^T(t),\vspace{0.1in}\\
\tilde{G}_{4j}(t, t)=\bar{u}_0(t)\mathcal{M},
\end{array}\right.
\ene
Similarly, the $4\times 4$ matrix-valued functions $\mathcal{L}(t, s)=(\mathcal{L}_{ij})_{4\times 4}$ and $\mathcal{G}(t, s)=(\mathcal{G}_{ij})_{4\times 4}$  satisfy the similar system (\ref{lge2g}) with $u_0(t)$ and $\chi$ replaced by $v_0(t)$ and
$\vartheta$, respectively.}

\vspace{0.1in}
\noindent
{\bf Proof.} Let us show that the linearizable boundary data correspond to the special cases of Proposition 5.1.

Case (i) The Dirichlet zero boundary data $q_j(x=0,t)=u_{0j}(t)=0$. It follows from the second one of system (\ref{lge}) that $\tilde{G}_{ij}(t,s)$ satisfy
\bee\label{linearg1}
\left\{\begin{array}{l}
\tilde{G}_{3\times 3t}+\tilde{G}_{3\times 3s}=iu_1^T(t)\tilde{G}_{4j}, \vspace{0.1in}\\
\tilde{G}_{j4t}-\tilde{G}_{j4s}=iu_1^T(t)\tilde{G}_{44}, \vspace{0.1in}\\
\tilde{G}_{4jt}-\tilde{G}_{4js}=-i \bar{u}_1(t)\mathcal{M}\tilde{G}_{3\times 3}, \vspace{0.1in}\\
\tilde{G}_{44t}+\tilde{G}_{44s}=-i \bar{u}_1(t)\mathcal{M}\tilde{G}_{j4},
\end{array}\right.
\ene
with the initial data (cf. Eq.~(\ref{lgi}))
\bee
\tilde{G}_{3\times 3}(t, -t)=0_{3\times 3},\quad \tilde{G}_{44}(t, -t)=0, \quad \tilde{G}_{j4}(t, t)=0_{j4},\quad \tilde{G}_{4j}(t, t)=0_{4j},
\ene

Thus the unique solution of Eq.~(\ref{linearg1}) is trivial, that is, $\tilde{G}_{3\times 3}(t,s)=0,\, \tilde{G}_{4j}(t,s)=0,\, \tilde{G}_{j4}(t,s)=0,\, \tilde{G}_{44}(t,s)=0$ such that Eq.~(\ref{psiglm}) reduces to Eq.~(\ref{psiglm1}) and the condition (\ref{lge}) with (\ref{lgi}) becomes  (\ref{lge2}) with (\ref{lgi2}). Similarly, for the Dirichlet zero boundary data $q_j(x=L,t)=v_{0j}(t)=0,\, j=1,2,3$, we can also show  Eq.~(\ref{phiglm1}).

(ii) Consider the Robin boundary data  $q_{jx}(x=0,t)-\chi q_j(x=0,t)=u_{1j}(t)-\chi u_{0j}(t)=0,\,(j=1,2,3)$, that is, the Dirichlet and Neumann boundary data have the linear relation
\bee\label{dnc}
 u_1(t)=\chi u_0(t).
 \ene

We introduce a $4\times 4$  matrix
\bee\no
 Q(t, s)=2L(t, s)-i\chi\sigma_4G(t, s)
 \ene
by the linear combinations of $L$ and $G$ such that we have
\bee\label{qm}
\left\{\begin{array}{l}
\d\tilde{Q}_{3\times 3}(t,s)=2\tilde{L}_{3\times 3}(t,s)-i\chi\tilde{G}_{3\times 3}(t,s), \vspace{0.1in}\\
\d\tilde{Q}_{j4}(t,s)=2\tilde{L}_{j4}(t,s)-i\chi\tilde{G}_{j4}(t,s), \vspace{0.1in}\\
\d\tilde{Q}_{4j}(t,s)=2\tilde{L}_{4j}(t,s)+i\chi\tilde{G}_{4j}(t,s), \vspace{0.1in}\\
\d\tilde{Q}_{44}(t,s)=2\tilde{L}_{44}(t,s)+i\chi\tilde{G}_{44}(t,s),
\end{array}\right.
\ene

It follows from Eq.~(\ref{lge}) and (\ref{qm}) with Eq.~(\ref{dnc}) that $\tilde{Q}_{ij}(t,s),\, \tilde{G}_{ij}(t,s),\, i,j=1,2$ satisfy
\bee\label{linearg2}
\left\{\begin{array}{l}
\tilde{Q}_{3\times 3t}+\tilde{Q}_{3\times 3s}
=\d\left[i\dot{u}_0^T(t)-u_0^T(t)\bar{u}_0(t)\mathcal{M}u_0^T(t)+\chi^2u_0^T(t)\right]\tilde{G}_{4j}, \vspace{0.1in}\\
\tilde{Q}_{j4t}-\tilde{Q}_{j4s}=\d\left[i\dot{u}_0^T(t)-u_0^T(t)\bar{u}_0(t)\mathcal{M}u_0^T(t)+\chi^2u_0^T(t)\right]\tilde{G}_{44}, \vspace{0.1in}\\
\tilde{Q}_{4jt}-\tilde{Q}_{4js}=\d\left[-\bar{u}_0(t)\mathcal{M}u_0^T(t)\bar{u}_0(t)\mathcal{M}-i\dot{\bar{u}}_0(t)\mathcal{M}
+\chi^2\bar{u}_0(t)\mathcal{M}\right]\tilde{G}_{3\times 3}, \vspace{0.1in}\\
\tilde{Q}_{44t}+\tilde{Q}_{44s}=\d\left[-\bar{u}_0(t)\mathcal{M}u_0^T(t)\bar{u}_0(t)\mathcal{M}-i\dot{\bar{u}}_0(t)\mathcal{M}
+\chi^2\bar{u}_0(t)\mathcal{M}\right]\tilde{G}_{j4}, \vspace{0.1in}\\
\tilde{G}_{3\times 3t}+\tilde{G}_{3\times 3s}=u_0^T(t) \tilde{Q}_{4j}, \vspace{0.1in} \\
\tilde{G}_{j4t}-\tilde{G}_{j4s}=u_0^T(t) \tilde{Q}_{44}, \vspace{0.1in}\\
\tilde{G}_{4jt}-\tilde{G}_{4js}= \bar{u}_0(t)\mathcal{M} \tilde{Q}_{3\times 3}, \vspace{0.1in}\\
\tilde{G}_{44t}+\tilde{G}_{44s}= \bar{u}_0(t)\mathcal{M} \tilde{Q}_{j4},
\end{array}\right.
\ene
with the initial data (cf. Eq.~(\ref{lgi}))
\bee\left\{\begin{array}{l}
\tilde{G}_{3\times 3}(t, -t)=0_{3\times 3},\quad \tilde{G}_{44}(t, -t)=0,\quad
\tilde{G}_{j4}(t, t)=u_0^T(t),\quad \tilde{G}_{4j}(t, t)=\bar{u}_0(t)\mathcal{M},\vspace{0.1in}\\
\tilde{Q}_{3\times 3}(t, -t)=0_{3\times 3},\quad \tilde{Q}_{44}(t, -t)=0,\quad
\tilde{Q}_{j4}(t, t)=0_{j4}, \quad \tilde{Q}_{4j}(t, t)=0_{4j},

\end{array}\right.
\ene

Thus the unique solution of Eq.~(\ref{linearg2}) is trivial, that is, $\tilde{Q}_{j4}(t,s)=\tilde{Q}_{4j}(t,s)
=\tilde{G}_{3\times 3}(t,s)=\tilde{G}_{44}(t,s)=0$  such that Eq.~(\ref{psiglm}) reduces to Eq.~(\ref{psiglm2}) and the condition
(\ref{lge}) with Eq.~(\ref{lgi}) becomes Eq.~(\ref{lge2g}) with Eq.~(\ref{lgi2g}). Similarly, for the Robin boundary data  $q_{jx}(x=L,t)-\vartheta q_j(x=L,t)=v_{1j}(t)-\vartheta v_{0j}(t)=0,\, j=1,2,3$, that is, $v_1(t)=\vartheta v_0(t)$, we can also show  Eq.~(\ref{phiglm2}). $\square$ \\

Based on the Theorem 5.3 and Proposition 5.4, we have the following Proposition.

\vspace{0.1in}
\noindent{\bf Proposition 5.5} {\it For the linearizable Dirichlet boundary data $u_0(t)=v_0(t)=0$, we have the Neumann boundary data $u_1(t)$ and $v_1(t)$:
\bee
 u_1^T(t)=\d\frac{4i}{\pi}\d\int_{\partial D_1^0}k\tilde{\Psi}_{j4}(\bar{\tilde{\phi}}_{44}-\mathbb{I})dk,\quad v_1^T(t)=\d\frac{4i}{\pi}\d\int_{\partial D_1^0}k\tilde{\phi}_{j4}(\bar{\tilde{\Psi}}_{44}-\mathbb{I})dk,
\ene
where
\bee\no
\left\{\begin{array}{l}
\tilde{\Psi}_{j4t}+4ik^2\tilde{\Psi}_{j4}=iu_1^T(t)(\tilde{\Psi}_{44}+\mathbb{I}), \vspace{0.1in}\\
\tilde{\Psi}_{44t}=-i\bar{u}_1(t)\mathcal{M}\tilde{\Psi}_{j4}, \vspace{0.1in}\\
\tilde{\phi}_{j4t}+4ik^2\tilde{\phi}_{j4}=iv_1^T(t)(\tilde{\phi}_{44}+\mathbb{I}), \vspace{0.1in}\\
\tilde{\phi}_{44t}=-i\bar{v}_1(t)\mathcal{M}\tilde{\phi}_{j4}.
\end{array}\right.
\ene }

\noindent {\bf Remark 5.6}. The analogous analysis of the Fokas unified method will use also to explore the IBV problems for other integrable nonlinear evolution PDEs with $4\times 4$ Lax pairs both on the the half-line and the finite interval, such as the three-component derivative-NLS equation and  the three-component higher-order NLS equation, which will be considered in other papers.

\vspace{0.2in}
\noindent
{\bf Acknowledgments}
\vspace{0.1in}

 This work was partially supported by the NSFC under Grant No.11571346 and the Youth Innovation Promotion Association, CAS.

\end{document}